\documentclass[11pt,a4paper]{article}
\usepackage{jheppub}
\usepackage{graphicx} 
\usepackage{bbold}
\usepackage[normalem]{ulem}
\newcommand{\be}{\begin{equation}}
\newcommand{\ee}{\end{equation}}
\newcommand{\bea}{\begin{eqnarray}}
\newcommand{\eea}{\end{eqnarray}}
\newcommand{\tr}{\mathrm{Tr}}

\usepackage{comment}
\usepackage{float}
\usepackage{soul}

\title{A bulk manifestation of Krylov complexity}

\author[a]{E. Rabinovici,} 

\author[b]{A. S\'{a}nchez-Garrido,}

\author[c]{R. Shir}

\author[b]{and J. Sonner}

\affiliation[a]{Racah Institute of Physics, The Hebrew University, Jerusalem 9190401, Israel}

\affiliation[b]{Department of Theoretical Physics, University of Geneva, 24 quai Ernest-Ansermet, 1214 Gen\`eve 4, Switzerland} 

\affiliation[c]{Department of Physics and Materials Science, University of Luxembourg, L-1511 Luxembourg}

\emailAdd{eliezer@mail.huji.ac.il}
\emailAdd{Adrian.SanchezGarrido@unige.ch}
\emailAdd{ruth.shir@uni.lu}
\emailAdd{Julian.Sonner@unige.ch}

\abstract{There are various definitions of the concept of complexity in Quantum Field Theory as well as for finite quantum systems. For several of them there are conjectured holographic bulk duals. In this work we establish an entry in the AdS/CFT dictionary for one such class of complexity, namely Krylov or K-complexity.
For this purpose we work in the double-scaled SYK model which is dual in a certain limit to JT gravity, a theory of gravity in AdS$_2$.  In particular, states on the boundary have a clear geometrical definition in the bulk.  We use this result to show that Krylov complexity of the infinite-temperature thermofield double state on the boundary of AdS$_2$ has a precise bulk description in JT gravity, namely the length of the two-sided wormhole. We do this by showing that the Krylov basis elements, which are eigenstates of the Krylov complexity operator, are mapped to length eigenstates in the bulk theory by subjecting K-complexity to the bulk-boundary map identifying the bulk/boundary Hilbert spaces. Our result makes extensive use of chord diagram techniques and identifies the Krylov basis of the boundary quantum system with fixed chord number states building the bulk gravitational Hilbert space.}

\begin{document}

\maketitle

\section{Introduction}
Notions of quantum complexity are of interest not only in the field of quantum computing but also in the field of quantum gravity, via the framework of the holographic duality or, more specifically, the AdS/CFT correspondence \cite{Maldacena:1997re, Witten:1998qj}. Any notion of quantum complexity is expected to have a very specific time evolution profile and to evolve for very long time scales (exponential in the number of degrees of freedom). Similar time evolution profiles have been found for certain diffeomorphism-invariant quantities of gravity in AdS space. In the spirit of the AdS/CFT correspondence, a conjecture has emerged \cite{Susskind:2014rva, Stanford:2014jda, Brown:2015bva}, namely that quantum complexity is dual (in the sense of the holographic duality) to diffeomorphism-invariant quantities such as the volume of the wormhole in two-sided AdS space.  

One particular definition of complexity in quantum systems is circuit complexity. Circuit complexity of a quantum state is defined as the minimal number of local unitary operations (known as ``gates") a quantum computer would need to apply to some reference state in order to create that state, up to some chosen precision set by a tolerance parameter. Analogously, circuit complexity can be defined for the operators themselves.  The complexity of a given operator would be the minimal number of local unitary operators whose composition equals the given operator to within the chosen tolerance parameter.

In a chaotic system the circuit complexity of a state evolves in the following manner: up to time scales of order the Hilbert space dimension, complexity grows, including a long period of linear growth (this was proven in \cite{Haferkamp:2021uxo}), and then, for a finite Hilbert space, complexity saturates at a value of order the Hilbert space dimension. After saturation, for a finite Hilbert space, at even larger time scales the value of complexity is expected to undergo endless Sisyphean Poincar\'e recurrences visiting again and again initial complexity values. 

A concrete notion of complexity which requires neither the introduction of a particular set of local unitary operations available to a quantum computer nor a choice of tolerance parameter, and under which the time evolution of complexity (for states evolving under a chaotic Hamiltonian) resembles the time-dependent profile described above, would be a candidate for holographic complexity. Such a candidate should be naturally defined on the quantum mechanical boundary as well as in the quantum gravity bulk, preferably relating the complexity evolution of a state on the boundary to some invariant observable's time evolution in the bulk.  

Krylov complexity, first defined in \cite{Parker:2018yvk} for operators evolving in time in the thermodynamic limit and later shown to satisfy the expected profile for operator quantum complexity at all time scales in \cite{Barbon:2019wsy, Rabinovici:2020ryf} in a finite system, is naturally defined for any quantum system without external (human) input. Originally defined for an operator evolving in time under a Hamiltonian in the Heisenberg picture, it was then defined also for states evolving under a Hamiltonian according to the Schr\"odinger equation \cite{Balasubramanian:2022tpr}. By its definition, Krylov complexity is bounded from above by the state/operator Hilbert space dimension and indeed can be seen to saturate in \cite{Rabinovici:2020ryf} in the case of operators, while for infinite Hilbert spaces it does not saturate in general.  

There have been several suggestions for the entry in the AdS/CFT dictionary for quantum complexity. 
In this work we find a precise one-to-one correspondence between Krylov complexity in a 1-dimensional quantum mechanical boundary and a specific bulk observable in a 2-dimensional quantum gravity theory.  
In particular, Krylov complexity of the thermofield double state (at infinite temperature) in the boundary theory, which is a particular limit of double-scaled SYK,
becomes equivalent to wormhole length in the bulk theory, which is 2-sided AdS$_2$ Jackiw-Teitelboim gravity \cite{Jackiw:1984je, Teitelboim:1983ux}. 

Previous works on the subject include \cite{Jian:2020qpp} which computed operator growth in SYK via Krylov complexity and compared it with the complexity=volume formulation in JT gravity; it was shown that the two have similar qualitative behaviours in time. The work \cite{Kar:2021nbm} studied the expected behaviour of K-complexity in low-dimensional gravity theories via relationships with random matrix theory.

This paper is organised as follows: in section \ref{Sec:Background} we review all the needed background for double-scaled SYK (DSSYK) and for Jackiw-Teitelboim (JT) gravity, and how they are related. In section \ref{Sect:KC} we quickly review the Krylov construction for states, introducing the Lanczos coefficients and the Krylov basis as well as the definition of state Krylov complexity (K-complexity).  We then identify the Lanczos coefficients and Krylov basis in DSSYK and discuss the results for Krylov complexity in DSSYK. 
In section \ref{Section_Gravity_matching} we present the main results of this work: we study Krylov complexity in the regime of DSSYK in which it is dual to JT gravity, and establish the correspondence of this quantity with the bulk wormhole length, adequately matching the parameters on either side of the duality.
Finally, in section \ref{Sec:Discussion} we summarize our results and discuss their interpretation and their impact on the understanding of higher dimensional cases.

\section{Background} \label{Sec:Background}
In this section we review the current state of knowledge for the double-scaled Sachdev-Ye-Kitaev (DSSYK) model and chord diagrams \cite{Berkooz:2018jqr, Berkooz:2018qkz}, the phase space of JT gravity \cite{Harlow:2018tqv}, and the equivalence between DSSYK (at a specific limit) and JT gravity \cite{Lin:2022rbf}. While the material in this section largely reviews results from the literature, it exposes the main ingredients necessary to match boundary Krylov complexity to its bulk counterpart. 

\subsection{Double-scaled SYK}\label{Subsection_Background_DSSYK}

The Sachdev-Ye-Kitaev (SYK) model \cite{Sachdev_1993, Kitaev_KITP_talk} is a strongly interacting, many-body quantum system described by all-to-all interactions between Majorana fermions $\psi_i$ with $1\leq i \leq N$ which satisfy $\{\psi_i,\psi_j\}=2\delta_{ij}$, grouped into $p$-body interactions $\psi_{i_1} \psi_{i_2} \dots \psi_{i_p}$ through random couplings $J_{i_1 i_2\dots i_p}$
in the form of the Hamiltonian
\begin{equation} \label{HSYK}
    H_{SYK} = i^{p/2} \sum_{1\leq i_1 < i_2 <\dots < i_p \leq N}\, J_{i_1 i_2\dots i_p} \, \psi_{i_1} \psi_{i_2} \dots \psi_{i_p} ~.
\end{equation}
The random couplings $J_{i_1 i_2\dots i_p}$ are usually taken from a Gaussian distribution with zero mean $\langle J_{i_1 i_2\dots i_p} \rangle = 0$ and an appropriate non-zero variance. The model (\ref{HSYK}) has been extensively studied for fixed $p$ and $N$, and at large $N$, as well as in the large $p$ and large $N$ limit (see e.g. \cite{Polchinski:2016xgd, Maldacena:2016hyu,  Cotler:2016fpe, Garcia-Garcia:2017pzl}).  
We will follow \cite{Berkooz:2018jqr} and take both $N$ and $p$ to infinity while keeping the ratio $2p^2/N$ fixed, a limit known as double-scaled SYK (or DSSYK), introducing the ratio parameter $\lambda$:
\begin{equation}
\label{lambda_def}
    \lambda \equiv \frac{2p^2}{N} ~,
\end{equation}
and we will take the variance $\langle J^2_{i_1 i_2\dots i_p} \rangle$ to be
\begin{equation}
\label{Variance}
    \langle J^2_{i_1 i_2\dots i_p} \rangle = \frac{1}{\lambda} \binom{N}{p}^{-1} J^2 ~.
\end{equation}
As discussed in appendix \ref{App:coupling_variance}, double-scaled SYK with this choice of coupling variance is, for small values of $\lambda$, compatible with the results for SYK where $N$ and $p$ are large and independent respecting the hierarchy $1\ll p \ll N$, which was studied in \cite{Maldacena:2016hyu}.

\begin{figure}
    \centering
    \includegraphics[scale=0.5]{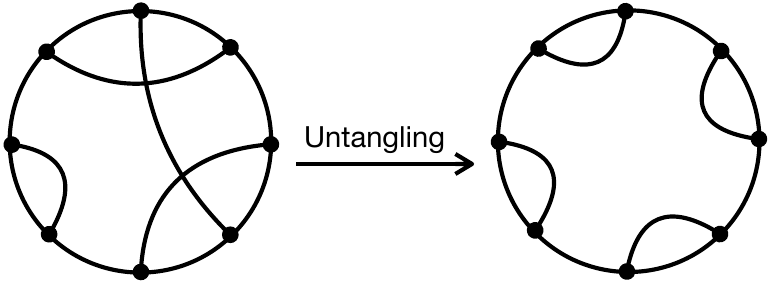}
    \caption{Untangling a chord diagram.}
    \label{fig:Chord_diagrams}
\end{figure}

\begin{figure}
    \centering
    \includegraphics[scale=0.5]{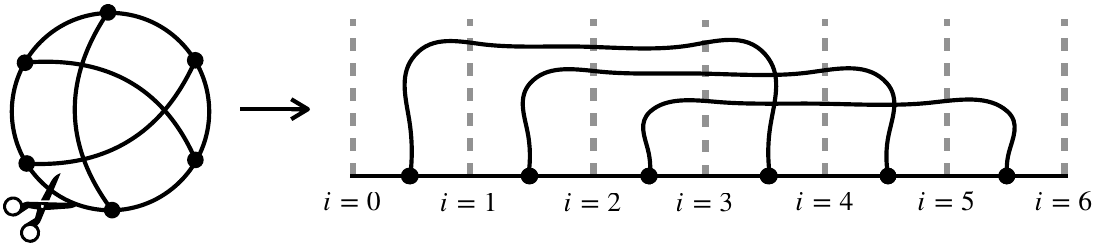}
    \caption{``Cutting open" a chord diagram \cite{Berkooz:2018jqr}. The cut-open chord diagram can be thought of as a process in which one starts with zero open chords and ends with zero open chords, while chords are created and annihilated in between.}
    \label{fig:cut_chord_diagram}
\end{figure}

In \cite{Berkooz:2018jqr} it was shown that in the limit where $N$ and $p$ go to infinity while keeping $\lambda$ fixed, the \textit{ensemble averaged} moments 
\begin{equation}
    \label{Moments_definition}
    M_{2k} \equiv \langle \, \tr (H^{2k}) \, \rangle
\end{equation}
of the Hamiltonian (\ref{HSYK}) are given by a sum over \textit{chord diagrams}. Chord diagrams represent pairwise Wick contractions among the couplings $J_{i_1...i_p}$ of the different Hamiltonians in the product inside the trace (\ref{Moments_definition}), which make the collective indices of the corresponding monomials $\psi_I \equiv \psi_{i_1} \psi_{i_2} \dots \psi_{i_p}$ coincide. For the $2k$-th moment, there are $2k$ paired monomials in each trace. 
To find the value of such a chord diagram, it needs to be ``untangled" to a configuration where all contracted pairs are sitting one beside the other (see figure \ref{fig:Chord_diagrams}). 
Once this is achieved, using the fact that $\psi_I^2 = i^p$ for $p$ even (it will be assumed that $p$ is even), one obtains a total phase of $i^{kp}$ coming from all the monomials squared, which exactly cancels the overall phase coming from the $i^{p/2}$ factor in the normalization of each Hamiltonian (\ref{HSYK}). Additionally, the system size dependence coming from the summation over the indices is mitigated by the binomial in the variance normalization (\ref{Variance}), yielding bounded results in the double-scaling limit. Therefore, the contribution of a given Wick contraction (i.e. a given chord diagram) boils down to the overall sign obtained after commuting the Majorana monomials through each other inside the trace until those with coincident indices are consecutive. In the double-scaling limit, it is found \cite{Berkooz:2018jqr} that the overall effect of such a commutation is captured by a multiplicative factor $q\equiv e^{-\lambda}$. The number of such commutations required to untangle a chord diagram is exactly the number of intersections of chords, and hence one is left with a simple rule for evaluating moments via chord diagrams:
\begin{equation}
     M_{2k} = \frac{J^{2k}}{\lambda^k} \sum_{\substack{\text{chord diagrams} \\ \text{ with $k$ chords}}} q^{\text{number of intersections}} ~.
\end{equation}
To evaluate the RHS, it is useful to ``cut open" each chord diagram and think of the process of constructing it as a transition from a state $|0\rangle$ with zero chords back to a state with zero chords, through a process of creating $k$ chords and eventually annihilating them all. An auxiliary Hilbert space $\{| 0 \rangle, | 1 \rangle, | 2 \rangle, \dots , | n \rangle \}$, which represents states with $0,1,2, \dots, n$ open chords, can now be introduced. An open chord is a chord that emanated from a given Hamiltonian insertion but which has not yet been closed by reaching another Hamiltonian insertion: Figure \ref{fig:cut_chord_diagram} depicts a chord diagram of six Hamiltonian insertions interpreted as a transition from zero open chords back to zero open chords through six intermediary steps (Hamiltonian insertions); if we look, for instance, at the step $i=4$, we note that there are two open chords. In the transition from $| 0 \rangle$ back to $| 0 \rangle$, chords can be `created' $|l\rangle \mapsto |l+1\rangle $ and `annihilated' $|l\rangle \mapsto |l-1\rangle $ at each step. 

All possible chord diagrams with $i$ steps (i.e. $i$ Hamiltonian insertions) are represented by a state $|\psi^{(i)}\rangle$ in the Hilbert space. We define such a state through its coordinates over the chord basis,
\begin{equation}
    \centering
    \label{State_psi_chord_basis}
    |\psi^{(i)}\rangle = \sum_{l\geq 0} \psi^{(i)}_l | l \rangle ~.
\end{equation}
The coordinates $\psi^{(i)}_l$ are defined to be the sum over all chord diagrams of $i$ steps with $l$ open chords, where each diagram is weighted by $q$ to the power of the number of intersections featured in it; that is, all diagrams with $i$ steps and $l$ open chords are represented by a state in the Hilbert space which is proportional to $|l\rangle$, with projection $\psi_{l}^{(i)}$, such that the sum of all diagrams of a fixed number of steps is represented by the linear combination $|\psi^{(i)}\rangle$ given in (\ref{State_psi_chord_basis}). The coordinates $\psi_l^{(i)}$ satisfy the following recurrence relation:
\begin{equation}\label{chord_recursion}
    \psi_l^{(i+1)} = \frac{J}{\sqrt{\lambda}}\psi_{l-1}^{(i)} +  \frac{J}{\sqrt{\lambda}}(1+q+\dots +q^{l})\psi_{l+1}^{(i)} ~,
\end{equation}
where the first term represents the possibility that a chord was created in the transition from step $i$ to step $i+1$, and the second term represents the possibility that a chord was annihilated in that transition. In the case in which the chord is annihilated at step $i+1$, there must have been $l+1$ open chords at step $i$ so that we are left with $l$ open chords at step $i+1$, and therefore the chord, in getting annihilated, can intersect up to $l$ other open chords, and in each case the diagram would get multiplied by $q$ to the number of such intersections. The term $1+q+...+q^l$ in (\ref{chord_recursion}) accounts therefore for all the possible ways to close a chord in the step $i+1$. 
This recursion (\ref{chord_recursion}) can be written concisely in a basis-independent way as

\begin{equation}
    |\psi^{(i+1)}\rangle = T |\psi^{(i)}\rangle ~,
\end{equation}
where, in coordinates over the $\{|l\rangle\}$ basis, the \textit{transfer matrix} $T$ takes the form:
\begin{equation} \label{T_non_sym}
    T \overset{*}{=} \frac{J}{\sqrt{\lambda}} \begin{pmatrix}
        0 & \frac{1-q}{1-q} & 0 & 0 & \dots  \\
        1 & 0 & \frac{1-q^2}{1-q} & 0 &  \dots  \\
        0 & 1 & 0 & \frac{1-q^3}{1-q} &  \dots \\
        0 & 0 & 1 & 0 &  \dots  \\
        \vdots & \vdots & \vdots & \vdots &  \ddots  \\
    \end{pmatrix},
\end{equation}
where the asterisk above the equality sign denotes that this is an expression in coordinates over a particular basis. 
At the first step $i=0$ there are no open chords and therefore $|\psi^{(0)}\rangle =|0\rangle$. Thus,
\begin{equation}
    \centering
    \label{psi_i_T_0}
    |\psi^{(i)}\rangle = T^i|0\rangle~.
\end{equation}
Since at each step we either create or annihilate a chord, to make a chord diagram with exactly $k$ chords we would need to apply the operator $T$ $2k$ times, creating $k$ chords and annihilating $k$ chords. The final state is $|\psi^{(2k)} \rangle= T^{2k}|0\rangle$, and thus
\begin{equation} \label{Moments_effective_Ham}
    M_{2k} = \langle 0 |\, T^{2k}\, |0\rangle = \langle 0 | \psi^{(2k)} \rangle = \psi^{(2k)}_0 ~.
\end{equation}

In view of this, $T$ can be interpreted as an effective Hamiltonian of DSSYK. In \eqref{T_non_sym} we give its expression in coordinates over the chord basis $\{ |l\rangle \}_{l=0}^\infty$, where it is represented by a non-symmetric matrix. In \cite{Berkooz:2018qkz, Berkooz:2018jqr} it was pointed out that there exists a diagonal similarity transformation that brings $T$ into a tridiagonal, symmetric form. Furthermore, in \cite{Lin:2022rbf} this transformation is explained in terms of the chord inner product defined in that article (and which was necessary to define in order to have an actual Hilbert space): according to such an inner product, the states $\{|l\rangle\}$ are orthogonal to each other, but they are not normalized (except for $|0\rangle$). The similarity transformation amounts to re-normalizing these states so that they form an orthonormal basis.
By a slight abuse of notation, from now on $\{|l\rangle \}$ will denote the \textit{normalized} fixed chord number states, i.e. $\langle l | l^\prime \rangle = \delta_{l,l^\prime}$. In this orthonormal basis, $T$ takes the form \cite{Berkooz:2018jqr,Lin:2022rbf}:
\begin{equation}
\label{T_sym}
 T    \overset{*}{=} \frac{J}{\sqrt{\lambda}} \begin{pmatrix}
        0 & \sqrt{\frac{1-q}{1-q} }& 0 & 0 & \dots  \\
        \sqrt{\frac{1-q}{1-q} } & 0 & \sqrt{\frac{1-q^2}{1-q}} & 0 &  \dots  \\
        0 & \sqrt{\frac{1-q^2}{1-q}} & 0 & \sqrt{\frac{1-q^3}{1-q}} &  \dots \\
        0 & 0 & \sqrt{\frac{1-q^3}{1-q}} & 0 &  \dots  \\
        \vdots & \vdots & \vdots & \vdots &  \ddots  
    \end{pmatrix} ~.
\end{equation} 
Algebraically, $T$ can be written as:
\begin{equation}
    \centering
    \label{T_alg}
    T = \frac{J}{\sqrt{\lambda}}\left( \alpha + \alpha^\dagger \right)~,
\end{equation}
where 
\begin{equation}
    \centering
    \label{q-annihilation_op}
    \alpha = \sum_{l\geq 0} \sqrt{[l+1]_q}\quad |l\rangle\langle l+1 |~,
\end{equation}
and where we have defined the q-number (or basic number) as:
\begin{equation}
    \centering
    \label{q_number}
    [l]_q \equiv \frac{1-q^l}{1-q} = \sum_{k=0}^{l-1} q^k~.
\end{equation}
In fact, the ladder operators $\alpha,\,\alpha^\dagger$, together with the chord number operator 
\begin{equation}
    \centering
    \label{Chord_number_op}
    \hat{l} = \sum_{l\geq 0} l |l\rangle \langle l| ~,
\end{equation}
form the algebra of a q-deformed oscillator \cite{Arik:1976}:
\begin{equation}
    \centering
    \label{q_oscillator_algebra}
    \begin{split}
        &[\alpha,\alpha^\dagger]_q\equiv \alpha \alpha^\dagger - q \alpha^\dagger \alpha = 1 \\
        &[\hat{l}, \alpha^\dagger] = \alpha^\dagger \\
        &[\hat{l},\alpha] = - \alpha
    \end{split}
\end{equation}
whose representation over the $|l\rangle$ basis satisfies
\begin{equation}
    \centering
    \label{q_oscillator_rep_chord_basis}
    \hat{l} |l\rangle = l |l\rangle, \qquad \alpha^\dagger |l\rangle = \sqrt{[l+1]_q} |l+1\rangle,\qquad\alpha |l\rangle = \sqrt{[l]_q} |l-1\rangle~.
\end{equation}
This algebra reduces to the usual oscillator algebra (the Heisenberg-Weyl algebra) in the limit $q\to 1$, since:
\begin{itemize}
    \item[(i)] $[A,B]_q\overset{q\to 1}{\longrightarrow} [A,B]$
    \item[(ii)] $[l]_q\overset{q\to 1}{\longrightarrow} l$.
\end{itemize}

From (\ref{T_sym}), the eigensystem equation for the components of the eigenvectors of $T$, $\psi_l(E)=\langle l|E\rangle$, is given by:
\begin{equation} \label{q-eigsys}
    E \,\psi_l(E) = \frac{J}{\sqrt{\lambda (1-q)}} \left( \sqrt{1-q^{l+1}}\,\psi_{l+1}(E) + \sqrt{1-q^{l}} \, \psi_{l-1}(E) \right)
\end{equation}
In \cite{Berkooz:2018jqr, Berkooz:2018qkz}, it was shown that the eigenvalues are a function of a continuous variable $\theta$:
\begin{equation} \label{Teigvals}
    E(\theta) = \frac{2J \, \cos \theta}{ \sqrt{\lambda(1-q)}}, \quad 0\leq \theta \leq \pi ~,
\end{equation}
and the normalized eigenvectors are given by
\begin{equation}
    \psi_l(\mu) = \sqrt{(q;q)_\infty} |(e^{2i\theta};q)_\infty| \frac{H_l(\mu|q)}{\sqrt{2\pi (q;q)_l}}, \quad \mu=\cos \theta
\end{equation}
where $(a;q)_n \equiv \prod_{k=0}^{n-1}(1-a q^k)$ is the \textit{q-Pochhammer} symbol\footnote{\url{https://en.wikipedia.org/wiki/Q-Pochhammer\_symbol}.} and $H_l(\mu|q)$ are the \textit{q-Hermite polynomials}.
Since
\begin{equation} \label{T0eigenvector}
    \psi_0(\mu)=   \sqrt{\frac{(q;q)_\infty}{2\pi}} |(e^{2i\theta};q)_\infty|~,
\end{equation}
$\psi_l(\mu)$ can be written as
\begin{equation} \label{Teigenvectors}
    \psi_l(\mu)= \psi_0(\mu) \frac{H_l(\mu|q)}{\sqrt{ (q;q)_l}} ~ , \quad \mu=\cos \theta~.
\end{equation}
See appendix \ref{Appx:EigSysDSSYK} for a detailed review of the derivation of these results.

\subsection{Phase space of JT gravity}\label{Subsect:JT}

We now turn to review some background on JT gravity and in particular its phase space, following \cite{Harlow:2018tqv, Harlow:2021dfp, Brown:2018bms}. This theory is dual to the low-energy regime of DSSYK, as we shall review in section \ref{Subsect:Bulk_Hilbert_Space}.

JT gravity is a 2-dimensional gravity theory with no propagating bulk degrees of freedom. Its action involves the metric $g_{\mu\nu}$ and the dilaton field $\Phi$ and is given by
\begin{equation} \label{S_JT}
    S_{JT} =   \int_\mathcal{M} d^2 x \sqrt{-g}\Big[\Phi_0R +  \Phi (R+2/l_{AdS}^2)\Big] +  2 \int_{\partial\mathcal{M}} dx \sqrt{\gamma} \Big[  \Phi_0 K + \Phi (K-1/l_{AdS}) \Big]
\end{equation}
where $\gamma_{\mu\nu}$ is the induced metric on the boundary, $K$ is the extrinsic curvature of the boundary and $l_{AdS}$ is a length scale of the 2-dimensional spacetime which turns out to be 2-dimensional Anti-de Sitter (AdS$_2$). This length is related to the cosmological constant. The terms involving the constant $\Phi_0$ are topological in nature and will not affect our discussion at the classical level. 
The boundary conditions
\begin{align} 
    ds^2 \big|_{\partial\mathcal{M}} &= -\frac{dt_b^2}{\epsilon^2} \label{bc_metric}\\
    \Phi \big|_{\partial\mathcal{M}} &= \frac{\phi_b}{\epsilon} \label{bc_Phi}
\end{align}
fix the induced metric $\gamma_{\mu\nu}$, with $t_b$ the time on the boundary, and fix the dilaton field $\Phi$ to a positive constant $\phi_b$ at the boundary. In the large volume limit we have $\epsilon \to 0$, i.e. $\epsilon$ plays the role of a boundary regulator.

\begin{figure}
    \centering
    \includegraphics[scale=0.5]{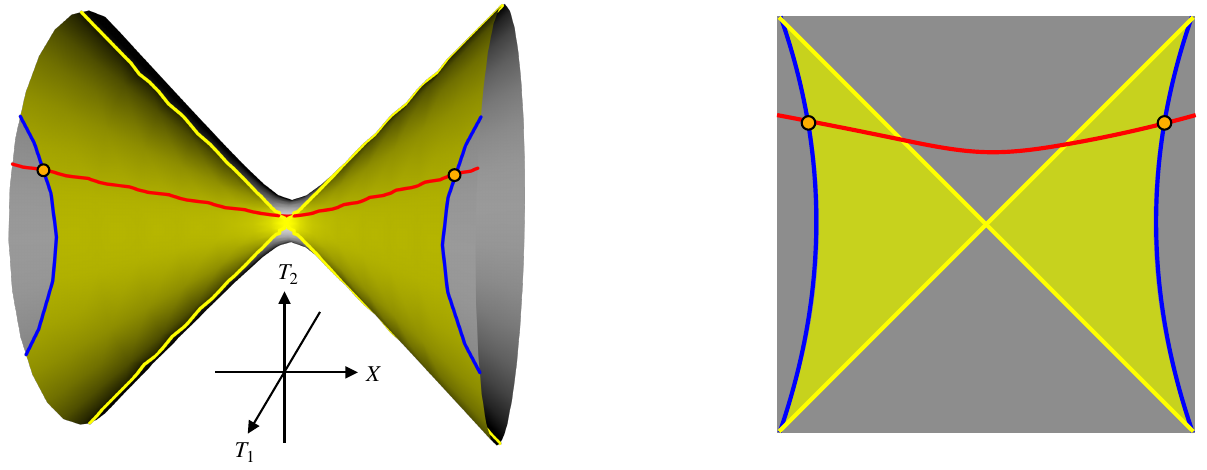}
    \caption{\textbf{Left:} The surface $-T_1^2-T_2^2+X^2=-l_{AdS}^2$ embedded in 3-dimensional space. 
    The wormhole length is represented by the red line, and the boundaries are represented by blue lines. The yellow dots represent the anchoring points at the boundary, between which the wormhole length is computed. The yellow lines represent the Schwarzschild horizon and the areas colored yellow represent the regions covered by the Schwarzschild coordinates. \textbf{Right:} Schematic Penrose diagram of the same geometry. }
    \label{fig:AdS}
\end{figure}

Variation of the action (\ref{S_JT}) gives the following equations of motion:
\begin{align}
    0 &= R+2/l_{AdS}^2 \label{metric_eq}\\
    0 &= (\nabla_\mu \nabla_\nu -g_{\mu\nu}/l_{AdS}^2)\Phi \label{Phi_eq}~.
\end{align}
The first equation of motion $R=-2/l_{AdS}^2$ tells us that the metric should have constant negative curvature, or in other words, it is described by AdS$_2$.  A metric for AdS$_2$ can be constructed via an embedding in a 3-dimensional Minkowski spacetime with two time directions and one space direction:
\begin{equation} \label{3d_metric}
    ds^2 = -dT_1^2-dT_2^2+dX^2.
\end{equation}
AdS$_2$ is the induced geometry on the hypersurface
\begin{equation} \label{AdS_embedding}
    -T_1^2-T_2^2+X^2=-l_{AdS}^2, 
\end{equation}
which has boundaries at $X\to \pm \infty$, see figure \ref{fig:AdS}.
A set of \textit{global} coordinates which cover the whole space is:
\begin{equation} \label{global_co}
    \begin{aligned}
    T_1 &= l_{AdS}\, \sqrt{x^2+1}\cos\tau \\
    T_2 &= l_{AdS}\, \sqrt{x^2+1}\sin\tau \\
    X &= l_{AdS}\, x
    \end{aligned}
\end{equation}
with $0\leq \tau <2\pi$ and $-\infty <x <\infty$. 
Plugging these coordinates into (\ref{3d_metric}) provides an AdS$_2$ metric on the surface (\ref{AdS_embedding}):
\begin{equation} \label{metric_AdS_global}
    ds^2/l_{AdS}^2 = -(1+x^2)d\tau^2 + \frac{1}{1+x^2} dx^2 ~.
\end{equation}
Allowing $\tau\in\mathbb{R}$ in \eqref{metric_AdS_global} yields the universal cover of AdS$_2$.
Turning to the second equation of motion, (\ref{Phi_eq}), its solution is given in the embedding coordinates $(T_1,T_2,X)$ by $\Phi = A T_1+ BT_2+C X$ \cite{Maldacena:2016upp, Harlow:2018tqv}. The unique solution, modulo $SO(1,2)$ rotations in embedding space, which respects the boundary condition (\ref{bc_Phi}), namely that $\Phi$ has the same (positive) value on both boundaries, can be expressed as $\Phi \propto T_1$ \cite{Harlow:2018tqv}.
That is:
\begin{equation} \label{Phi_xtau}
    \Phi(x,\tau)  
    =   \Phi_h  \sqrt{1+x^2} \, \cos\tau ~.
\end{equation}

Another set of coordinates, with a well-defined time on the boundary, is given by the \textit{Schwarzschild} coordinates
\begin{equation} \label{Sch_co}
    \begin{aligned}
    T_1 &= l_{AdS}\, r/r_s \\
    T_2 &= l_{AdS}\, \sqrt{(r/r_s)^2-1}\sinh(r_s t/l_{AdS}^2)\\
    X &= l_{AdS}\, \sqrt{(r/r_s)^2-1}\cosh(r_s t/l_{AdS}^2)~,
\end{aligned}
\end{equation}
with $r>r_s$ and $-\infty < t < \infty$. These coordinates do not cover the whole space, as shown in figure \ref{fig:AdS}. The induced metric in Schwarzschild coordinates is given by
\begin{equation} \label{metric_AdS_Sch}
    ds^2 = -\frac{r^2-r_s^2}{l_{AdS}^2}dt^2 +\frac{ l_{AdS}^2}{r^2-r_s^2}dr^2~.
\end{equation}
The solution to (\ref{Phi_eq}) with boundary condition (\ref{bc_Phi}) acquires a simple form in terms of the Schwarzschild coordinates:
\begin{equation} \label{Phi_rt}
    \Phi(r,t)= \phi_b\, r /l_{AdS}~.
\end{equation}
Note that when $r=r_s$, the two coordinate sets (\ref{global_co}, \ref{Sch_co}) give the relationship
$1= \sqrt{1+x^2} \cos\tau$
which from (\ref{Phi_xtau}) means that at $r=r_s$ we have 
$\Phi = \Phi_h$.
On the other hand, from (\ref{Phi_rt}) we have at $r=r_s$ that
$\Phi = \phi_b r_s/l_{AdS} $.
We thus have the relationship
\begin{equation} \label{rs_eq}
    r_s = l_{AdS} \frac{\Phi_h}{\phi_b}~.
\end{equation}

At the boundary, $r= l_{AdS}/\epsilon$ and so the metric (\ref{metric_AdS_Sch}) and dilaton (\ref{Phi_rt}) become
\begin{align}
    ds^2 \big|_{\text{boundary}} &= -\frac{dt^2}{\epsilon^2}(1-\epsilon^2 (r_s/l_{AdS})^2)\\
    \Phi \big|_{\text{boundary}} &= \phi_b/\epsilon
\end{align}
which from the boundary conditions (\ref{bc_metric}, \ref{bc_Phi}) shows that the Schwarzschild time $t$ becomes the boundary time when $\epsilon \to 0$.  
At the boundary, the relationship between $\tau$ and $t$ can be worked out by considering the relationship between the two sets of coordinates (\ref{global_co}, \ref{Sch_co}):
\begin{equation}
    \left( \frac{r}{r_s} \frac{1}{\cos \tau} \right)^2-1= \Big[\left( \frac{r}{r_s} \right)^2 -1 \Big] \cosh^2(r_s t/l_{AdS}^2)~.
\end{equation}
On the boundary $r=\frac{l_{AdS}}{\epsilon}$, and in the limit $\epsilon \to 0$ the relationship between the global time $\tau$ and the boundary time $t_b$ is found to be
\begin{equation}
    \cos \tau = \frac{1}{\cosh(r_s t_b/l_{AdS}^2)} ~.
\end{equation}

\subsubsection{The length of the wormhole and its canonical conjugate}
In \cite{Harlow:2018tqv} it was shown that the phase space coordinates of JT gravity consist of: geodesic distance between the two boundaries, at a time slice defined by boundary time; and its canonical conjugate.
The geodesic distance between two points on the two asymptotic boundaries of AdS$_2$ can be identified as the length of a (Lorentzian) \textit{wormhole}, see figure \ref{fig:AdS}. In order to compute it for a given boundary time $t_b$, we need the metric (\ref{metric_AdS_global}) and to set $\tau=\text{const}$.  We also need the value of $x$ at the boundary points labeled by time $t_b$.  Again, the relationship between the two sets of coordinates (\ref{global_co}, \ref{Sch_co}) gives
\begin{equation}
    x = \sqrt{(r/r_s)^2-1}\, \cosh(r_s t/l_{AdS}^2) ~.
\end{equation}
At the boundary $ r=l_{AdS}/\epsilon$ so setting $t=t_b$ in the above equation and taking $\epsilon \to 0$ gives
\begin{equation}
    x_b = \frac{l_{AdS}}{\epsilon \, r_s} \cosh (r_s t_b/l_{AdS}^2) ~.
\end{equation}
The wormhole length is then
\begin{align}
    l/l_{AdS} &= \int_{-x_b}^{x_b} \frac{dx}{\sqrt{1+x^2}} = 2\, \textrm{arcsinh}( x_b) = 2 \log \left(x_b+\sqrt{x_b^2+1} \right) \\
    &= 2 \log \left(  \frac{2 l_{AdS}}{\epsilon \, r_s} \cosh (r_s t_b/l_{AdS}^2)\right) + O(\epsilon) ~.
\end{align}
Using $r_s= l_{AdS} \Phi_h /\phi_b$ from (\ref{rs_eq}), the above result can be expressed as
\begin{equation}
    l/l_{AdS} = 2\log\left( \frac{2\phi_b}{ \epsilon}\right) + 2 \log \left[  \frac{1}{\Phi_h} \cosh \left(\frac{\Phi_h}{ l_{AdS}\, \phi_b}\, t_b \right)\right] + O(\epsilon) ~.
\end{equation}
The first term is infinite in the limit $\epsilon\to 0$ (since in the large volume limit the boundary of AdS is at infinity), and this result needs to be renormalized. The renormalized length of the wormhole is given by
\begin{equation}
\label{Bulk_length}
    \tilde{l}/l_{AdS}= l-2\log\left( \frac{2\phi_b}{ \epsilon}\right)  
    = 2 \log \left[\cosh \left(\frac{\Phi_h}{l_{AdS}\,\phi_b}\, t_b \right)\right] - 2 \log \Phi_h ~.
\end{equation}
In terms of the value of the Hamiltonian (or energy) on the boundary \cite{Harlow:2018tqv}
\begin{equation}\label{energy_config_bulk}
    H = \frac{2\Phi_h^2}{ l_{AdS}\,\phi_b}=E,
\end{equation}
result \eqref{Bulk_length} becomes:
\begin{equation} \label{Ren_Wormhole_length_E}
    \tilde{l}/ l_{AdS} = 2 \log \left[\cosh \left(\sqrt{\frac{E}{2  l_{AdS}\, \phi_b}}\, t_b \right)\right] -  \log \left(\frac{ l_{AdS}\,E\phi_b}{2} \right) ~.
\end{equation}
Together with the canonical conjugate of $\tilde{l}$, given by \cite{Harlow:2018tqv}
\begin{equation}
    P\,  l_{AdS} = \sqrt{2  l_{AdS}\, E\phi_b } \tanh \left( \sqrt{\frac{E}{2 l_{AdS}\,\phi_b}}t_b\right), 
\end{equation}
the Hamiltonian of JT gravity takes the form \cite{Harlow:2018tqv}
\begin{equation}\label{JT_Liouville_Ham}
    H = \frac{1}{ l_{AdS}\,\phi_b} \left( \frac{ l_{AdS}^2 \, P^2}{2} + 2 e^{-\tilde{l}/ l_{AdS}}\right) ~.
\end{equation}
This Hamiltonian is written in terms of a two-sided\footnote{ We call $\tilde{l}$ a two-sided length because it joins the two disconnected (and regularized) boundaries of AdS$_2$.} phase space variable $\tilde{l}$ and its conjugate momentum, and thus the quantum description of the system will consist of a Hilbert space spanned by the eigenfunctions of this two-sided length.  This Hilbert space is \textit{not} factorizable as a product of other Hilbert spaces describing the degrees of freedom of each side separately, as was discussed in \cite{Harlow:2018tqv}.

\subsection{Bulk Hilbert space} \label{Subsect:Bulk_Hilbert_Space}

This section reviews how the chord Hilbert space of DSSYK can be understood equivalently as the bulk Hilbert space of JT gravity. The guiding principle in \cite{Lin:2022rbf} is the following: in a given chord diagram, it is possible to identify a \textit{left} and a \textit{right} region by arbitrarily choosing two points on the circumference that will separate them; such points are identified with (Euclidean) future and past infinity, and one can define a constant (Euclidean) time slice by a line whose anchoring points are on different regions: the state on such a slice is given by the number of open chords intersecting it (defined in a unique way \cite{Lin:2022rbf}). This is nothing but a re-interpretation of the discussion around (\ref{State_psi_chord_basis}), but the picture is now very reminiscent of the non-factorizable, two-sided Hilbert space of JT gravity reviewed in the preceding section. The connection can be made even more explicit by showing that the effective DSSYK Hamiltonian becomes, in the suitable limit, the Liouville Hamiltonian of JT gravity \cite{Berkooz:2018jqr,Lin:2022rbf}, where the length operator is in fact proportional to the chord number operator. We review this below.

We will now denote the chord basis by $\{|n\rangle \}_{n=0}^\infty$. The effective Hamiltonian of the averaged theory for DSSYK, given by (\ref{T_alg}), may be written as:
\begin{equation}
    \centering
    \label{Ham_semi_inf_chain}
    T=\frac{J}{\sqrt{\lambda}}\left( \alpha + \alpha^\dagger \right) = \frac{J}{\sqrt{\lambda (1-q)}}\left( D\sqrt{1-q^{\hat{n}}} + \sqrt{1-q^{\hat{n}}} D^\dagger \right),
\end{equation}
where $\alpha,\,\alpha^\dagger$ are the ladder operators of the $q$-deformed harmonic oscillator and $\hat{n} | n
\rangle = n | n \rangle$, while $D^\dagger$ is a non-normalized version of the creation operator, which therefore acts as a unit-displacement operator on the semi-infinite ordered chord basis:
\begin{equation}
    \centering
    \label{Displacement_op}
    D^\dagger |n\rangle = |n+1\rangle,\qquad D|n\rangle = |n-1\rangle,\qquad D|0\rangle = 0.
\end{equation}
On this chain, $\hat{n}$ plays the role of a position operator. Despite the inherent discreteness of the semi-infinite lattice at hand, it is possible to define a \textit{conjugate} canonical momentum $p$ for $\hat{n}$ as the generator of translations\footnote{If the discrete lattice had a finite length $K$, its Hilbert space would be finite and therefore it would be impossible to fulfill the canonical commutation relation $[\hat{n},p]=i$, but we would still call $p$ a \textit{canonical} momentum in the sense that it satisfies (\ref{Canonical_momentum_discrete}). Taking the suitable limit $K\to \infty$ one recovers the semi-infinite and discrete lattice and, restricted to a physically relevant domain, the commutator $[n,p]$ does tend to $i$, as explained in detail in \cite{Cannata:1991I,Cannata:1991II}.}:
\begin{equation}
    \centering
    \label{Canonical_momentum_discrete}
    D^\dagger \equiv e^{-ip}.
\end{equation}
Observing (\ref{Canonical_momentum_discrete}) we note that the discreteness of the lattice manifests itself in the fact that only displacements of unit length in $n$-space are allowed. A continuum limit would be achieved by allowing displacements of arbitrary length $\Delta n$, of the form $e^{-i\, \Delta n\,p}$. In this limit we would have that $p=-i\partial_n$.

Before proceeding further, it is convenient to redefine the position variable so that it acquires dimensions of length and also so that it is well-behaved in the small-$\lambda$ limit that we shall shortly take. Introducing a \textit{fundamental length scale} $l_f$ about which we don't need to be specific\footnote{It will eventually become the AdS length. It is not a parameter of the DSSYK theory, so one may equivalently say that chord number is related to bulk length \textit{normalized by AdS units.}}, we define a new length variable $l$ as:
\begin{equation}
    \centering
    \label{Length_def}
    l = l_f \lambda n,
\end{equation}
whose conjugate canonical momentum is therefore $k=\frac{p}{l_f \lambda}$, becoming $k=-i\partial_l$ in the continuum limit. With this all, the Hamiltonian becomes:
\begin{equation}
    \centering
    \label{Ham_l_k_before_inverting}
    T = \frac{J}{\sqrt{\lambda (1-q)}}\left( e^{i\lambda l_f k}\sqrt{1-e^{-\frac{l}{l_f}}} + \sqrt{1-e^{-\frac{l}{l_f}}} e^{-i\lambda l_f k} \right).
\end{equation}
The spectrum of this Hamiltonian is bounded (for $\lambda>0$) and symmetric about zero, as can be seen in (\ref{Teigvals}), so we shall do the harmless replacement $T\to -T$ that ensures that the ground state corresponds to the minimal momentum $k$ \cite{Lin:2022rbf}. This choice is arbitrary at this point but will be important later on, as it will ensure that the triple-scaled Hamiltonian is bounded from below. Let us write the explicit form of the Hamiltonian after the replacement $T\mapsto -T\equiv \tilde{T}$:
\begin{equation}
    \centering
    \label{Ham_l_k}
    \tilde{T} = -\frac{J}{\sqrt{\lambda (1-q)}}\left( e^{i\lambda l_f k}\sqrt{1-e^{-\frac{l}{l_f}}} + \sqrt{1-e^{-\frac{l}{l_f}}} e^{-i\lambda l_f k} \right).
\end{equation}

We can now take the remaining limit that brings us to the so-called \textit{triple-scaling limit}. It is a small-$\lambda$ limit in which $l$ is taken to be accordingly large, as follows\footnote{Our definition of the triple-scaling limit differs by a factor of $2$ from that in \cite{Lin:2022rbf}. See appendix \ref{Appx:triple_scaling} for a discussion on this choice, which eventually does not imply any fundamental modification of the bulk theory.}:
\begin{equation}
    \centering
    \label{Trple-scaling-limit}
    \lambda\to 0,\qquad l\to\infty,\qquad\frac{e^{-\frac{l}{l_f}}}{(2\lambda)^2} \equiv e^{-\frac{\tilde{l}}{l_f}}\;\text{fixed.}
\end{equation}
The last condition in (\ref{Trple-scaling-limit}) can be rewritten as:
\begin{equation}
    \centering
    \label{renormalized_length}
    \frac{\tilde{l}}{l_f} = \frac{l}{l_f} -2\log\left(\frac{1}{2\lambda}\right).
\end{equation}
We shall call $\tilde{l}$ the \textit{renormalized length}\footnote{It will become the actual renormalized bulk length eventually.}. In this triple-scaled limit, the Hamiltonian (\ref{Ham_l_k}) becomes\footnote{This can be shown in the continuum limit, using $k=-i\partial_l$, it still holds for the discrete, semi-infinite chain using the commutation relation $[l,k]=i$, which is satisfied within the relevant physical domain as argued in \cite{Cannata:1991I,Cannata:1991II}.}:
\begin{equation}
    \centering
    \label{Triple_scaled_Hamiltonian}
    \tilde{T}-E_0 = 2\lambda J \left( \frac{l_f^2 k^2}{2} + 2e^{-\frac{\tilde{l}}{l_f}} \right)\;+\mathit{O}\left(\lambda^2\right),
\end{equation}
where $E_0=\frac{-2J}{\lambda}$ is a constant energy shift. The moral of this triple-scaling is now clear: we are zooming in near the ground state $E_0$ of the Hamiltonian (\ref{Ham_l_k}), and in this regime it takes the form of the Hamiltonian of Liouville quantum mechanics, whose spectral density is proportional to $\sinh(2\pi\sqrt{E})$ \cite{Bagrets:2016cdf}, which is in turn the Hamiltonian describing the dynamics of the single pair of phase space variables of JT gravity, as seen in (\ref{JT_Liouville_Ham}). This establishes the correspondence between \textit{triple}-scaled SYK and JT gravity: their Hilbert spaces are identified and the Hamiltonian generating dynamics is the same. Incidentally, we note that instances of DSSYK with different values of $\lambda$ and $J$ may collapse, in the triple-scaling limit, onto the same Liouville Hamiltonian if their product $\lambda J$ is equal, since this is the only parameter controlling the Liouville Hamiltonian (\ref{Triple_scaled_Hamiltonian}), and hence the gravity dual. In other words, the following parameter identification connects Hamiltonians (\ref{JT_Liouville_Ham}) and (\ref{Triple_scaled_Hamiltonian}):
\begin{equation}
    \label{Param_identification_Hamiltonians}
    l_f = l_{AdS},\qquad\qquad\qquad 2 \lambda J = \frac{1}{l_{AdS}\phi_b}.
\end{equation}
We will come back to these identifications in section \ref{Section_Gravity_matching}, where boundary K-complexity is matched to the corresponding bulk length computation.

\section{Krylov complexity and chords}\label{Sect:KC}
In this section we quickly review the definition of Krylov complexity and then provide exact results for it in DSSYK. Although the double-scaling limit of SYK is not the limit of SYK which corresponds directly to JT gravity, we study K-complexity in this limit as a warm-up and as an interesting result in itself which has not been studied before. 

Krylov complexity measures the spreading of states or operators in a quantum system under Hamiltonian time evolution, over a special ordered basis (sometimes called the \textit{Krylov chain}) adapted to the state/operator's time evolution. Here, we will focus on the definition of K-complexity for states. 
Given an initial state $|0 \rangle$ at $t=0$, its time evolution generated by the Hamiltonian $H$ in the Schrödinger picture is given by
\begin{equation}
    |\phi(t)\rangle = e^{-iHt}|0 \rangle = \sum_{n=0}^\infty \frac{(-it)^k}{k!} H^k |0\rangle 
\end{equation}
written as a linear combination over the basis $\{ |0\rangle, H|0\rangle, H^2|0\rangle,\dots \}$. The survival amplitude, which is the overlap of the time-evolving state with the state at $t=0$ (also known as \textit{fidelity}), is given by
\begin{equation} \label{survival_prob}
    \langle 0|\phi(t)\rangle = \langle 0| e^{-iHt}|0 \rangle = \sum_{n=0}^\infty \frac{(-it)^k}{k!} \langle0| H^k |0\rangle \equiv \sum_{n=0}^\infty \frac{(-it)^k}{k!} M_k
\end{equation}
where we define the moments $M_k \equiv \langle0| H^k |0\rangle$.  If the survival probability $\langle 0|\phi(t)\rangle$ is an even function of $t$ then all odd moments $M_{2k+1}$ are zero.  This, in particular, is the case for the effective Hamiltonian of DSSYK with the initial state being the zero-chord state $|0\rangle$.

The Krylov basis is obtained via the Lanczos algorithm which is essentially a Gram-Schmidt orthonormalization procedure over the basis $\{ |0\rangle, H|0\rangle, H^2|0\rangle,\dots \}$. For an even survival probability function (with odd moments being zero) the Lanczos algorithm takes the form\footnote{In the general case, where odd moments are not necessarily zero, the Lanczos algorithm is a little more involved. See e.g. \url{https://en.wikipedia.org/wiki/Lanczos\_algorithm}. }:
\begin{enumerate}
    \item $|A_1\rangle = H |0 \rangle$, compute $b_1 = |A_1|$, if $b_1=0$ stop. Otherwise define $|1 \rangle = |A_1\rangle /b_1 $.
    \item For $n\geq 1$: $|A_{n+1}\rangle = H |n \rangle - b_{n} |n-1 \rangle$, compute $b_{n+1}=|A_{n+1}|$, if $b_{n+1}=0$ stop. Otherwise define $|n+1\rangle = |A_{n+1}\rangle/b_{n+1}$.
\end{enumerate}
Here it was assumed that the initial state $|0\rangle$ is normalized.
In this way the ordered \textit{Krylov basis} (or Krylov chain) $\{|n\rangle \}_{n=0}^{K-1}$ is constructed and the \textit{Lanczos coefficients} $\{b_n\}_{n=1}^{K-1}$ are defined, where $K$ is the dimension of the Krylov space. $K$ can be infinite if the original Hilbert space has infinite dimension, and must be finite otherwise.
By construction, the Krylov basis vectors form an orthonormal basis, satisfying $\langle m |n\rangle =\delta_{mn}$. In the Krylov basis, the Hamiltonian acquires the following tridiagonal form determined by the Lanczos coefficients: 
\begin{equation} \label{HKB}
    H|n\rangle = b_{n+1}|n+1\rangle  +b_n|n-1\rangle~.
\end{equation}
The Lanczos coefficients can be determined from the moments of the survival amplitude \eqref{survival_prob}, via the iterative procedure \cite{viswanath1994recursion}:
\begin{align} 
    M_{2k}^{(n)} &= \frac{M_{2k}^{(n-1)}}{b_{n-1}^2}-\frac{M_{2k-2}^{(n-2)}}{b_{n-2}^2}, \quad n=1,2,\dots,L \quad k=n,n+1,\dots, L \nonumber\\
    b_n^2 &= M_{2n}^{(n)}   \label{recursion_moments} 
\end{align}
where $M_{2k}^{(0)} \equiv M_{2k}$ as well as $M_{2k}^{(-1)}=0$ and $b_{-1}=1=b_0$. Using this algorithm, the first $L$ Lanczos coefficients can be determined from the moments $M_{2k}$ with $k=0,\dots,L$. 

The time-evolving state can now be expanded over the Krylov basis:
\begin{equation}\label{phi_t_phi_n}
    |\phi(t)\rangle = \sum_{n=0}^{K-1} \phi_n(t)|n\rangle
\end{equation}
where $\phi_n(t) \equiv \langle n|\phi(t)\rangle$ can be thought of as a ``wavefunction" spreading over the ordered Krylov basis with initial condition $\phi_n(t=0)=\langle n|0\rangle =\delta_{n0}$. From unitarity of the time evolution, the norm is preserved, i.e. $\sum_{n=0}^{K-1}|\phi_n(t)|^2=1$ for all $t$.
Note that $\phi_0(t)$ is the survival probability defined in (\ref{survival_prob}).

A `position' operator, $\hat{n}$, over the ordered Krylov basis can be defined as
\begin{eqnarray} \label{nOperator}
    \hat{n} = \sum_{n=0}^{K-1} n \, |n \rangle \langle n|,
\end{eqnarray}
and \textit{Krylov complexity} is then defined as the expectation value of $\hat{n}$ as a function of time, or equivalently, the average position of the wavefunction $|\phi(t)\rangle$ over the ordered Krylov basis:
\begin{equation} \label{KC_definition}
    C_K(t) = \langle \hat{n}(t)\rangle = \langle \phi(t)| \hat{n} | \phi(t) \rangle = \sum_{n=0}^{K-1} n \, |\phi_n(t)|^2 ~.
\end{equation}

From (\ref{HKB}) we can obtain a recurrence relation for the eigenvectors of $H$, $H|E\rangle=E|E\rangle$,
since the components of each eigenvector $|E\rangle$ over the Krylov basis, defined as $\psi_n(E)\equiv \langle n|E\rangle$, satisfy:
\begin{equation}
    E\, \psi_n(E) = b_{n+1}\psi_{n+1}(E) + b_n \psi_{n-1}(E) ~.
\end{equation}
The components of the eigenvectors of $H$ over the Krylov basis are useful in the determination of the wavefunctions $\phi_n(t)$, and if known, together with the eigenvalues of $H$, they provide the following closed-form expression for the wavefunction as a function of time:
\begin{equation} \label{Phi_EnergyB}
    \phi_n(t) \equiv \langle n|\phi(t)\rangle = \langle n|e^{-iHt} |0\rangle = \sum_E e^{-iEt}\langle n|E\rangle \langle E|0\rangle =  \sum_E e^{-iEt}\, \psi_n(E) \, \psi_0^*(E)~.
\end{equation}

\subsection{Lanczos coefficients in DSSYK}

Observing the steps leading to the expression of the effective Hamiltonian in coordinates over the basis of \textit{fixed chord number} states, given in matrix form in (\ref{T_sym}), the connection to Lanczos coefficients is argued as follows. The fixed chord number states $\{|n\rangle\}$ are an orthonormal set of linear combinations of states with up to $n$ Hamiltonian insertions $T^n|0\rangle$; and using them as a basis, they bring the transfer matrix (or effective Hamiltonian) to a tridiagonal form. They thus turn out to be Krylov basis elements, with the matrix elements of the symmetric, tridiagonal $T$ giving the Lanczos coefficients:
\begin{equation}
    \centering
    \label{Lanczos}
    b_n = J \sqrt{\frac{[n]_q}{\lambda}}=J\sqrt{\frac{1-q^n}{\lambda(1-q)}}~.
\end{equation}
Consequently, the chord number operator $\hat{n}$, which gives the position in the chord number basis, gives equivalently the position in the Krylov basis, so it \textit{is} the K-complexity operator.

\begin{figure}
    \centering
    \includegraphics[scale=0.3]{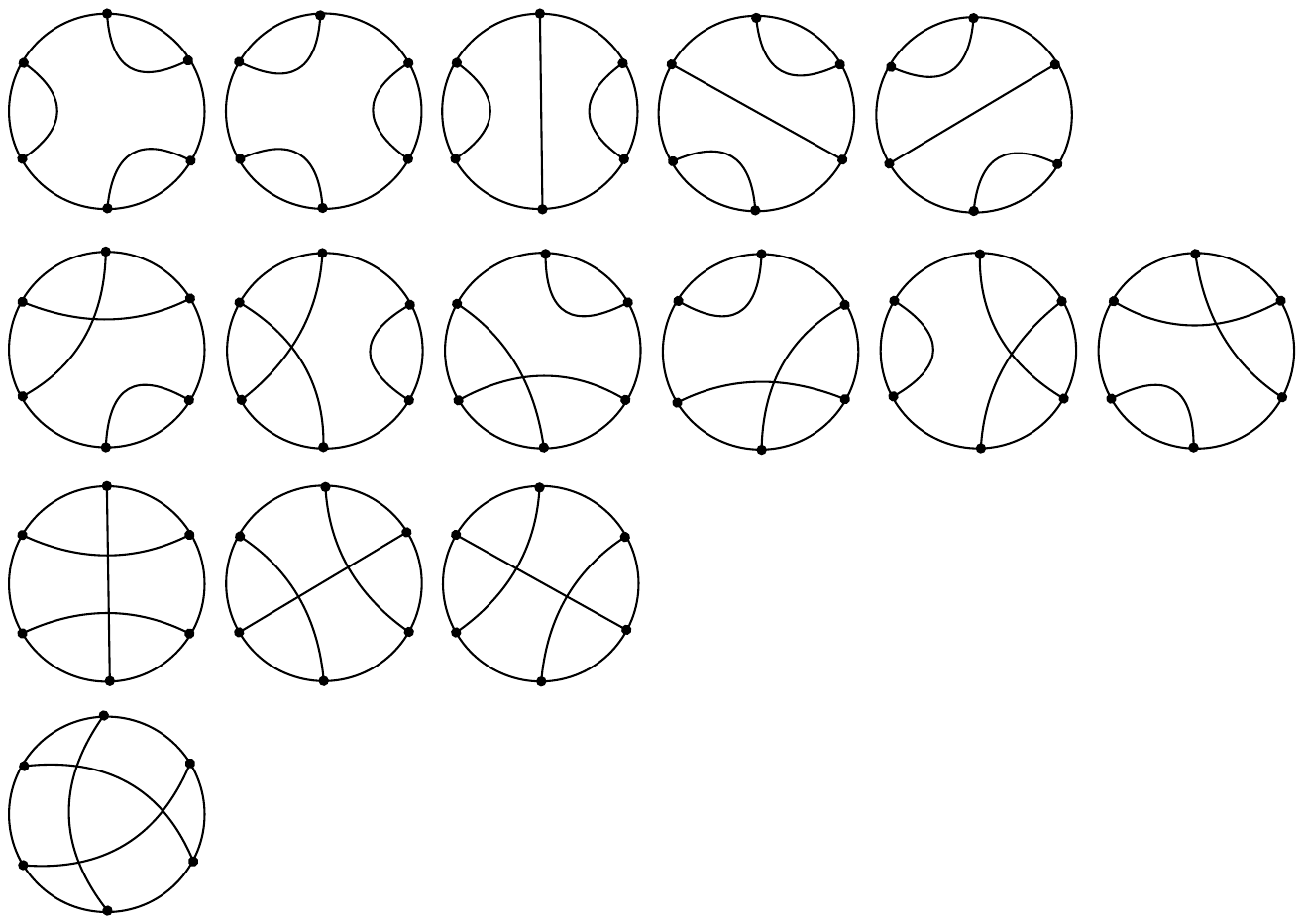}
    \caption{The 15 chord diagrams contributing to $M_6$: the \textbf{top row} shows chord diagrams with zero intersections (5 diagrams), the \textbf{second row} shows chord diagrams with 1 intersection (6 diagrams), the \textbf{third row} shows diagrams with 2 intersections (3 diagrams) and the \textbf{fourth row} shows the only chord diagram with 3 intersections. These numbers correspond to the calculation of $M_6$ in (\ref{First_three_even_moments}).}
    \label{fig:chord_diagrams_M6}
\end{figure}

There is certainly more than one orthonormal basis that can bring the effective Hamiltonian $T$ to tridiagonal form. However, given a seed state $|0\rangle$, there is a unique (up to global phases) orthonormal basis $\{|n\rangle\}$ such that the state $|n\rangle$ is a linear combination of $\{T^k |0\rangle\}_{k=0}^n$, as can be shown inductively. Such a basis can therefore be identified, up to global phases, with the Krylov basis adapted to the state $|0\rangle$ and the Hamiltonian $T$, which can be built efficiently through the Lanczos algorithm and brings the Hamiltonian to tridiagonal form with real entries. The fact that the matrix expression (\ref{T_sym}) for the effective Hamiltonian in the fixed chord number basis has positive entries indicates that such a basis coincides exactly with the Krylov basis (i.e. they do not differ even by a global phase). This argument proves that the fixed chord number states are Krylov elements and therefore that the entries of the tridiagonal version of $T$ in (\ref{T_sym}) are the Lanczos coefficients. In addition, we can perform a non-trivial check of the fact that the sequence (\ref{Lanczos}) is actually the Lanczos coefficients sequence by taking the moments $M_{2k}$ computed via chord-diagram combinatorics and showing that the usual transformation (\ref{recursion_moments}) will indeed yield a sequence of Lanczos coefficients in agreement with expression (\ref{Lanczos}). Using chord diagrams, we compute explicitly the first three even moments:
\begin{equation}
    \centering
    \label{First_three_even_moments}
    \begin{split}
        &M_2 = \frac{J^2}{\lambda} \\
        &M_4 = \frac{J^4}{\lambda^2}(2+q)\\
        &M_6=\frac{J^6}{\lambda^3}(5+6q+3q^2+q^3)
    \end{split}
\end{equation}
where $M_6$ involves evaluating the 15 chord diagrams shown in figure \ref{fig:chord_diagrams_M6}.
With this, we can apply the recursion method (\ref{recursion_moments}):
\begin{equation}
    \centering
    \label{b_n_from_Mn_first}
    \begin{split}
        &b_1^2=M_2=\frac{J^2}{\lambda}=\frac{J^2}{\lambda}[1]_q \\
        &b_2^2 = \frac{M_4}{M_2}-M_2 = \frac{\frac{J^4}{\lambda^2}(2+q)}{\frac{J^2}{\lambda}}-\frac{J^2}{\lambda}=\frac{J^2}{\lambda}(1+q) = \frac{J^2}{\lambda} [2]_q \\
        &b_3^2 = \frac{\frac{M_6}{M_2}-M_4}{\frac{M_4}{M_2}-M_2}-\frac{M_4}{M_2}=(...)=\frac{J^2}{\lambda}(1+q+q^2) = \frac{J^2}{\lambda}[3]_q~.
    \end{split}
\end{equation}
Equations (\ref{b_n_from_Mn_first}) illustrate that the Lanczos coefficients (\ref{Lanczos}) are consistent with the Hamiltonian moments (\ref{First_three_even_moments}).

\subsubsection{State dependence of the Lanczos coefficients}\label{Subsubsect_Krylov_TFD}

An important discussion is in order: what initial state are these Lanczos coefficients being computed for? They are the Lanczos coefficients associated to the evolution under the effective Hamiltonian $T$ of the initial state $|0\rangle$, given by the state
\begin{equation}
    \centering
    \label{state_T_0}
    |\phi(t)\rangle = e^{-itT}|0\rangle,
\end{equation}
as the moments to which the coefficients (\ref{Lanczos}) are in one-to-one correspondence are the coefficients of the Taylor series of the survival amplitude of (\ref{state_T_0}):
\begin{equation}
    \centering
    \label{Survival_amplitude_zero_ket}
    \langle 0 | e^{-itT}|0\rangle =\sum_{k=0}^{+\infty} \frac{(-it)^{2k}}{(2k)!} M_{2k},
\end{equation}
where the moments $M_{2k}$ are given in (\ref{Moments_effective_Ham}).

Furthermore, the state $|0\rangle$ that seeds the evolution can be thought of as an effective version of the infinite-temperature thermofield ``double'' state in the ensemble-averaged theory, and the survival amplitude (\ref{Survival_amplitude_zero_ket}) is in fact the ensemble-averaged analytic continuation of the partition function, which is itself known to give the survival amplitude of the thermofield double state \cite{delCampo:2017bzr,Balasubramanian:2022tpr}. We shall now elaborate on this. The ensemble-averaged analytic continuation of the partition function, evaluated at infinite temperature, reads:
\begin{equation}
    \centering
    \label{Averaged_partition_fn_beta_zero}
    \left.\left\langle Z(\beta + i t) \right\rangle\right|_{\beta=0} = \left\langle \text{Tr}\left[ e^{-itH} \right] \right\rangle.
\end{equation}

Now, one can check that the trace in (\ref{Averaged_partition_fn_beta_zero}) can be rewritten as the expectation value of the evolution operator in a certain state $|\Omega\rangle$, as follows:
\begin{equation}
    \centering
    \label{Trace_omega_state}
    \text{Tr}\left[ e^{-itH} \right]=\langle \Omega | e^{-itH} |\Omega \rangle,\qquad\qquad\qquad |\Omega\rangle \equiv \frac{1}{\sqrt{\mathcal{N}}}\sum_E |E\rangle,
\end{equation}
where $\mathcal{N}\to\infty$ denotes the Hilbert space dimension, and the factor $1/\sqrt{\mathcal{N}}$ is required for consistency with the convention $\text{Tr}[\mathbb{1}]=1$, ensuring normalization of the state $|\Omega\rangle$. By inspection, we note that this state is an ``unconventional'' version of the infinite-temperature thermofield ``double'' state, in the sense that it does not belong to the tensor product of two identical Hilbert spaces but rather to the only Hilbert space available in the problem. This feature will be consistent with the bulk interpretation of this Hilbert space as the non-factorizable two-sided Hilbert space of JT gravity \cite{Harlow:2018tqv}, as discussed in \cite{Lin:2022rbf}.

Finally, taking the ensemble average  amounts to replacing the non-averaged Hamiltonian $H$ by the effective Hamiltonian $T$, and the state $|\Omega\rangle$ by the zero-chord state $|0\rangle$~:
\begin{equation}
    \centering
    \label{Ensemble_average_survival}
      \left\langle \text{Tr}[e^{-itH}] \right\rangle = \Big\langle \langle \Omega | e^{-itH} |\Omega \rangle \Big\rangle = \langle 0 | e^{-it T} | 0 \rangle,
\end{equation}
which connects, as promised, the averaged analytic continuation of the partition function (\ref{Averaged_partition_fn_beta_zero}) to the survival amplitude (\ref{Survival_amplitude_zero_ket}) and provides the justification for considering the zero-chord state $|0\rangle$ as an effective version of the infinite-temperature thermofield ``double'' state in the averaged theory.

\subsubsection{Regimes of the Lanczos sequence}
In order to better understand the different regimes of K-complexity as a function of time, it will be useful to analyze the regimes that can be identified in the Lanczos sequence (\ref{Lanczos}).
At this point, we recall that $q=e^{-\lambda} = e^{-\frac{2p^2}{N}}$. For $\lambda>0$ we have $0<q<1$ and the Lanczos coefficients $b_n$ in (\ref{Lanczos}) are bounded. They have a horizontal asymptote at $b_\infty=\frac{J}{\sqrt{\lambda(1-q)}}>0$. In the limit $\lambda\to+\infty$ (i.e. $q\to 0$) the Lanczos sequence is constant $b_n\sim \frac{J}{\sqrt{\lambda}}\to 0$, while in the limit $\lambda\to0$ we have $q\to1$ and the Lanczos coefficients grow indefinitely as $b_n\sim J\sqrt{\frac{n}{\lambda}}$. The limit $\lambda \to 0$ is of particular importance since in \cite{Lin:2022rbf} the connection to gravity is done in a \textit{triple} scaling limit in which $\lambda \ll 1$ (i.e. $q\to 1$). 

In order to identify regimes in the Lanczos sequence, the above analysis needs to be performed more carefully. The arguments above have considered limits of $\lambda$ keeping $n$ fixed, whereas below we shall study the behavior of the $b_n$ sequence given a fixed parameter $\lambda>0$ for $n$ either sufficiently big or small compared to $\lambda$.

Inspecting (\ref{Lanczos}) we note that, given $0<q<1$, for sufficiently large $n$, the factor $q^n$ will be small and therefore we can use the approximation:
\begin{equation}
    \centering
    \label{Lanczos_large_n}
    b_n\approx \frac{J}{\sqrt{\lambda(1-q)}}\left(1-\frac{q^n}{2}\right),\qquad\text{for large }n. 
\end{equation}
Conversely, if $|n\log(q)|$ is sufficiently small, we can approximate $1-q^n=1-e^{-n \log(1/q)}\approx n\log(1/q)=\lambda n>0$ (for $n>0$), so that:
\begin{equation}
    \centering
    \label{Lanczos_small_n}
    b_n\approx J\sqrt{\frac{n}{1-q}},\qquad\text{for small }n.
\end{equation}
We can estimate the critical value $n_*(q)$ that marks the transition between the regimes (\ref{Lanczos_small_n}) and (\ref{Lanczos_large_n}) by taking the value of $n$ for which $n\log(1/q)$ becomes of order $1$, which yields:

\begin{equation}
    \centering
    \label{nstar_good_estimate}
    n_*(q) \approx \frac{1}{\log\left(\frac{1}{q}\right)}=\frac{1}{\lambda}.
\end{equation}

\begin{figure}
    \centering
    \includegraphics[width=7.4cm]{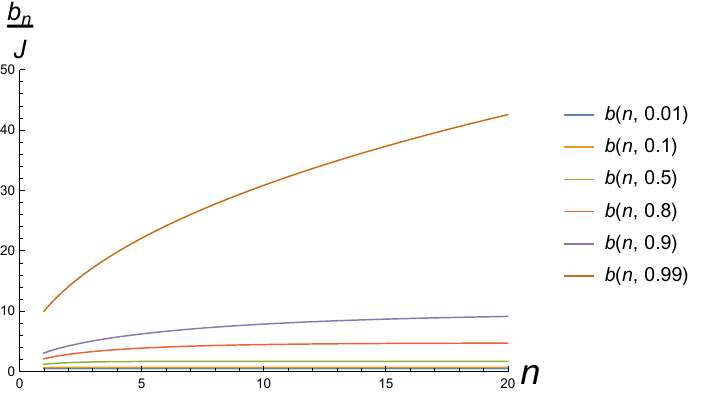} \includegraphics[width=7.4cm]{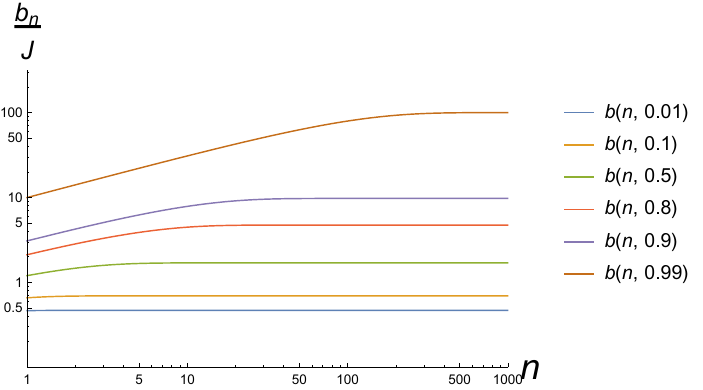}
    \caption{Lanczos coefficients in DSSYK for different values of $q$, denoted $b_n\equiv b(n,q) $. \textbf{Left:} Linear scale along both axes, focusing at small values of $n$. \textbf{Right:} Log-log plot. In this scale, the initial linear shape is compatible with a square-root behavior of $b_n$; figure \ref{fig:Lanczos_example} illustrates this in more detail.}
    \label{fig:Lanczos_various_qvals}
\end{figure}

\begin{figure}
    \centering
    \includegraphics[width=7.4cm]{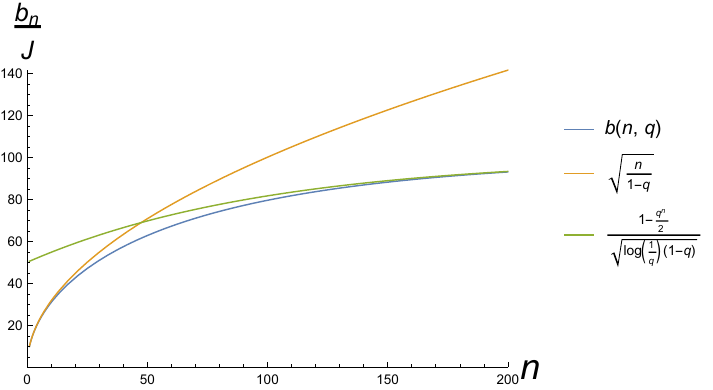} \includegraphics[width=6.5cm]{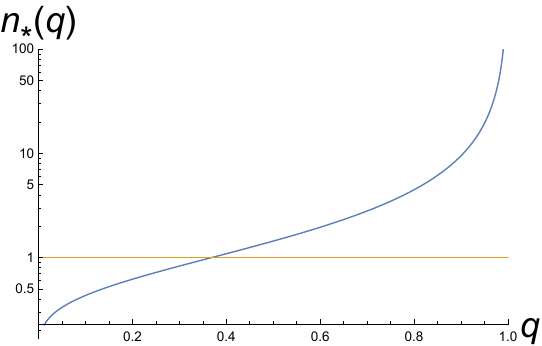}
    \caption{\textbf{Left:} Lanczos coefficients of DSSYK for $\lambda=0.01$ (i.e. $q\approx0.99005$), together with the limiting regimes for small and large $n$.
    \textbf{Right:} Transition value $n_*(q)$ as a function of $q$ (in blue). The constant $n_*=1$ has been marked in orange, for reference. Note the logarithmic scale along the vertical axis. In particular, we note that $n_*(q)$ has a vertical asymptote at $q=1$, implying that in the limit $q\to 1$, which is of interest in this paper, there will be a clear separation of scales in the Lanczos sequence.}
    \label{fig:Lanczos_example}
\end{figure}

Summarizing, we have found that the Lanczos coefficients start growing as $\propto \sqrt{n}$ before $n_*(q)$, after which they saturate at a horizontal asymptote following $\propto 1-\frac{q^n}{2}$:
\begin{equation}
    \centering
    \label{Lanczos_regimes_summary}
    b_n =  \begin{cases}
        J\sqrt{\frac{n}{1-q}},&  n\lesssim \frac{1}{\lambda}\\
        \frac{J}{\sqrt{\lambda(1-q)}}\left(1-\frac{q^n}{2}\right),& n\gtrsim \frac{1}{\lambda}
    \end{cases} 
\end{equation}
Note that $\lim_{q\to 1^{-}} n_*(q)=\lim_{\lambda \to 0^{+}}n_*(q(\lambda))=+\infty$, consistent with the fact that when $q\to1$ we have $b_n\propto\sqrt{n}$, i.e. only the square-root behavior is featured. Additionally, inspecting (\ref{nstar_good_estimate}) we realize that for $\lambda>1$ (or, equivalently, $q<e^{-1}\approx 0.37$) the transition value of $n$ is $n_*<1$, which implies that the first regime in (\ref{Lanczos_regimes_summary}) will not be visible at all in the actual Lanczos sequence, since strictly speaking $n\in \mathbb{N}$. In the cases of interest for us we will need to consider both regimes, as the connection to gravity occurs in a limit where $\lambda$ is small. Figure \ref{fig:Lanczos_various_qvals} depicts several Lanczos sequences for different values of $q$, and figure \ref{fig:Lanczos_example} shows the different regimes in the Lanczos sequence for a particular value of $q$ close to $1$, together with the $q$-dependence of the transition value $n_*(q)$.

For reference, we can rewrite (\ref{Lanczos_regimes_summary}) for small $\lambda$ (i.e. $q$ close to $1$):
\begin{equation}
    \centering
    \label{Lanczos_regimes_summary_smallLambda}
    b_n \approx  \begin{cases}
        J\sqrt{\frac{n}{\lambda}},&  n\lesssim \frac{1}{\lambda}\\
        \frac{J}{\lambda}\left(1-\frac{e^{-n\lambda}}{2}\right),& n\gtrsim \frac{1}{\lambda}
    \end{cases} 
\end{equation}
where the corrections are subleading in an expansion in powers of $\sqrt{\lambda}$.

\subsection{K-complexity regimes in DSSYK}

The regimes of the Lanczos coefficients (\ref{Lanczos_regimes_summary}) allow us to distinguish different regimes in the growth of K-complexity. The initial condition of the wave packet is $\phi_n(0)=\delta_{n0}$, where $\phi_n(t)$ are the coordinates of the state $|\phi(t)\rangle$, defined in (\ref{state_T_0}), over the Krylov basis (i.e. the basis of fixed-chord-number states); see (\ref{Phi_EnergyB}) for the precise definition. Therefore, at sufficiently early times, the state wave packet only probes the first regime of (\ref{Lanczos_regimes_summary}), where the Lanczos coefficients exhibit a square-root behavior. Following \cite{Caputa:2021sib,Balasubramanian:2022tpr} we can recycle the results for the Heisenberg-Weyl algebra. Namely, the Hamiltonian is given by
\begin{equation}
    \centering
    \label{T_Heisenberg_Weyl}
    T = \gamma \left(a + a^\dagger\right),
\end{equation}
where $\gamma= \frac{J}{\sqrt{1-q}} = \frac{J}{\sqrt{1-e^{-\lambda}}}$ and $a,\,a^\dagger$ act over the Krylov basis as ladder operators of a simple bosonic harmonic oscillator, satisfying the usual algebra
\begin{equation}
    \centering
    \label{Heisenberg_Weyl_ladder_ops}
    [a,a^\dagger]=1,
\end{equation}
i.e. in this regime the q-deformation of the algebra (\ref{q_oscillator_algebra}) is not important as long as we restrict to the subspace of states $\{|n\rangle\;/\;n\lesssim n_*(q)\}$. Using the Baker-Campbell-Hausdorff formula and the simple commutation relation (\ref{Heisenberg_Weyl_ladder_ops}), the exact wave function $\phi_n(t)$ is computed in \cite{Caputa:2021sib}, as this is nothing but the evolution along a one-parameter family of coherent states:
\begin{equation}
    \centering
    \label{Heisenberg_Weyl_Wave_Fn}
    \phi_n(t) = e^{-\frac{\gamma^2 t^2}{2}}\frac{(-i\gamma t)^n}{\sqrt{n!}} ~.
\end{equation}
The result (\ref{Heisenberg_Weyl_Wave_Fn}) for the Krylov space wave function can equivalently be obtained using the spectral decomposition of the tridiagonal Hamiltonian. Using this method, in section~\ref{subsect:formal_KC} we shall derive a formal expression for the wave functions $\phi_n(t)$ at arbitrary $q$ and show that it reduces to (\ref{Heisenberg_Weyl_Wave_Fn}) in the $q\to 1$ limit in section \ref{Sec:qto1Limit}. Likewise, in appendix \ref{App:Wavefunctions} we present an independent diagonalization of the tridiagonal Hamiltonian at $q=1$.

From this wave function K-complexity is given by:
\begin{equation}
    \centering
    \label{KC_early_times}
    C_K(t)=\sum_{n=0}^{+\infty}n|\phi_n(t)|^2=\gamma^2\,t^2 = \frac{(tJ)^2}{1-q}=\frac{(tJ)^2}{1-e^{-\lambda}}\,.
\end{equation}
For small $\lambda$, we can approximate (\ref{KC_early_times}) as $C_K(t)\approx\frac{(tJ)^2}{\lambda}$, up to subleading corrections in a $\lambda$-expansion.

We can use $C_K(t)$ as an estimate for the position of the peak of the coherent packet $\phi_n(t)$. Thus, the packet will start to probe the second region of (\ref{Lanczos_regimes_summary}), where the Lanczos coefficients approach a horizontal asymptote, when $C_K(t)$ becomes of the order of $n_*(q)$. We therefore define a transition time scale $t_*(q)$ by the relation
\begin{equation}
    \centering
    \label{tstar_defining_equation}
    C_K\Big(t_*(q)\Big) = n_*(q),
\end{equation}
giving
\begin{equation}
    \centering
    \label{tstar}
    t_*(q) = \frac{1}{J}\sqrt{\frac{1-q}{\log\left(\frac{1}{q}\right)}}=\frac{1}{J}\sqrt{\frac{1-e^{-\lambda}}{\lambda}},
\end{equation}
which behaves as $t_*\approx J^{-1}$ for small $\lambda$.
We note that, as $\lambda$ goes to zero, the transition value for $n$ goes to infinity, since $n_*=\frac{1}{\lambda}$. On the other hand, K-complexity at early times, $C_K(t)\sim \frac{1}{\lambda}(tJ)^2$, has an accordingly increasing growth rate, so that the $\lambda$-dependence cancels out in the expression of the transition time $t_*$, which is only controlled by $J^{-1}$.

\begin{figure}
    \centering
    \includegraphics[width=7.4cm]{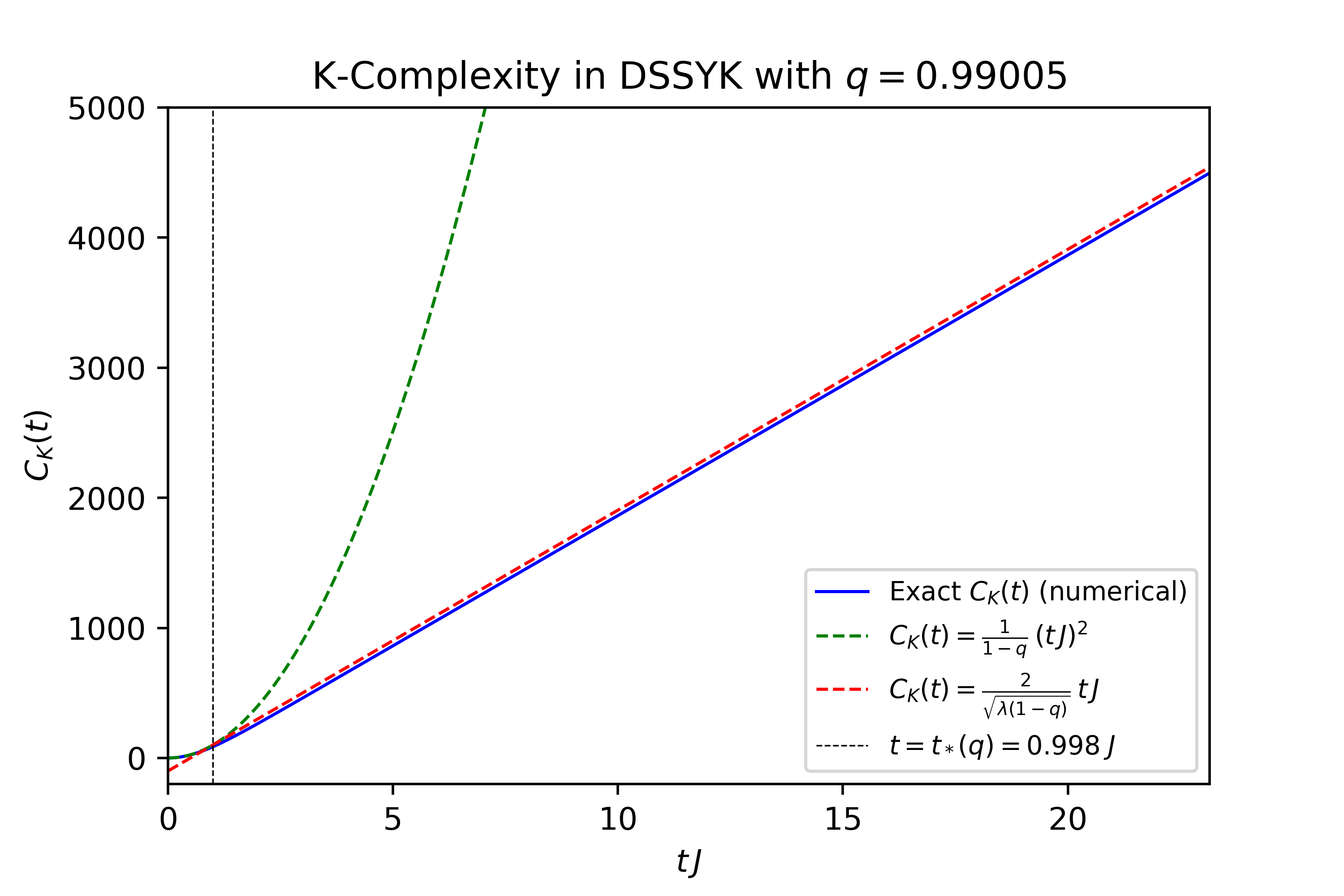} \includegraphics[width=7.4cm]{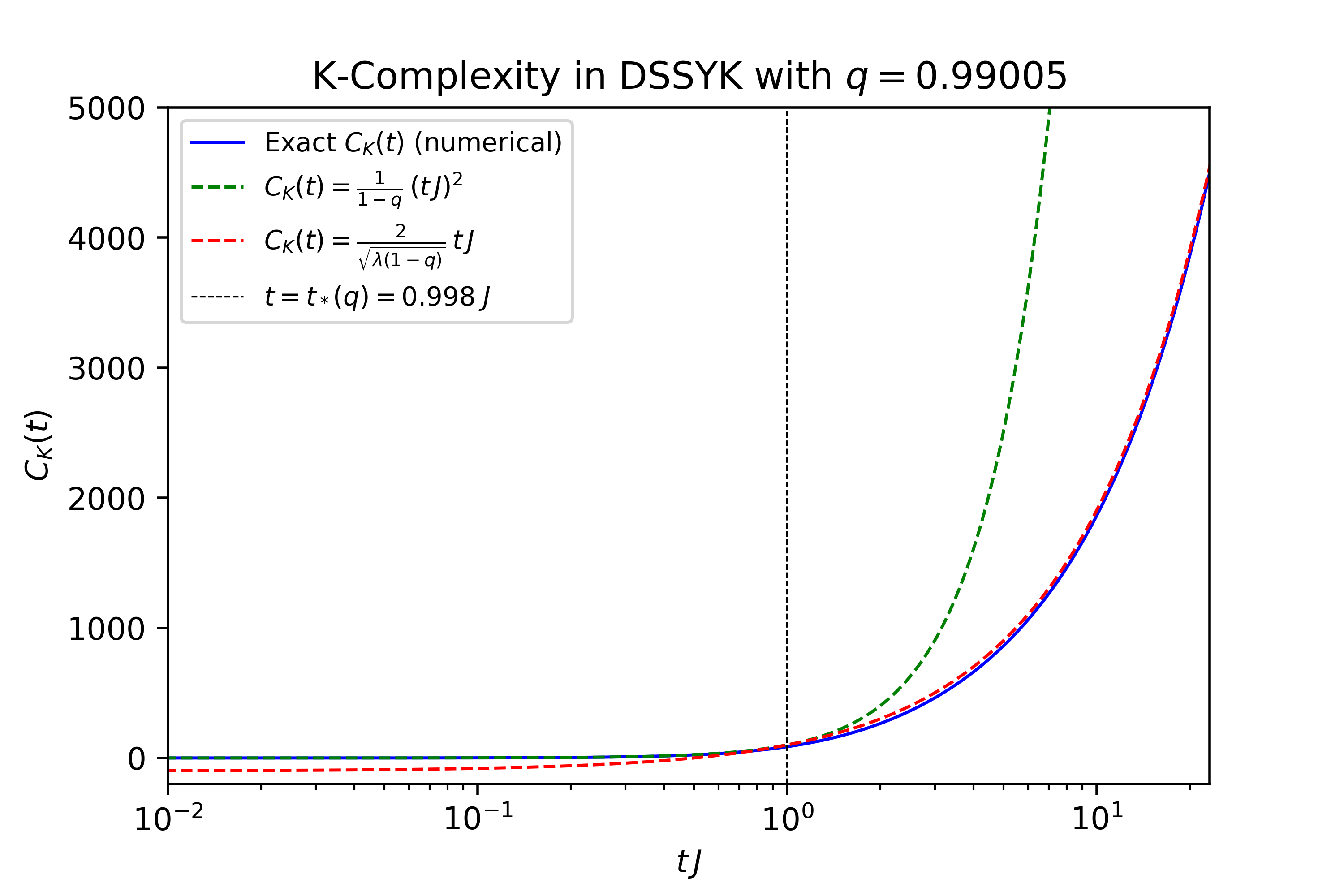}
    \caption{K-complexity regimes in DSSYK. The plots compare the exact $C_K(t)$ computed numerically (in blue) with the early (green) and late (red) time regimes described in (\ref{KC_regimes_summary}). The numerical computation imposed an artificial cutoff $K=5000$ in the length of the Lanczos sequence, but the plot only depicts a time range during which the wave packet doesn't yet probe the right edge of the artificially finite Krylov chain, hence not featuring spurious finite-size effects. \textbf{Left:} Linear scale on both axes. Note that the curve for the late-time approximation is parallel to the exact result but it doesn't exactly overlap with it; this is because of the lack of knowledge of the constant $c(\lambda)$ in (\ref{KC_regimes_summary}), which in the case of this plot was chosen to make the early- and late-time regimes match at $t=t_*(q)$. \textbf{Right:} Logarithmic scale along the horizontal axis. This allows us to note the good agreement at early times between the exact result and the corresponding approximation. Additionally, the logarithmic scale washes away the effect of the constant $c$, and the red and blue curves do overlap at late times, confirming the expected linear growth of $C_K(t)$ at times well after $t_*(q)$.}
    \label{fig:KC_numerics}
\end{figure}

Well after $t_*(q)$ the wave packet probes a region where the Lanczos coefficients are effectively constant, $b_n\approx b_{\infty}\equiv \frac{J}{\sqrt{\lambda(1-q)}}$. According to \cite{Barbon:2019wsy}, the wave functions $\phi_n(t)$ in this case are given by Bessel functions, the position of whose front-most peak evolves in time as $\sim 2 b_\infty \,t$. Using the peak position as an estimate for K-complexity, we obtain in this case:
\begin{equation}
    \centering
    \label{KC_well_after_tstar}
    C_K(t)\approx \frac{2\,tJ}{\sqrt{\lambda(1-q)}}\,. 
\end{equation}
In summary:
\begin{equation}
    \centering
    \label{KC_regimes_summary}
    C_K(t)= \begin{cases}
        \frac{(tJ)^2}{1-q}\,,& t\lesssim t_*(q) \\
        \frac{2\,tJ}{\sqrt{\lambda(1-q)}}\,+c(\lambda)\,,&  t\gg t_*(q)
    \end{cases} 
\end{equation}
where $c(\lambda)$ is some $\lambda$-dependent constant that should, strictly speaking, be there in order to ensure the matching of the two regimes.
For small $\lambda$ this becomes:
\begin{equation}
    \centering
    \label{KC_regimes_summary_small_Lambda}
    C_K(t)\approx \begin{cases}
        \frac{(tJ)^2}{\lambda}\,,& t\lesssim \frac{1}{J} \\
        \frac{2\,tJ}{\lambda}+c(\lambda)\,,& t\gg \frac{1}{J}
    \end{cases} .
\end{equation}
Figure \ref{fig:KC_numerics} depicts the different regimes of K-complexity, compared with the numerical result, for a value of $q$ close to $1$.

\subsection{Continuum approximation to K-complexity in DSSYK}\label{subsect_continuum_approx}

In this section we obtain an analytic expression for the K-complexity of the infinite-temperature thermofield ``double'' state in DSSYK (i.e. the zero-chord-number state) making use of a continuum approximation of Krylov space. Generically, the approximation developed in \cite{Barbon:2019wsy}, reviewed in appendix \ref{Appx:Cont_approx}, consists of promoting $n$ to a continuous variable in such a way that the recursion relation describing the evolution of the wave function $\phi_n(t)$ becomes a differential equation which is, at leading order, a chiral wave equation with a velocity field given by $v(n)\equiv 2 b_n$. As a result of this equation, the wave packet propagation is ballistic and one can estimate the value of K-complexity by the position of the peak of the wave function, which behaves as a point particle propagating through the above-mentioned velocity field $v(n)$. As reviewed in appendix \ref{Appx:Cont_approx}, for a finite Krylov space of dimension $K$, this continuum approximation is controlled by the small parameter $\varepsilon=\frac{1}{K-1}$, but it turns out that, due to some divisions by $\varepsilon$ in intermediate steps of the discussion, the applicability of this approximation in the strict thermodynamic limit $K\to\infty$ is not well under control, even if strictly speaking $\varepsilon=0$ in that limit. 

In the particular model at hand in this article, however, the parameter $\lambda$ in (\ref{lambda_def}) can be used to control the continuum approximation even in the $K\to\infty$ limit, as we shall argue here. Inspecting the expression of the Lanczos coefficients (\ref{Lanczos}) we note that we can define a dimensionless position variable as $x \equiv \lambda n$ in such a way that its spacing goes to zero when $\lambda\to 0$, becoming a continuous variable. More precisely, a continuum limit for the Krylov space of DSSYK can be defined as the limit in which $\lambda\to 0$, $n\to\infty$ with $x=\lambda n$ fixed. In this limit, the Lanczos coefficients adopt the limiting form of a continuous function of $x$:
\begin{equation}
    \centering
    \label{Lanczos_cont_lambda}
    b_n=J\sqrt{\frac{1-e^{-\lambda n}}{\lambda(1-q)}}\longrightarrow \frac{J}{\lambda}\sqrt{1-e^{-x}}+\mathit{O}(\lambda^0)\equiv b(x).
\end{equation}
As in the argument reviewed in appendix \ref{Appx:Cont_approx}, the starting point for studying this continuum limit is the recursion relation satisfied by the wave function. Defining a real $\varphi_n(t)$ such that $\phi_n(t)=i^{n}\varphi_n(t)$, we have that (\ref{HKB}) and (\ref{phi_t_phi_n}) imply the recursion:
\begin{equation}
    \centering
    \label{recurrence_varphi_n_maintext}
    \dot{\varphi}_n(t) = b_n\varphi_{n-1}(t)-b_{n+1}\varphi_{n+1}(t).
\end{equation}
Since in the above-mentioned $\lambda\to 0$ limit the Lanczos coefficients $b_n$ are given by a continuous function loosely denoted by $b(x)$, we can assume that the wave functions $\varphi_n(t)$ are also described by a continuous function $f(t,x)$ such that $\varphi_n(t)=f(t,n\lambda)$. With this, the recursion (\ref{recurrence_varphi_n_maintext}) reads:
\begin{equation}
    \centering
    \label{rec_cont_lambda}
    \partial_t f(t,x) = b(x)f(t,x-\lambda)-b(x+\lambda) f(t,x+\lambda).
\end{equation}
Expanding in powers of $\lambda$ we find:
\begin{equation}
\label{rec_cont_lambda_chiral}
\partial_t f(t,x) = -v(x)\partial_x f(t,x)-\frac{v^\prime(x)}{2}f(t,x)+\mathit{O}(\lambda), 
\end{equation}
which at leading order becomes a first-order wave equation for $f(t,x)$ with velocity field $v(x) \equiv 2\lambda b(x)$. Thanks to the fact that $b(x)$ scales like $\frac{1}{\lambda}$ in \eqref{Lanczos_cont_lambda}, the velocity field $v(x)$ is of order $\lambda^0$:
\begin{equation}
    v(x)=2\lambda b(x) = 2J\sqrt{1-e^{-x}}+\mathit{O}(\lambda)\overset{\lambda\to 0}{\longrightarrow} 2J\sqrt{1-e^{-x}}.
\end{equation}
In order to manipulate \eqref{rec_cont_lambda_chiral} further, we change the position variable to $y$ such that $x=0\Rightarrow y=0$ and $dy = \frac{dx}{v(x)}$. This, together with the wave function redefinition
\begin{equation}
    \centering
    \label{g_f_cont_lambda}
    g(t,y) \equiv \sqrt{v\left(x(y)\right)}f\left( t,x(y) \right), 
\end{equation}
yields the equation
\begin{equation}
    \label{g_chiral_lambda}
    \left(\partial_t+\partial_y\right)g(t,y)=0+\mathit{O}(\lambda).
\end{equation}
Equation (\ref{g_chiral_lambda}) becomes exact for $\lambda=0$. This reasoning is analogous to the analysis in appendix \ref{Appx:Cont_approx}, where the small parameter is $\varepsilon=\frac{1}{K-1}$ instead of $\lambda$; however, in that case redefinitions like (\ref{g_f_cont_lambda}) and $dy=\frac{dx}{v(x)}$ turn out to be problematic because they imply multiplication or division by zero when $\varepsilon=0$, which does not occur in the present case because $v(x)$ remains finite in the $\lambda\to 0$ limit.

The initial condition $\phi_n(t)=\delta_{n0}$ on the Krylov chain translates, in the continuum limit, into $f(0,x)=\delta(x)$, and therefore $g(0,y)$ is also proportional to $\delta(y)$. We note that, given an initial condition $g(0,y)\equiv g_0(y)$, the solution of \eqref{g_chiral_lambda} simply propagates the initial condition as $g(t,y)=g_0(y-t)$, and hence in our case we have $g(t,y)\propto\delta(y-t)$. Consequently, the position expectation value of the wave packet, which gives K-complexity (up to the factor $\lambda$ used to define $x=\lambda n$ in the continuum limit) is simply given by the position of the peak $x_p(t)$, which behaves as a point particle evolving in time following the velocity field $v(x)$:
\begin{equation}
    \centering
    \label{peak_position_cont_lambda}
    t = \int_0^{y_p(t)}dy = \int_0^{x_p(t)}\frac{dx}{v(x)} ~.
\end{equation}
This approximation can therefore be thought of as a classical approximation where the evolution of the wave packet is replaced by the propagation in $x$-space of a point particle. In fact, it is possible to show that the relation $\dot{x}=v(x)=2J\sqrt{1-e^{-x}}$ is a classical solution of the equation of motion generated by a Hamiltonian with an exponential potential:
\begin{equation}
    \centering
    \label{Liouville_Ham_classical}
    H^\prime \equiv E_0 + 2\lambda J \left( \frac{l_f^2k^2}{2} +\frac{2}{(2\lambda)^2} e^{-\frac{l}{l_f}} \right),
\end{equation}
where we have written $x$ as $\frac{l}{l_f}$.
$H^\prime$ is a Liouville Hamiltonian, but it is not quite equal to the effective Hamiltonian $\tilde{T}$ of DSSYK given in \eqref{Ham_l_k} at small $\lambda$. This is a manifestation of the fact that these two Hamiltonians are only classically equivalent. The classical limit\footnote{As discussed in \cite{Lin:2022rbf}, the classical limit is defined as the limit in which $\lambda\to 0$ and $\lambda k$ is held fixed. In this way, $[l,k]=i$ but $[l,\lambda k]=i\lambda\to 0$.} of \eqref{Ham_l_k} is:
\begin{equation}
    \centering
    \label{T_tilde_class}
    \tilde{T}_{\text{class}}=-\frac{2J}{\lambda}\cos(\lambda l_f k) \sqrt{1-e^{-\frac{l}{l_f}}}\qquad+\qquad \text{subleading} ~,
\end{equation}
and one can verify that Hamiltonians \eqref{Liouville_Ham_classical} and \eqref{T_tilde_class} yield the same Euler-Lagrange equation of motion $\ddot{x}=2J^2e^{-x}$, which produces the above-mentioned trajectory $v(x)$ upon choosing the initial conditions $x(0)=0$ and $\dot{x}(0)=0$. However, the two Hamiltonians differ at the quantum level. In order to retrieve a (quantum) Liouville Hamiltonian from DSSYK it is necessary, as described in section \ref{Subsect:Bulk_Hilbert_Space}, to take a triple-scaling limit in which $\lambda$ is taken to be small but $\frac{l}{l_f}=\lambda n$ is taken to be sufficiently large such that $\frac{e^{-l/l_f}}{(2\lambda)^2}=e^{-\tilde{l}/l_f}$ is fixed, in order to remain close to the ground state of the system. We will revisit this in section \ref{Section_Gravity_matching}, where we will compute K-complexity in the regime in which DSSYK is dual to JT gravity.

We now resume the computation of K-complexity in the continuous $\lambda\to 0$ limit of DSSYK. Since formally $x= \lambda n$ and $v(x)=2\lambda b(x)$, we can use \eqref{peak_position_cont_lambda} and give an approximation for K-complexity in DSSYK where we just promote $n$ to be a continuous variable and keep $\lambda$ fixed:

\begin{equation}
    \centering
    \label{n_preak_ballistic}
    \int_0^{n_p(t)}\frac{dn}{v(n)}=t,\qquad v(n)\equiv2b_n=2J\sqrt{\frac{1-q^n}{\lambda(1-q)}}.
\end{equation}
The integral (\ref{n_preak_ballistic}) can be performed analytically, yielding an implicit equation for $n_p(t)$:
\begin{equation}
    \centering
    \label{n_peak_implicit_eqn}
    t J = \sqrt{\frac{1-q}{\lambda}}\text{arctanh} \sqrt{1-q^{n_p(t)}},
\end{equation}
which can be solved for $n_p(t)$. We reach:
\begin{equation}
    \centering
    \label{KC_continuum_ballistic}
    C_K(t) \approx n_p(t) = \frac{2}{\lambda}\log \left\{ \cosh \left[ tJ\sqrt{\frac{\lambda}{1-q}} \right] \right\}.
\end{equation}
This approximation to K-complexity, which results from promoting $n$ (instead of $x=\lambda n$) to a continuous variable, is expected to be good at small $\lambda$, which is when the continuum limit (for the variable $x$) applies. The numerical analysis presented below confirms this expectation.

In section \ref{Section_Gravity_matching} we will recover this functional behavior of K-complexity from a gravity computation, being more specific about the bulk-boundary matching.
Furthermore, expression (\ref{KC_continuum_ballistic}) recovers exactly the early- and late-time regimes identified in (\ref{KC_regimes_summary}):

\begin{itemize}
    \item At early times, using $\cosh(x)=1+\frac{x^2}{2}+\mathit{O}(x^4)$ and $\log(1+x)=x + \mathit{O}(x^2)$, we find
    \begin{equation}
        \centering
        \label{KC_cont_ball_early}
        C_K(t)\approx \frac{2}{\lambda}\log \left\{ 1 + (tJ)^2 \frac{\lambda}{2(1-q)} \right\}\approx \frac{(tJ)^2}{1-q}\,,
    \end{equation}
    which matches perfectly the first line in (\ref{KC_regimes_summary}).
    \item At late times we just make use of the asymptotic behavior of $\cosh{x}\sim \frac{e^x}{2}$, giving
    \begin{equation}
        \centering
        \label{KC_cont_ball_late}
        C_K(t)\approx \frac{2}{\lambda}\log \left\{ \frac{1}{2} \exp \left[ tJ \sqrt{\frac{\lambda}{1-q} }\right] \right\} = \frac{2\,tJ}{\sqrt{\lambda(1-q)}}\,-\frac{2\log{2}}{\lambda},
    \end{equation}
    which also agrees with the second line of (\ref{KC_regimes_summary}) and even provides an estimate for the constant $c(\lambda)$.
\end{itemize}
Finally, the transition time between (\ref{KC_cont_ball_early}) and (\ref{KC_cont_ball_late}) can be estimated as the value of $t$ for which the argument of the $\cosh$ in (\ref{KC_continuum_ballistic}) becomes of order 1. This yields
\begin{equation}
    \centering
    \label{tstar_cont_ball}
    t_*(q) \approx \frac{1}{J}\sqrt{ \frac{1-q}{\lambda}},
\end{equation}
which agrees with our previous estimate (\ref{tstar}), and therefore it also behaves as $t_*\sim J^{-1}$ for small $\lambda$.

The fact that the regimes (\ref{KC_cont_ball_early}) and (\ref{KC_cont_ball_late}) match (\ref{KC_regimes_summary}) is a non-trivial check, since the analysis leading to (\ref{KC_regimes_summary}) did not assume the continuum approximation: it was achieved by analysing separately the different sectors of the Lanczos sequence and using known solutions for the discrete recurrence relation for such cases. We elaborate on this below:

\begin{itemize}
    \item At early times $t<t_*(q)$, both the continuum approximation and the exact discrete result derived using the Heisenberg-Weyl algebra yield $C_K(t)\approx \frac{(tJ)^2}{1-q}$. We may intuitively understand this agreement by noting that the continuum approximation can be seen as the classical propagation of a point particle in Krylov space, while the discrete result consists of the propagation of coherent states, which are known to behave semiclassically. 
    \item At late times $t\gg t_*(q)$ the exact, discrete solution for $\phi_n(t)$ consists of Bessel functions, see appendix \ref{App:WFq0}, whose front-most peak propagates in $n$-space at a constant velocity equal to $2b_\infty = \frac{2J}{\sqrt{\lambda(1-q)}}$. Approximating K-complexity by the position of this peak yields the estimate $C_K(t)\approx \frac{2\,tJ}{\sqrt{\lambda(1-q)}}$, which agrees with the point-particle propagation from the continuum approximation at late times. However, such ballistic approximation does not always accurately apply to the discrete, exact solution, which features a wave packet that develops a tail as it propagates through Krylov space. 
\end{itemize}

\begin{figure}
    \centering
    \includegraphics[width=7.4cm]{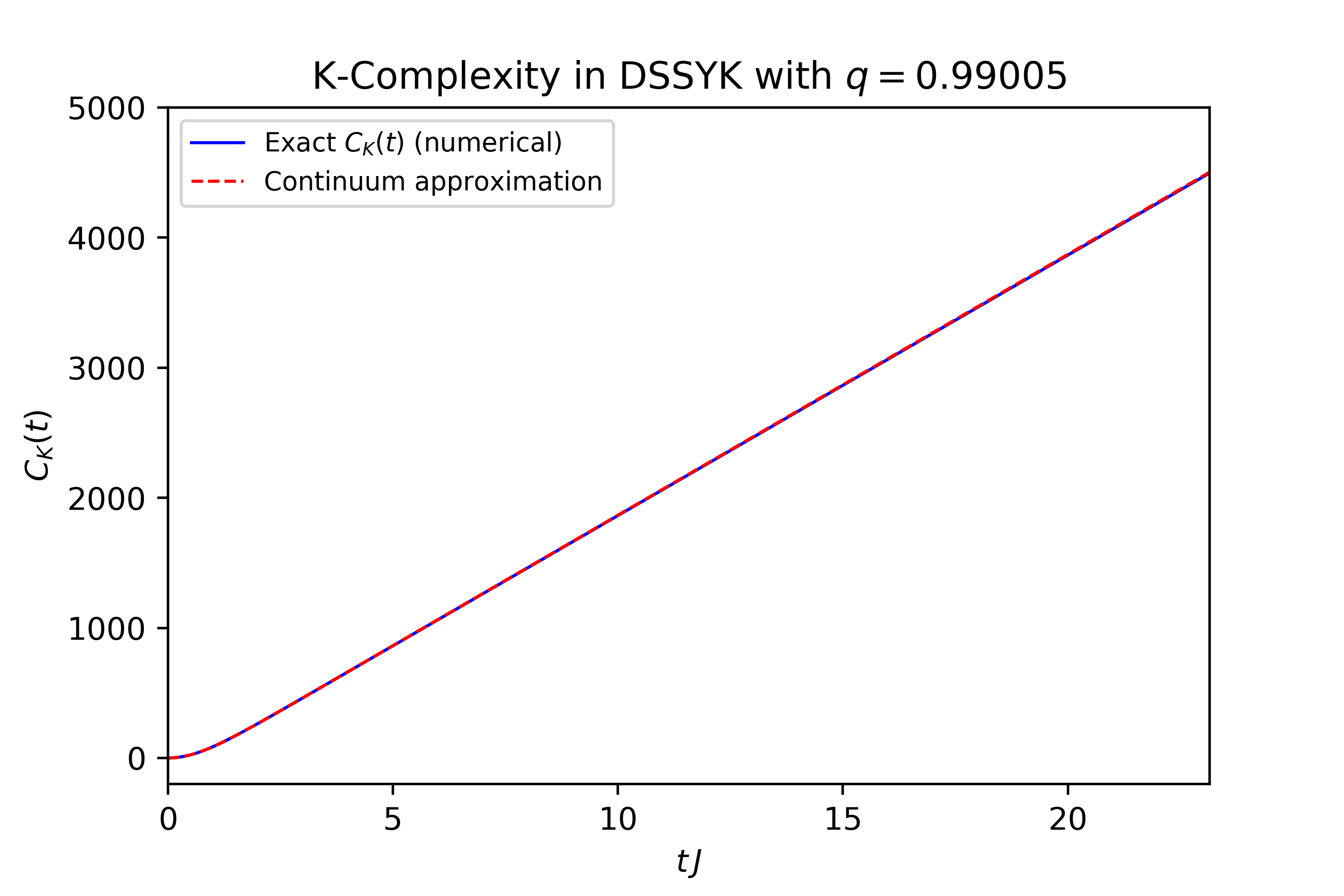} \includegraphics[width=7.4cm]{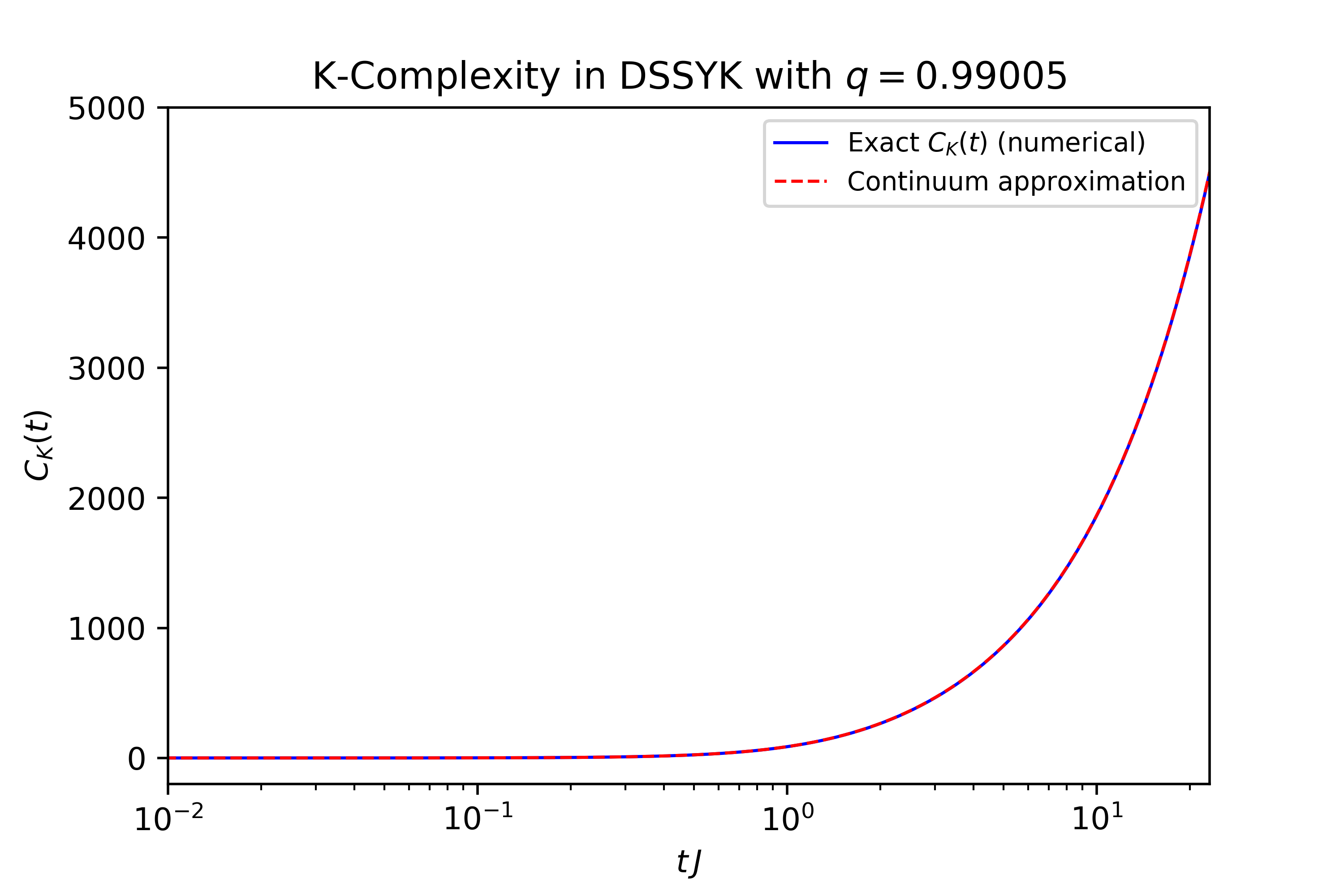} \\
    \includegraphics[width=7.4cm]{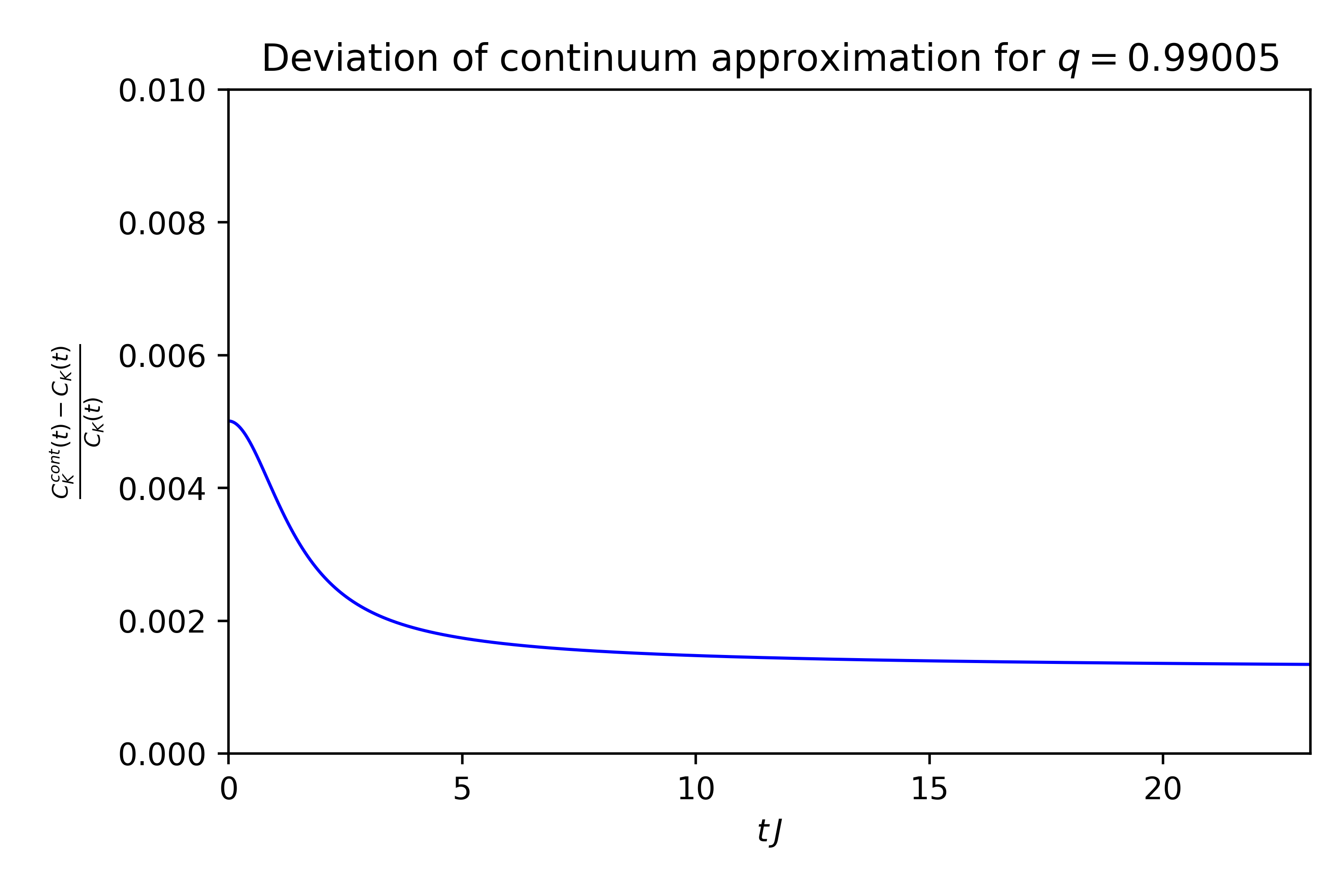}
    \caption{Comparison between the exact $C_K(t)$ computed numerically and the continuum approximation (\ref{KC_continuum_ballistic}) for DSSYK with $\lambda=0.01$ ($q\approx0.99005$), i.e. the same value as in figure \ref{fig:KC_numerics}. The agreement is excellent. \textbf{Top left:} Linear scale along both axes. \textbf{Top right:} Logarithmic scale along the horizontal axis. \textbf{Bottom:} Normalized deviation as a function of time.}
    \label{fig:KC_exact_vs_cont_approx_q0pt9905}
\end{figure}

\begin{figure}
    \centering
    \includegraphics[width=7.4cm]{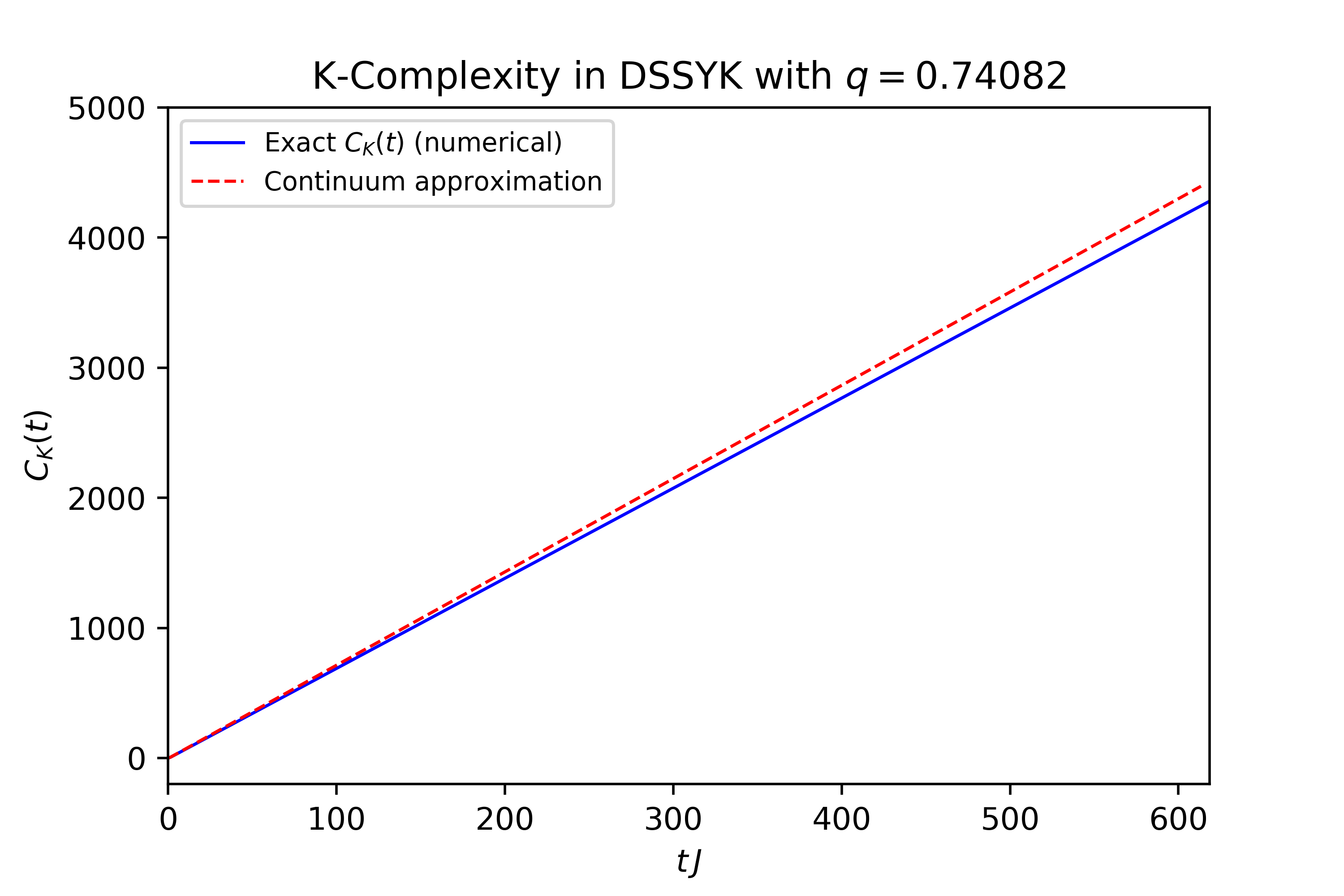} \includegraphics[width=7.4cm]{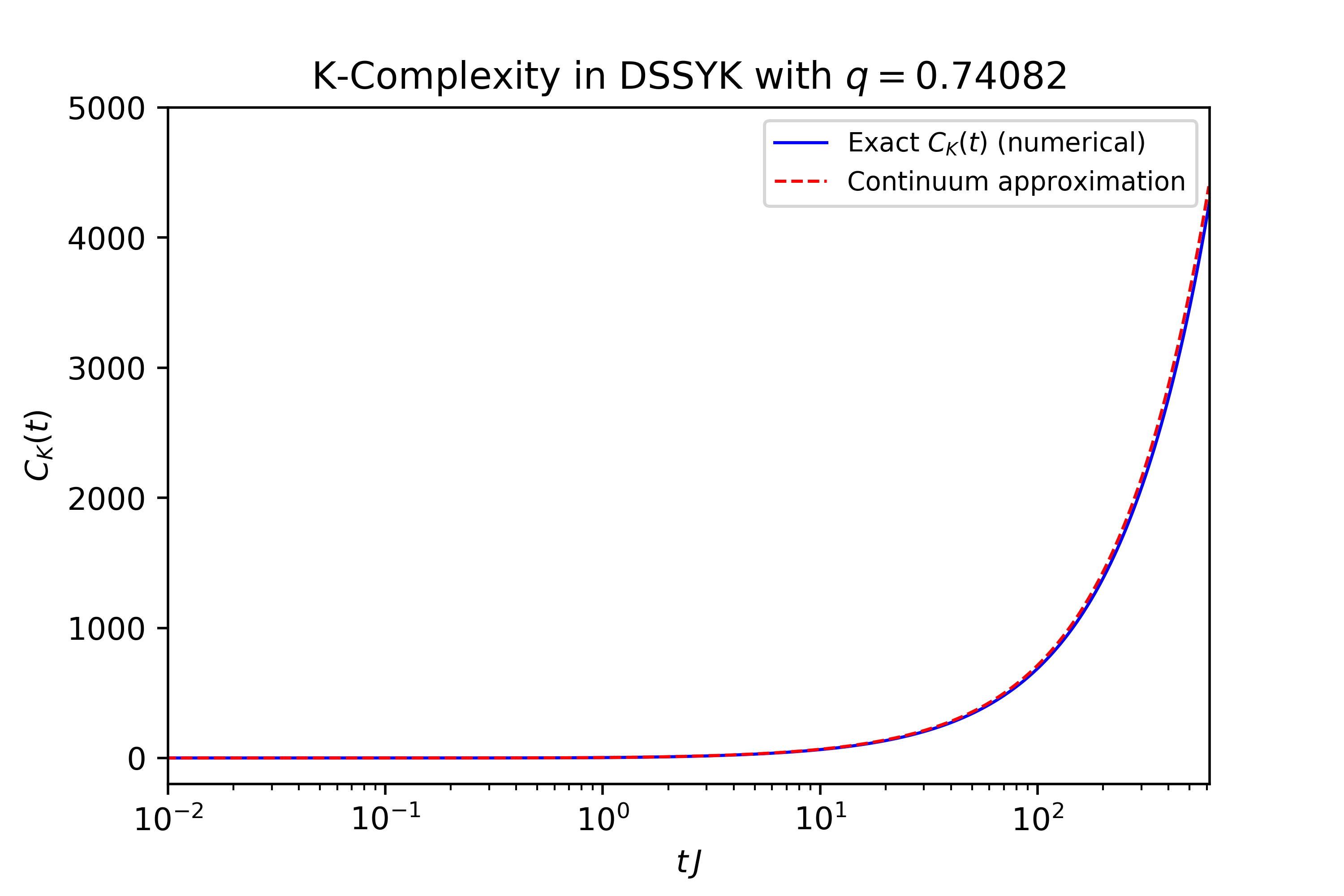} \\
    \includegraphics[width=7.4cm]{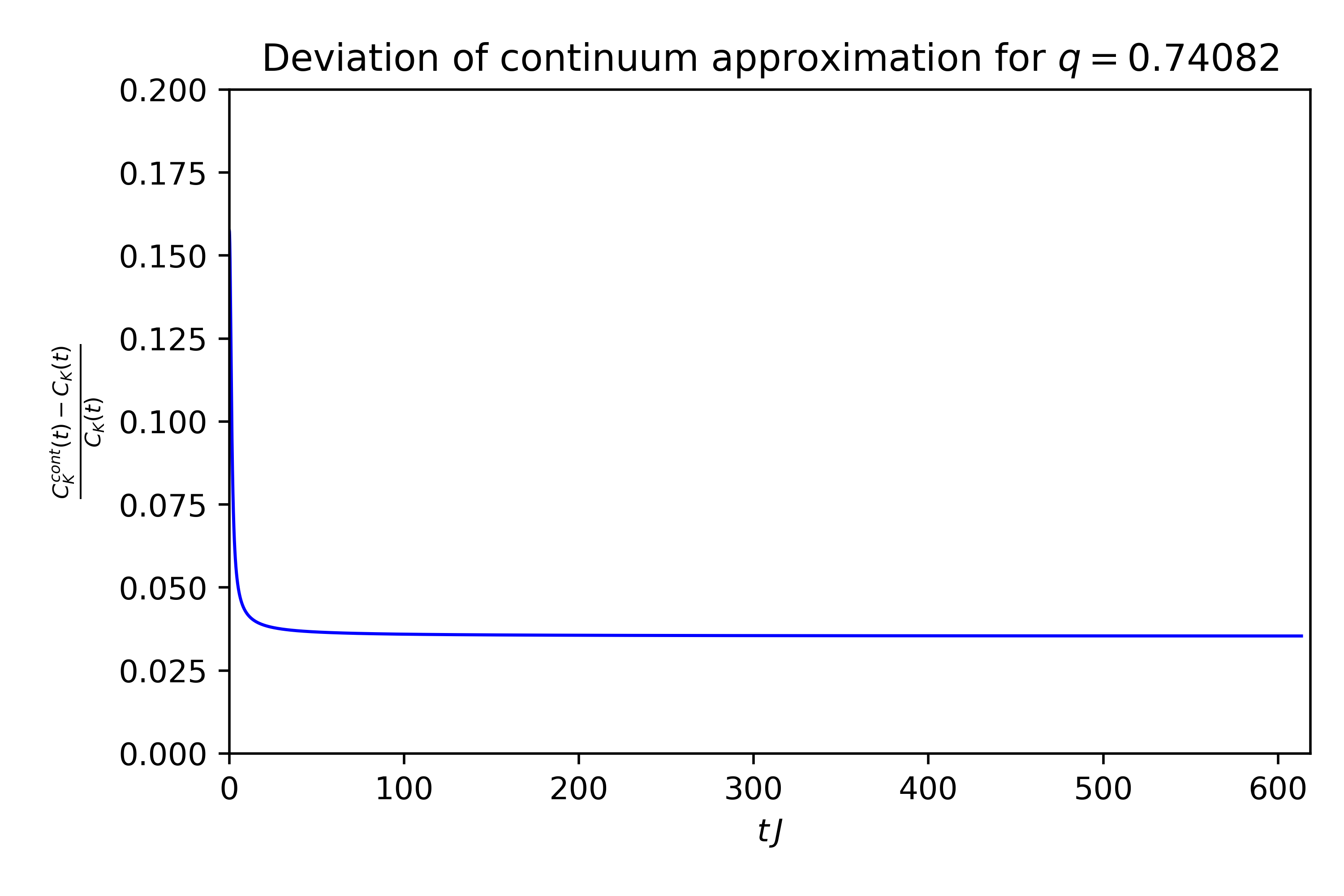}
    \caption{Continuum approximation vs exact (numerical) $C_K(t)$ for $\lambda=0.3$ (i.e. $q\approx 0.74082$).}
    \label{fig:KC_exact_vs_cont_approx_q0pt74082}
\end{figure}

\begin{figure}
    \centering
    \includegraphics[width=7.4cm]{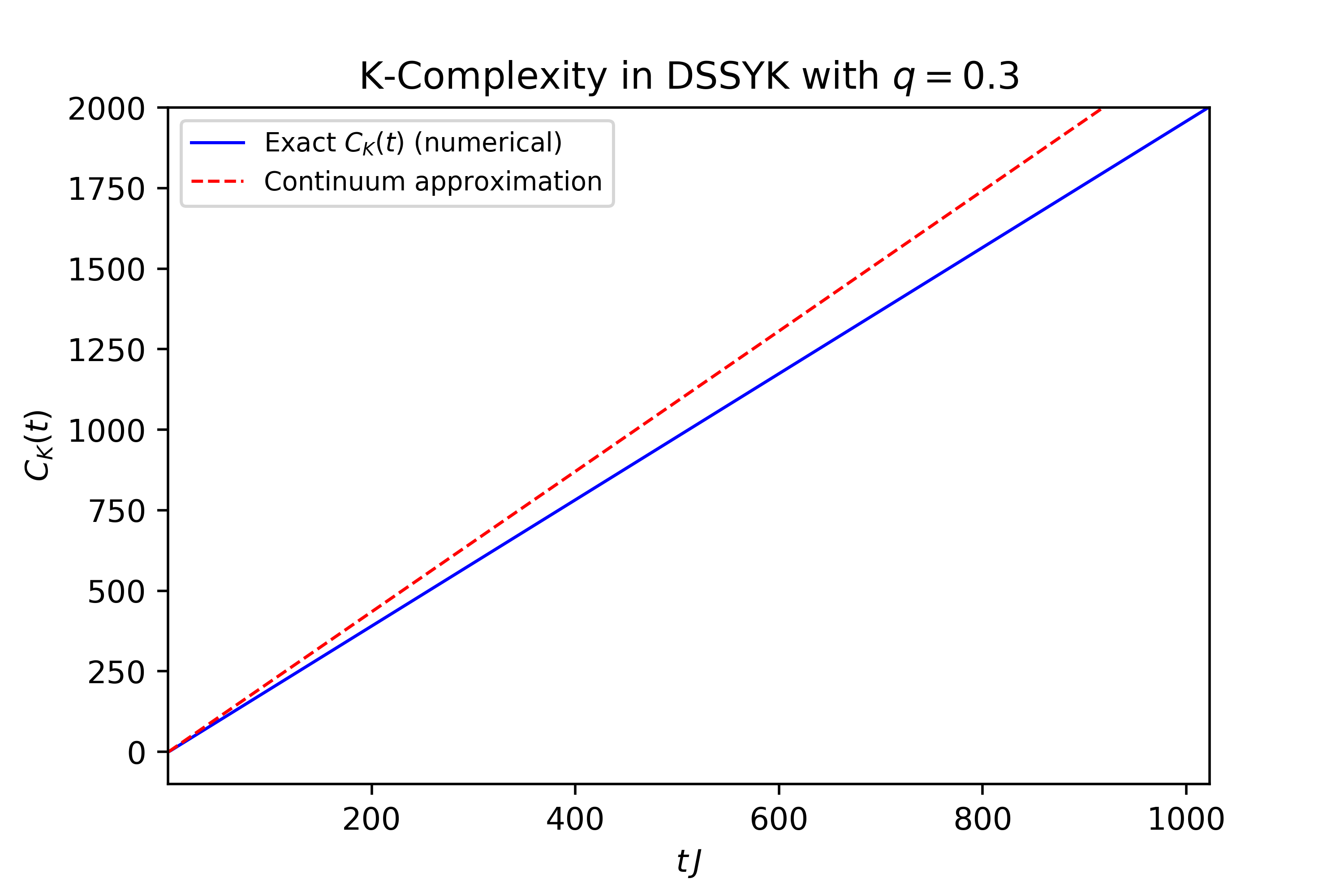} \includegraphics[width=7.4cm]{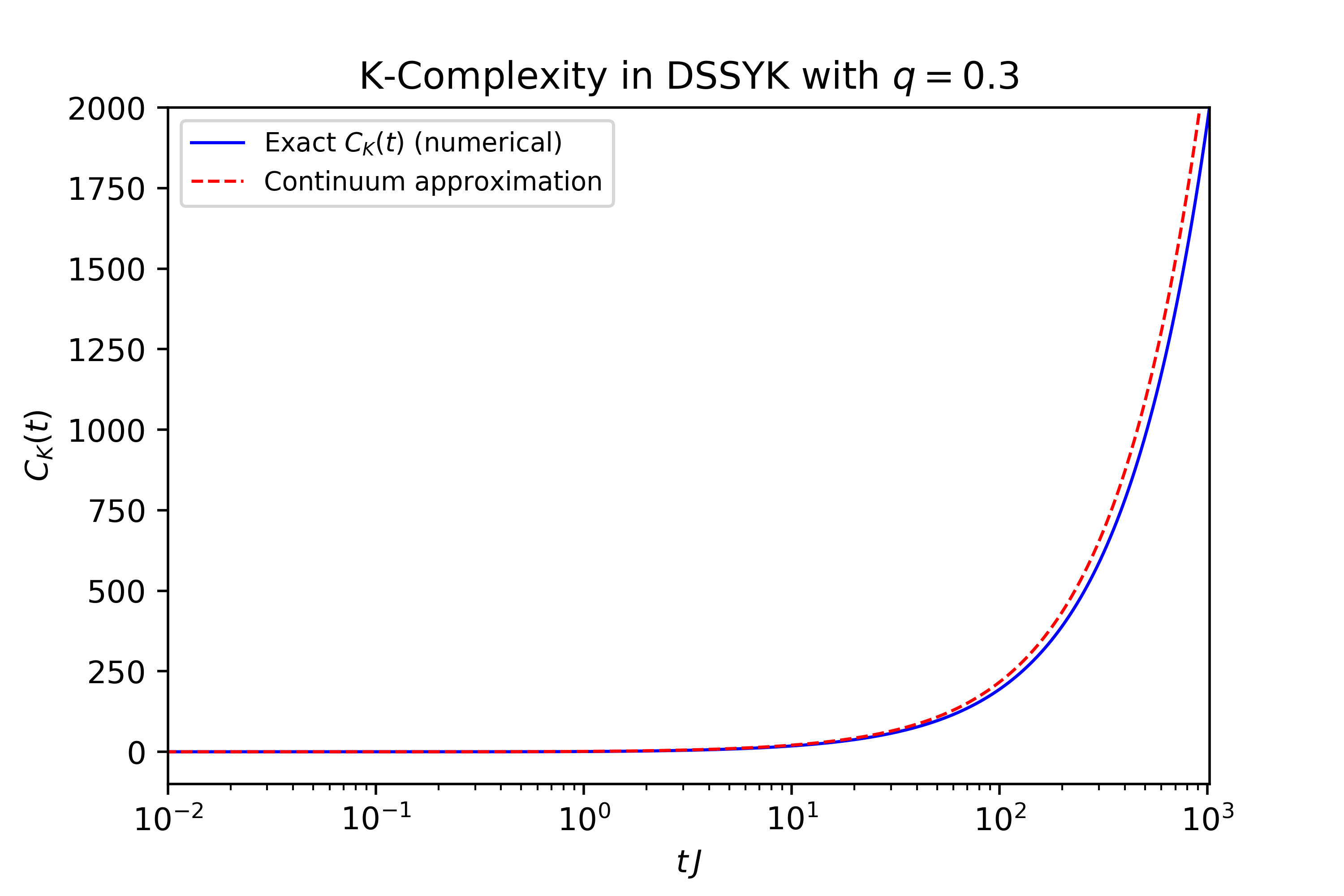} \\
    \includegraphics[width=7.4cm]{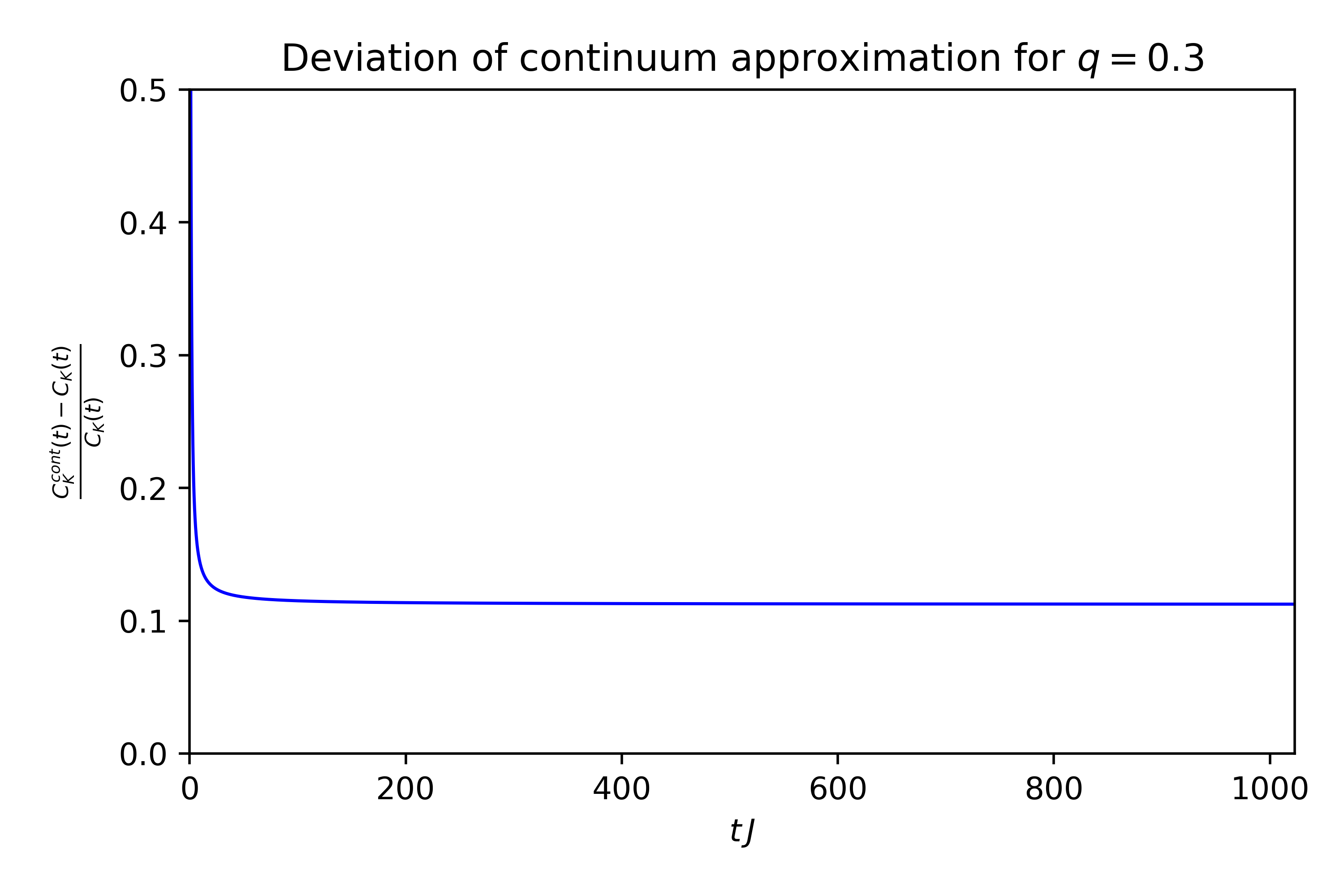}
    \caption{Continuum approximation vs exact (numerical) $C_K(t)$ for $\lambda\approx 1.20397$ (i.e. $q=0.3$).}
    \label{fig:KC_exact_vs_cont_approx_q0pt3}
\end{figure}

\begin{figure}
    \centering
    \includegraphics[width=7.4cm]{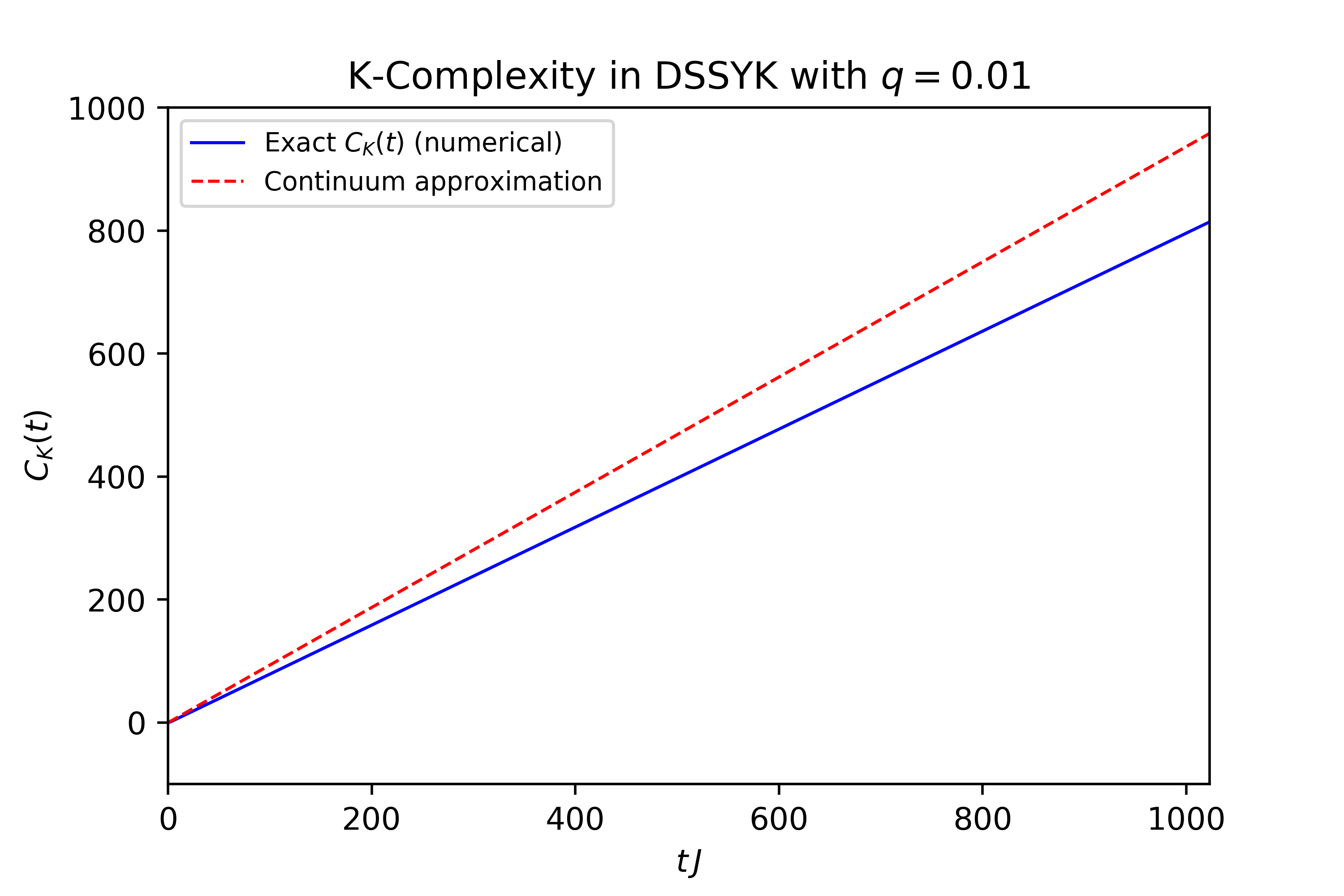} \includegraphics[width=7.4cm]{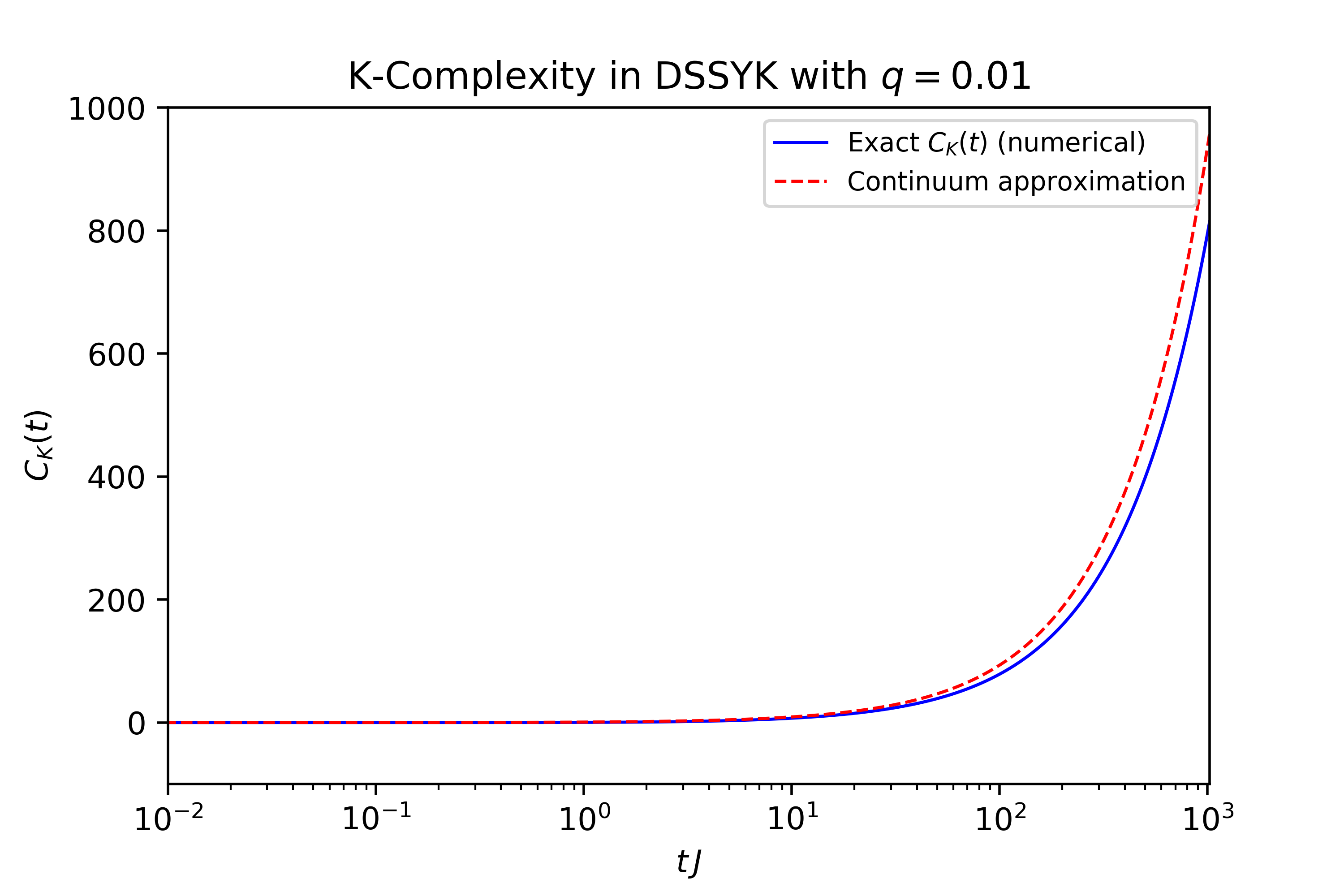}\\
    \includegraphics[width=7.4cm]{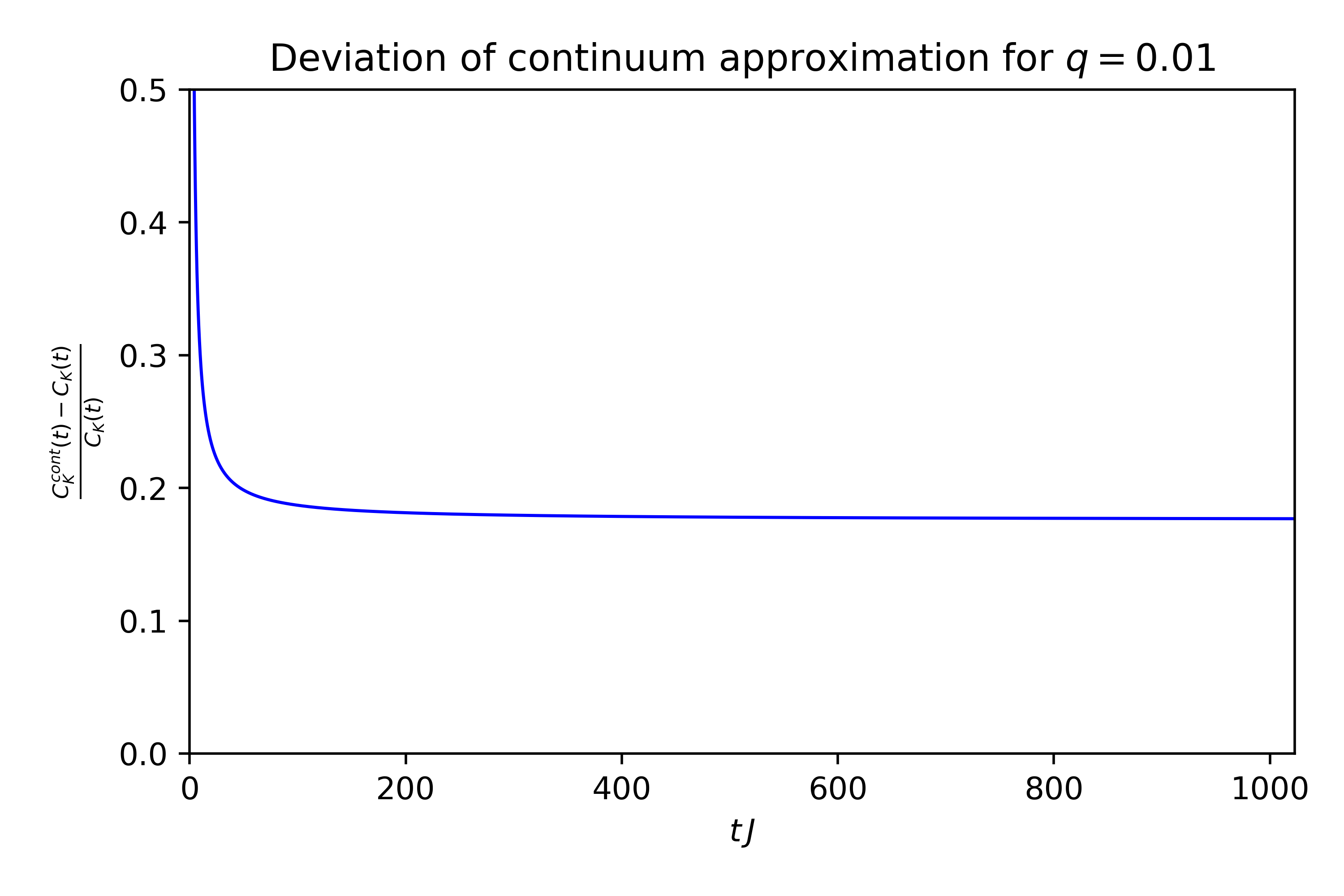}
    \caption{\textbf{Left:} Exact $C_K(t)$ computed numerically for $\lambda\approx 4.61$ (i.e. $q=0.01$), when the Lanczos sequence (\ref{Lanczos}) becomes very close to a constant $b_n\approx\frac{J}{\sqrt{\lambda}}$, vs the result of the continuum approximation. \textbf{Right:} Logarithmic scale along the horizontal axis. \textbf{Bottom:} Relative deviation. Its long-time average is $\sim 0.177$, very close to the value expected for the relative deviation of the ballistic estimate with respect to the exact result in the discrete analysis.}
    \label{fig:KC_exact_vs_cont_approx_q0}
\end{figure}

Numerics show that the continuum (and therefore ballistic) approximation is better the closer $q$ is to $1$, as illustrated in figures \ref{fig:KC_exact_vs_cont_approx_q0pt9905}, \ref{fig:KC_exact_vs_cont_approx_q0pt74082} and \ref{fig:KC_exact_vs_cont_approx_q0pt3}: we studied the values $q=0.99005,\,0.74082,\,0.3$, for which the continuum approximation induces errors of around $0.1\%$, $3\%$ and $10\%$, respectively. In fact, the exact solution for the discrete problem with constant Lanczos coefficients $b_n=b$ yields $C_K(t)=\frac{16}{3\pi}bt+o(t)\approx1.7\,bt+o(t)$, as shown in (\ref{KC_linear}). An approximation estimating $C_K(t)$ from the position of the front-most peak of the wave packet would give, even in the discrete case, $C_K(t)\sim 2bt$, resulting in a relative deviation (at late times) of $\frac{2-\frac{16}{3\pi}}{\frac{16}{3\pi}}\approx 0.18$. In figure \ref{fig:KC_exact_vs_cont_approx_q0} we observe that the relative deviation between the continuum approximation and the numerical result for K-complexity at $q=0.01$ tends precisely to this value, confirming that the error is (mainly) due to the ballistic approximation, which misses the development of a tail behind the wavefront that the exact discrete solution features. In all numerical computations we artificially truncated the Krylov chain, giving it a finite dimension $K=5000$: time-dependent results are therefore only reliable during the time interval in which the wave-packet does not yet probe this artificial edge of Krylov space, and this is the time range to which the above-mentioned plots are restricted.

For reference, we can rewrite (\ref{KC_continuum_ballistic}) for small $\lambda$ (i.e. $q$ close to $1$):
\begin{equation}
    \centering
    \label{KC_continuum_ballistic_small_lambda}
    C_K(t)\approx \frac{2}{\lambda}\log\left\{ \cosh \left(tJ\right) \right\},
\end{equation}
which in turn recovers the early- and late-time regimes listed in (\ref{KC_regimes_summary_small_Lambda}). We emphasize that this result is valid strictly in double-scaled SYK (and with small $\lambda$). In evaluating this result we have not yet taken the triple-scaling limit (\ref{Trple-scaling-limit}), for which taking $\lambda$ to be small is not enough. The derivation of K-complexity in the triple-scaling limit is deferred to Section \ref{Section_Gravity_matching}.

Understood as a classical description, the results of this continuum approximation of Krylov space can be useful for the gravity interpretation: \cite{Lin:2022rbf} discusses that, since chord number (which we now understand as K-complexity) is a discrete quantity, it provides a discretized, or quantized, version of the bulk length. At this point, we propose to make the converse argument and claim that this continuum approximation of Krylov space should match gravity results obtained from a classical, geometric description. We will provide further evidence in favor of this point of view in section \ref{Section_Gravity_matching}. A further argument in favor of interpreting the continuum approximation (\ref{KC_continuum_ballistic_small_lambda}) as a classical description of Krylov space dynamics consists in noting, as we have explained, that it coincides with the classical trajectory generated by the Hamiltonian \eqref{Liouville_Ham_classical}.

\subsection{Exact formal results for K-complexity in DSSYK}\label{subsect:formal_KC}
In this section we will present the formal result for Krylov complexity in DSSYK as a function of $q$, following the prescription for the wavefunctions in (\ref{Phi_EnergyB}).  We will use the results for the eigenvectors and eigenvalues of the effective Hamiltonian $T$ (shown at the end of section \ref{Subsection_Background_DSSYK}) directly in (\ref{Phi_EnergyB}). The sum over energies will be replaced by an integral over $\theta$, as discussed in \cite{Berkooz:2018jqr, Berkooz:2018qkz} and reviewed after equation (\ref{eigvals1}) in appendix \ref{Appx:EigSysDSSYK}. Thus:
\begin{align} 
    \phi_n(t) &= \int_0^\pi d\theta \, e^{-itE(\cos \theta)} \, \psi_n(\cos \theta)\, \psi_0(\cos \theta)\\
    & = \int_0^\pi \frac{d\theta}{2\pi} \, e^{-2iJt\frac{\cos\theta}{\sqrt{\lambda(1-q)}}} \, \frac{(q;q)_\infty}{\sqrt{(q;q)_n}}\, |(e^{2i\theta};q)_\infty|^2 \, H_n(\cos\theta|q). \label{phin_closedform}
\end{align}
In the second line we used the expressions (\ref{Teigvals}), (\ref{T0eigenvector}) and (\ref{Teigenvectors}) for the eigenvalues and eigenvectors of $T$. The small $\lambda$ limit, $q\to 1$, of this expression is discussed in section \ref{Sec:qto1Limit}.
Exact results for the wavefunctions for $q=0$ and for $q=1$ are given in appendix \ref{App:Wavefunctions}.  

It can be checked that the wavefunction is normalized, i.e. $\sum_{n=0}^\infty |\phi_n(t)|^2=1$, as follows:
\begin{align*}
    \sum_{n=0}^\infty |\phi_n(t)|^2 &= [(q;q)_\infty]^2 \int_0^\pi \frac{d\theta}{2\pi} \int_0^\pi \frac{d\phi}{2\pi} e^{-\frac{2iJt}{\sqrt{\lambda(1-q)}}(\cos\theta-\cos\phi)}|(e^{2i\theta};q)_\infty|^2 |(e^{2i\phi};q)_\infty|^2 \nonumber\\ 
    &\times \sum_{n=0}^\infty \frac{1}{(q;q)_n} 
      H_n(\cos \theta|q)H_n(\cos \phi|q) =\frac{(q;q)_\infty}{2\pi} \int_0^\pi d\theta \, |(e^{2i\theta};q)_\infty|^2 = 1
\end{align*}
where (\ref{qH_orthogonality}) was used in the second equality and (\ref{exp_qH_identity}) with $n=m=0$ was used in the final equality.

We are now ready to write down a formal expression for K-complexity in double-scaled SYK as a function of $q=e^{-\lambda}$:
\begin{align}
    C_K(t) =\sum_{n=0}^\infty n |\phi_n(t)|^2 =& [(q;q)_\infty]^2 \int_0^\pi \frac{d\theta}{2\pi} \int_0^\pi \frac{d\phi}{2\pi} e^{-\frac{2iJt}{\sqrt{\lambda(1-q)}}(\cos\theta-\cos\phi)}|(e^{2i\theta};q)_\infty|^2 |(e^{2i\phi};q)_\infty|^2 \nonumber\\ 
    &\times \sum_{n=0}^\infty \frac{n}{(q;q)_n} 
      H_n(\cos \theta|q)H_n(\cos \phi|q) ~.
\end{align}

\subsubsection{The $q\to 1$ limit} \label{Sec:qto1Limit}

In the $q\to 1$ limit we can reproduce the result (\ref{Heisenberg_Weyl_Wave_Fn}) by carefully taking the $\lambda\to 0$ limit in the general expression for the wavefunction (\ref{phin_closedform}). We begin by finding expressions\footnote{See \url{https://mathworld.wolfram.com/q-PochhammerSymbol.html}.} for $(x;q)_\infty$ and $(q;q)_\infty$ for $q=e^{-\lambda}$ in the limit $\lambda \to 0$, as well as for $H(\cos \theta|q)$ and $(q;q)_n$:
\begin{align}
    (x;q)_\infty &\approx \exp\Big[ -\frac{1}{\lambda} \mathrm{Li}_2(x)+\frac{1}{2}\log(1-x)+O(\lambda) \Big]\\
    (q;q)_\infty &\approx \sqrt{\frac{2\pi}{\lambda}} \exp \Big[ -\frac{\pi^2}{6\lambda}+\frac{\lambda}{24} +O(\lambda^2) \Big] \\
    (q;q)_n &\approx n! \lambda^n \\
    H(\cos \theta; q) &\approx 2^n \cos^n(\theta) ~.
\end{align}
From the first identity above we find that  for $q=e^{-\lambda}$ in the limit $\lambda \to 0$,
\begin{align}
    |(e^{2i\theta};q)_\infty|^2 &=(e^{2i\theta};q)_\infty(e^{-2i\theta};q)_\infty \approx  \sqrt{1-e^{2 i \theta}}\sqrt{1-e^{-2 i \theta}}\; e^ {-\frac{1}{\lambda}[\mathrm{Li}_2(e^{2i\theta})+\mathrm{Li}_2(e^{-2i\theta})]} \nonumber\\
    &=2|\sin \theta| \; e^ {-\frac{1}{\lambda}[\mathrm{Li}_2(e^{2i\theta})+\mathrm{Li}_2(e^{-2i\theta})]}~.
\end{align}

Plugging these expansions into (\ref{phin_closedform}) while keeping only first order terms in $\lambda$ and expanding $\sqrt{1-q}\approx \sqrt{\lambda}$, we have:
\begin{align}
    \phi_n(t) & \approx  2\sqrt{\frac{2\pi}{\lambda}} \frac{2^n}{\sqrt{n! \lambda^n}} \int_0^\pi \frac{d\theta}{2\pi} e^{-\frac{1}{\lambda}\left[\frac{\pi^2}{6} + \mathrm{Li}_2(e^{2i\theta})+\mathrm{Li}_2(e^{-2i\theta}) +2itJ\cos \theta \right] } |\sin \theta| \cos^n(\theta).
\end{align}
To proceed we will use the identity\footnote{See for example \url{https://dlmf.nist.gov/25.12}.}
\begin{equation}
    \mathrm{Li}_2(e^{2i\theta})+\mathrm{Li}_2(e^{-2i\theta}) = \frac{\pi^2}{3}-2\pi \theta +2\theta^2 
\end{equation}
which gives us $-\frac{1}{\lambda}[\pi^2/2 -2\pi \theta +2\theta^2] =-\frac{2}{\lambda}(\theta-\pi/2)^2$ in the exponential term. We change variables to $x = \theta -\pi/2$ to get
\begin{align}
    \phi_n(t) \approx 2\sqrt{\frac{1}{2\pi\lambda}} \frac{2^n}{\sqrt{n! \lambda^n}}\int_{-\pi/2}^{\pi/2} dx \,  e^{-\frac{2}{\lambda}(x^2-itJ\sin x)} |\cos x| (-\sin x)^n ~.
\end{align}

For $\lambda \to 0$, and $tJ\ll 1$, the main contribution to this integral is for $|x|\ll 1$, hence we expand $\sin(x) \approx x$ and $ \cos x \approx 1$:
\begin{align}
    \phi_n(t) \approx \sqrt{\frac{2}{\pi\lambda}} \frac{2^n}{\sqrt{n! \lambda^n}}\int_{-\pi/2}^{\pi/2} dx \, e^{-\frac{2}{\lambda}(x^2-2itJ x)} (-x)^n ~.
\end{align}
Completing the square, we find another Gaussian integral:
\begin{align}
    \phi_n(t) \approx \sqrt{\frac{2}{\pi\lambda}} \frac{(-2)^n}{\sqrt{n! \lambda^n}} e^{-\frac{(tJ)^2}{2\lambda}}\int_{-\pi/2}^{\pi/2} dx \, e^{-\frac{2}{\lambda}\left(x-\frac{itJ}{2}\right)^2} x^n ~.
\end{align}
Using a saddle-point approximation for $\lambda \to 0$, we set $x\approx \frac{itJ}{2}$ and perform the Gaussian integral:
\begin{equation}
     \phi_n(t) \approx \sqrt{\frac{2}{\pi\lambda}} \frac{(-2)^n}{\sqrt{n! \lambda^n}} e^{-\frac{(tJ)^2}{2\lambda}} \sqrt{\frac{\pi \lambda}{2}}\left( \frac{itJ}{2}\right)^n = e^{-\frac{(tJ)^2}{2\lambda}} \frac{(-itJ/\sqrt{\lambda})^n}{\sqrt{n!}} ~,
\end{equation}
arriving at the result (\ref{Heisenberg_Weyl_Wave_Fn}) which is indeed expected for $\lambda \to 0 $ and $tJ\ll 1$. Interestingly, this analysis agrees with our previous estimate (\ref{tstar_cont_ball}) and tells us that in the limit $q\to 1$ the transition time remains finite, $\lim_{q \to 1^{-}} t_*(q)=J^{-1}$, and hence the Heisenberg-Weyl result still only applies at early times.

\section{Bulk wormhole length and Krylov complexity} \label{Section_Gravity_matching}
We have now assembled all the ingredients necessary to carry out the main task we set ourselves in this work: giving a precise match of bulk and boundary Krylov complexity in the context of DSSYK. Let's recall the two main building blocks for the argument:

\begin{itemize}
    \item In section \ref{Subsect:Bulk_Hilbert_Space} we reviewed the identification between the bulk JT Hilbert space and the boundary Hilbert space of double-scaled SYK, whose dynamics in the triple-scaling limit are generated by the Liouville Hamiltonian. From this identification one concludes that fixed chord-number states are bulk length eigenstates.
    \item In section \ref{Sect:KC} we showed that fixed chord-number states are equal to the Krylov basis elements associated to the zero chord-number state (or the infinite-temperature thermofield ``double'' state). 
\end{itemize}

Putting these two items together it follows that \textit{the Krylov basis elements are bulk length eigenstates. Therefore, the position expectation value on the Krylov chain (i.e., K-complexity, as expressed in (\ref{KC_definition}) via the position operator (\ref{nOperator})) gives the length expectation value. This establishes the correspondence between K-complexity and two-sided bulk length}.

\begin{figure}
    \centering
    \includegraphics[scale=0.6]{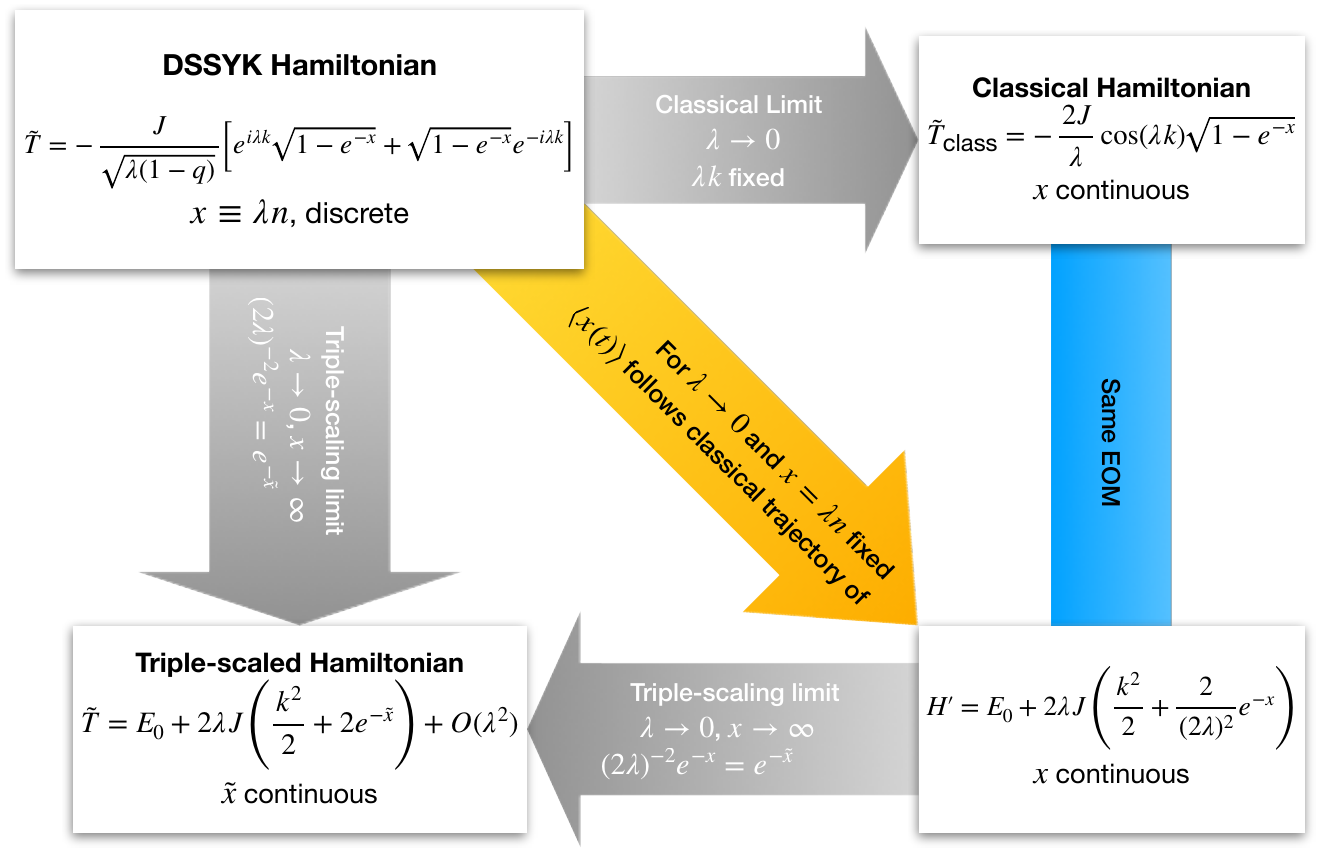}
    \caption{Diagram relating the various limiting forms of the Hamiltonian studied in this paper. The \textbf{top left} expression gives the effective Hamiltonian for the averaged theory of DSSYK, written in \eqref{Ham_l_k}. Its classical limit is the \textbf{top right} equation, stated in \eqref{T_tilde_class}. As explained in section \ref{subsect_continuum_approx}, this Hamiltonian yields the same Euler-Lagrange equation of motion as the Liouville Hamiltonian depicted in the \textbf{bottom right} block and presented in \eqref{Liouville_Ham_classical}. As indicated by the yellow arrow, this Hamiltonian is relevant because the continuous analysis of the DSSYK Hamiltonian presented in \ref{subsect_continuum_approx} yields a position expectation value that follows a classical trajectory generated by this bottom-right Liouville Hamiltonian. Finally, the \textbf{bottom left} box contains the triple-scaled Hamiltonian \eqref{Triple_scaled_Hamiltonian}: a Liouville Hamiltonian to which the full DSSYK Hamiltonian reduces in the triple-scaling limit. This limit has the effect of zooming in near the ground state, and therefore this is an effective low-energy Hamiltonian. This is the regime where DSSYK is dual to JT gravity. More technically, the introduction of the regularized position variable is required in order to obtain potential and kinetic terms of the same order in a $\lambda$-expansion, as outlined by the horizontal arrow at the bottom of the diagram. For simplicity and despite the notational abuse, in this figure $k$ denotes the (dimensionless) conjugate momentum of $x$.}
    \label{fig:Hamiltonians}
\end{figure}

In order to be more specific, 
we need to go to the regime of DSSYK in which its Hamiltonian becomes that of JT gravity. We argued in section \ref{subsect_continuum_approx} that only taking the continuous $\lambda\to 0$ limit is not sufficient for this, because in that case the DSSYK Hamiltonian is only classically equivalent to a Liouville Hamiltonian.
As reviewed in section \ref{Subsect:Bulk_Hilbert_Space}, in order to find a quantum Liouville Hamiltonian as some limiting form of the full DSSYK Hamiltonian, it is necessary to perform a triple-scaling limit \eqref{Trple-scaling-limit} that effectively zooms in near the ground state of the spectrum. In terms of the regularized length $\tilde{l}$ defined in this limit, the resulting Hamiltonian is
\eqref{Triple_scaled_Hamiltonian}, which we restate here:
\begin{equation}
    \centering
    \label{Triple_scaled_Hamiltonian_restatement_gravity_discussion}
    \tilde{T} = E_0 + 2\lambda J \left( \frac{l_f^2 k^2}{2} + 2e^{-\frac{\tilde{l}}{l_f}} \right)\;+\mathit{O}\left(\lambda^2\right).
\end{equation}
Incidentally, we note that \eqref{Triple_scaled_Hamiltonian_restatement_gravity_discussion} can be reached as the result of applying the triple-scaling limit to (\ref{Liouville_Ham_classical}). In this procedure, the regularized length introduced through the triple-scaling limit becomes necessary in order to get a well-behaved limiting Hamiltonian in which kinetic and potential terms appear at the same order. For the sake of clarity, figure~\ref{fig:Hamiltonians} depicts a diagram relating the various limiting forms of the Hamiltonian discussed in this paper.

In short, in this section we will perform a K-complexity calculation analogous to that in section \ref{subsect_continuum_approx}, specialized to the regime of DSSYK in which it becomes dual to JT gravity. For this, let's first argue that it is still possible to relate the evolution of \textit{regularized} length to K-complexity in the triple-scaled system.
Eigenstates of $\tilde{l}$ can be regarded as the continuum version of position eigenstates on a lattice with labels $\tilde{n}$ such that $\frac{\tilde{l}}{l_f}=\lambda\tilde{n}$. We note that $\tilde{n}$ differs from $n$ in a $\lambda$-dependent shift, necessary to zoom in near the ground state of the original $n$-lattice. Via an argument analogous to that in section \ref{Subsubsect_Krylov_TFD} relating the infinite-temperature thermofield ``double'' state of the full Hamiltonian to the $|n=0\rangle$ state in the averaged theory, we can associate the infinite-temperature thermofield ``double'' state of the low-energy Hamiltonian to the $|\tilde{n}=0\rangle$ state in the averaged theory.
Finally, from the fact that $\frac{\widehat{\tilde{l}}}{l_f}=\lambda \widehat{\tilde{n}}$ holds as an operator identity, taking expectation value on both sides yields the relation $\frac{\tilde{l}(t)}{l_f}=\lambda \widetilde{C_K}(t)$, between bulk regularized length in JT and the K-complexity $\widetilde{C_K}(t)$ of the infinite-temperature thermofield double state of triple-scaled SYK in the averaged theory. 

We can elaborate further on K-complexity from the point of view of the $\tilde{n}$-lattice. Its Lanczos coefficients are obtained by performing the triple-scaling protocol on (\ref{Lanczos}):
\begin{equation}
    \centering
    \label{Triple-scaled-Lanczos}
    b_{\tilde{n}}=\frac{J}{\lambda}\sqrt{1-(2\lambda)^2 q^{\tilde{n}}}+\mathit{O}(\lambda^0)=b -2\lambda J q^{\tilde{n}}+\mathit{O}(\lambda^2),
\end{equation}
where the constant $b = \frac{J}{\lambda}+\mathit{O}(\lambda^0)$ is related to the ground-state energy. We note that $q^{\tilde{n}}=e^{-\lambda \tilde{n}}$ is fixed in the triple-scaling limit in such a way that $\lambda \tilde{n}\equiv \frac{\tilde{l}}{l_f}$ does not scale with $\lambda$. In this limit, the variable $\tilde{l}$ becomes continuous; however, the analysis of the recurrence equation in Krylov space in this continuum limit is slightly different from the case in section~\ref{subsect_continuum_approx}: here, one does not reach the conclusion that $2\lambda b(\tilde{x})$ (where $\tilde{x}\equiv\frac{\tilde{l}}{l_f}$) plays the role of a velocity, because the orders of the $\lambda$-expansion of $b(\tilde{x})$ get mixed with those coming from the displacement operator. Instead, it can be checked that the differential equation that this continuous analysis yields is nothing but the Schrödinger equation dictated by the triple-scaled Hamiltonian (\ref{Triple_scaled_Hamiltonian_restatement_gravity_discussion}), consistently. Furthermore, the small-$\lambda$ prefactor of this Hamiltonian allows for a classical approximation in which the expectation value of $\tilde{l}$ is given by the solution to the equation of motion dictated by the Liouville Hamiltonian \eqref{Triple_scaled_Hamiltonian_restatement_gravity_discussion} derived on the boundary. We perform such a classical evaluation using the initial conditions $\frac{\widetilde{l}(0)}{l_{AdS}}=\widetilde{x}_0$, $\dot{\widetilde{l}}(0)=0$ and, setting $l_f=l_{AdS}$ as announced in \eqref{Param_identification_Hamiltonians}, we reach:
\begin{equation}
    \centering
    \label{Length-KC-tilde}
    \lambda \widetilde{C_K}(t)=\frac{\widetilde{l}(t)}{l_{AdS}}=\widetilde{x}_0+2\log\left\{ \cosh \left( 2\lambda J e^{-\widetilde{x}_0/2}\, t \right) \right\},
\end{equation}
which is a solution with energy\footnote{For the sake of notational simplicity, we define the energy $E$ as the difference between the total energy and the ground-state energy, i.e. $E\equiv \langle \tilde{T} \rangle-E_0$, where $\tilde{T}$ is given in \eqref{Triple_scaled_Hamiltonian_restatement_gravity_discussion}.} 
\begin{equation}
    \centering
    \label{Energy_solution_init_cond}
    E=4\lambda J e^{-\tilde{x}_0}.
\end{equation}
For fixed $\widetilde{x}_0$, this is consistent with the fact that the Hamiltonian \eqref{Triple_scaled_Hamiltonian_restatement_gravity_discussion} describes configurations that constitute excitations of energies of order $\lambda$, i.e. close to the ground state. This allows us to rewrite \eqref{Length-KC-tilde} as:
\begin{equation}
    \centering
    \label{Length-KC-tilde_E}
    \lambda \widetilde{C_K}(t)=\frac{\widetilde{l}(t)}{l_{AdS}}=2\log\left\{ \cosh \left( t\, \sqrt{E \lambda J} \right) \right\} -\log\left(\frac{E}{4\lambda J}\right).
\end{equation}
Recalling that the identification between the bulk JT Hamiltonian (\ref{JT_Liouville_Ham}) and the triple-scaled boundary Hamiltonian \eqref{Triple_scaled_Hamiltonian_restatement_gravity_discussion} reviewed in section \ref{Subsect:Bulk_Hilbert_Space} implies $2\lambda J = \frac{1}{l_{AdS} \phi_b}$, as stated in \eqref{Param_identification_Hamiltonians}, we have that \eqref{Length-KC-tilde_E} reads:
\begin{equation}
    \centering
    \label{Length-KC-tilde_E_grav}
    \lambda \widetilde{C_K}(t)=\frac{\widetilde{l}(t)}{l_{AdS}}=2\log\left\{ \cosh \left( t\, \sqrt{\frac{E}{2l_{AdS}\phi_b}} \right) \right\} -\log\left(\frac{l_{AdS}E\phi_b}{2}\right),
\end{equation}
which coincides exactly with the gravity computation of the regularized length \eqref{Ren_Wormhole_length_E}. 

To summarize, we have succeeded in demonstrating a direct match between the calculated bulk length (\ref{Ren_Wormhole_length_E}) and the boundary K-complexity in the triple-scaled limit \eqref{Length-KC-tilde_E}.  This precise match is expressed in (\ref{Length-KC-tilde_E_grav}) after making the appropriate identification between bulk and boundary parameters.

\subsection{Some remarks on the bulk-boundary matching}
We can elaborate further on the role of the initial condition $\tilde{x}_0$ in the bulk.
Comparing the expression for the configuration energy \eqref{Energy_solution_init_cond} with that in the gravity computation \eqref{energy_config_bulk}, we find that:
\begin{equation}
    \centering
    \label{init_cond_is_phi_h}
    \widetilde{x}_0=-2\log\Phi_h~.
\end{equation}
That is: from the bulk point of view, the choice of $\Phi_h$
amounts to the choice of a coordinate patch adapted to an accelerated (Rindler) observer in the boundary \cite{Spradlin:1999bn,Maldacena:2016upp} that sees the bulk as a black hole whose horizon lies at the locus where the dilaton takes the value $\Phi_h$. From the perspective of the triple-scaled Hamiltonian derived from the boundary theory, this is equivalent to choosing the initial condition for the solution of the equation of motion. The relation between such an initial condition and the dilaton field at the horizon is precisely \eqref{init_cond_is_phi_h}. In particular, the K-complexity of the infinite-temperature thermofield ``double'' state starts at zero when $t=0$, corresponding to the choice $\Phi_h = 1$.

To summarize, in this section we have revisited the need to perform a triple-scaling limit in order to find a (quantum) Liouville Hamiltonian near the ground state of DSSYK, hence retrieving the regime in which the bulk theory is described by JT gravity. This limit defines a notion of regularized length in terms of which the Krylov problem can be posed in a manner analogous to that in section \ref{Sect:KC}, where we studied DSSYK away from the triple-scaling limit. Making use of the bulk/boundary Hilbert space identification reviewed in section \ref{Subsect:Bulk_Hilbert_Space}, we were able to identify the K-complexity of the infinite-temperature thermofield ``double'' state of the triple-scaled Hamiltonian with the expectation value of bulk regularized length, which we could evaluate classically thanks to the smallness of $\lambda$, which plays the role of an $\hbar$ parameter controlling the semiclassical expansion. This correspondence is summarized in \eqref{Length-KC-tilde_E}, where the excitation energy is given by
\begin{equation}
    \centering
    \label{energy_bulk_bdry}
    E = \frac{2\Phi_h^2}{l_{AdS}\phi_b} = 4\lambda J e^{-\tilde{x}_0}.
\end{equation}
Plugging this result into either \eqref{Length-KC-tilde_E} or \eqref{Length-KC-tilde_E_grav} one recovers \eqref{Bulk_length}.

For completeness, we may now elaborate on the features of the emergent bulk described by the triple-scaled SYK Hamiltonian, together with the accelerated observer patch specified by the initial condition $\tilde{x}_0$. This discussion is based on the parameter identifications \eqref{Param_identification_Hamiltonians} and \eqref{init_cond_is_phi_h}.
The bulk consists of a black hole with a horizon radius given by\footnote{Since $l_f=l_{AdS}$, we shall avoid notational cluttering by simply denoting this length scale by $L$.} \eqref{rs_eq}:
\begin{equation}
    \centering
    \label{rs_J}
    r_s = \frac{\Phi_h}{\phi_b}L = 2\lambda JL^2e^{-\tilde{x}_0/2}~.
\end{equation}
Similarly, the black hole temperature can be computed through \cite{Brown:2018bms}:
\begin{equation}
    \centering
    \label{BH_temp_J}
    T_{\text{BH}}=\frac{r_s}{2\pi L^2} = \frac{\lambda J}{\pi}e^{-\tilde{x}_0/2}~.
\end{equation}
For the particular case of the infinite-temperature thermofield ``double'' state of triple-scaled SYK, we have $\tilde{x}_0=0$ and therefore $T_{BH}=\frac{\lambda J}{\pi}$, that is, a temperature of order $\lambda$. This is consistent with the fact that it is the infinite temperature thermofield ``double'' state of the low-energy Hamiltonian, rather than that of the full DSSYK Hamiltonian, with respect to which it is actually a low-temperature state because it zooms in near the ground state of the system.

As we have stressed, neither $r_s$ nor $T_{\text{BH}}$ are a property of global AdS$_2$, but of the coordinates adapted to an observer that sees a horizon in the same way that a Rindler observer in flat space would perceive a horizon due to its acceleration\footnote{They may be understood as an actual black hole radius and temperature whenever the 2-dimensional gravity setup comes from the near-horizon limit of a higher dimensional near-extremal black hole \cite{Brown:2018bms,Sarosi:2017ykf}. In such a case, $\Phi_h$ can also be understood as a measure of the black hole entropy.}. Accordingly, the bulk theory (\ref{Triple_scaled_Hamiltonian}) is solely controlled by the parameters $L$ and $2 \lambda J$, while the choice of a patch (specified by $\Phi_h$) is equivalent to the choice of the initial condition $\widetilde{x}_0$ for the equation of motion of \eqref{Triple_scaled_Hamiltonian_restatement_gravity_discussion}.

\section{Summary and discussion} \label{Sec:Discussion}
The main result of this work is a precise match between the (renormalized) wormhole length in JT gravity \eqref{Ren_Wormhole_length_E} and Krylov complexity in the triple-scaling limit of SYK \eqref{Length-KC-tilde_E}, for the infinite temperature TFD state. To achieve this we related the fixed-chord-number states in DSSYK to Krylov basis elements and used the fact that Krylov complexity is the expectation value of the position operator on the Krylov chain.  We then used the result of \cite{Lin:2022rbf} that in the triple-scaled limit of DSSYK, fixed-chord-number states are fixed wormhole length states in JT gravity.  Together with the identification of fixed-chord-number states as Krylov basis elements, this provides a direct match between K-complexity -- or the expectation value of position -- in the triple-scaling limit, and the wormhole length in JT gravity.   We now provide a more detailed summary and some discussion of our conclusions.

\subsection{Results for DSSYK}
We began by establishing that the fixed chord-number states, in terms of which the Hilbert space of DSSYK can be constructed, are nothing but the Krylov basis elements for the effective Hamiltonian of the averaged theory and for the infinite-temperature thermofield ``double'' state, understood as the linear combination with equal weights of all the eigenstates of the theory. The associated sequence of Lanczos coefficients was read off the matrix elements of the effective Hamiltonian expressed in coordinates over the chord basis, and they were used to compute the K-complexity profile as a function of time following various procedures involving increasingly higher levels of sophistication and detail, which we summarize below.

In a first approach, we used the different regimes (in $n$-space) of the Lanczos coefficients in order to estimate the K-complexity profile in its different time regimes. These regimes are defined as the intervals of time during which the wave packet defined over Krylov space remains mostly contained within the corresponding region where the Lanczos coefficients take a particular limiting form. We found that the $b_n$ sequence of the infinite-temperature thermofield ``double'' state transitions from a square-root growth, $b_n \sim \sqrt{n}$, to saturation at a plateau in $n$-space, which implied a K-complexity profile that transitions from quadratic growth at early times to linear growth at times greater than $J^{-1}$, the inverse of the DSSYK coupling strength. We did not obtain a saturation of K-complexity at late times as, due to the fact that we are studying the double-scaled SYK, the system is already in the thermodynamic limit, where the Hilbert space has infinite-dimension: K-complexity saturation is a non-perturbative effect that appears at finite late times that are exponentially large in the number of degrees of freedom \cite{Barbon:2019wsy, Rabinovici:2020ryf}, and therefore this time scale gets pushed away to infinity in the limit under consideration. Note that in this limit the spectrum of the effective Hamiltonian is continuous. Even for finite systems with a continuous spectrum non-perturbative behaviors require a case-by-case consideration: in some cases the behavior associated with the discreteness of the spectrum survives (e.g. in averaged but finite SYK the plateau behavior persists in the spectral form factor), while calculating a two-point function in a black hole background \cite{Maldacena:2001kr, Barbon:2003aq} yields a result that tends asymptotically to zero as a function of time. 

A further refinement consists in using a continuum approximation to Krylov space dynamics in order to obtain a smooth function $C_K(t)$ for the K-complexity profile. As a result of this approximation, the evolution of the wave packet in Krylov space was reduced to the propagation of a point particle driven by a space-dependent velocity field whose profile is given by the (continuous limit of the) Lanczos coefficients. We found that $C_K(t)$ grows in time following a log-cosh functional form that recovers the quadratic- and linear-growth behaviors in the corresponding regimes, consistent with the previous analysis. The accuracy of this continuum approximation is determined by $\lambda$, the `` 't Hooft coupling'' in DSSYK, which was found to play the role of an $\hbar$-like parameter controlling the semiclassical expansion.

We also provided a formal expression of K-complexity obtained by introducing a resolution of the identity in the definition of the Krylov space wave packet in terms of the spectral decomposition of the effective Hamiltonian of DSSYK. (The spectral decomposition of DSSYK was worked out in \cite{Berkooz:2018jqr,Berkooz:2018qkz}.) The resulting expression involves integrals and sums of q-deformed special functions. We analyzed the exact analytical K-complexity profile for specific limits of the 't Hooft coupling of DSSYK, $\lambda\to\infty$ and $\lambda\to 0$, verifying that it reduces to the expected K-complexity profiles for a flat Lanczos sequence and for a sequence dictated by the Heisenberg-Weyl algebra, respectively.

\subsection{Boundary-bulk correspondence}
As argued in \cite{Lin:2022rbf}, taking a controlled small-$\lambda$ limit in double-scaled SYK (the so-called triple-scaling limit), one zooms in near the ground state of the system. In this limit, where one expects dynamics to be governed by the Schwarzian action \cite{Maldacena:2016hyu}, one finds, consistently, that the Hamiltonian takes the form of a Liouville Hamiltonian, which is the ADM Hamiltonian of JT gravity \cite{Harlow:2018tqv}. \cite{Lin:2022rbf} exploited this further and noted that the Hilbert spaces of JT and DSSYK can be identified in such a way that fixed chord-number states are bulk length eigenstates. We combined this with our result on chord states being Krylov basis elements associated to the infinite-temperature thermofield double (TFD) state, and concluded that Krylov elements are, in turn, bulk length eigenstates. Therefore, K-complexity agrees quantitatively with bulk length through the relation (\ref{Length-KC-tilde_E_grav}), which matches the K-complexity of the TFD state of the triple-scaled Hamiltonian to the time evolution of bulk regularized length, previously computed in \cite{Harlow:2018tqv}. This matching was established upon a consistent identification of parameters in the bulk and boundary theories given in (\ref{Param_identification_Hamiltonians}) and \eqref{init_cond_is_phi_h}. The log-cosh behavior previously obtained for $C_K(t)$ is responsible for the well-known log-cosh growth of bulk length in the unperturbed geometry, which we matched including numerical pre-factors. From the bulk point of view, it is natural to argue that the lack of late-time saturation of the length as a function of time is due to the absence of (doubly) non-perturbative contributions of higher-genus, which lead to the plateau behaviour of the spectral form factor \cite{Altland:2020ccq,Altland:2022xqx} and correlation functions \cite{Altland:2021rqn}, as we are only considering the disk topology\footnote{See for example \cite{Iliesiu:2021ari} for a bulk computation of a notion of bulk volume associated to complexity which does saturate at late times due to contributions of non-trivial topology.}; this is consistent with the fact that the boundary theory is considered in the thermodynamic limit. For finite-size systems, the K-complexity late-time saturation value for the TFD state can be related to the late-time plateau of the spectral form factor, as proposed in \cite{Balasubramanian:2022tpr} and explored further in \cite{Erdmenger:2020iko}. It is worth noting, at this point, that the growth of the bulk length in a back-reacted geometry in the presence of a matter shock wave may be studied by computing the K-complexity of an operator insertion in DSSYK\footnote{We thank E. Witten for this suggestion. Work in progress.}.

\subsection{Discussion}
The agreement between the DSSYK Hamiltonian and JT gravity occurs in the triple-scaling limit (which implies a small value of $\lambda$) because this is the regime in which the SYK Hamiltonian takes a Liouville form. In this regime, the theories are dual to each other, both classically and quantum-mechanically. In this work we were able to explore this duality by a semiclassical computation of both K-complexity on the boundary and the geometric description in the bulk. We expect that quantum corrections will still match by construction, since both the Hilbert spaces and the Hamiltonians are identified \cite{Berkooz:2018jqr,Lin:2022rbf}. On top of this, going away from the small-$\lambda$ limit in DSSYK would amplify the corrections at higher orders in $\lambda$ in the effective Hamiltonian (\ref{Triple_scaled_Hamiltonian}). Hence, the gravity theory gradually deviates from JT towards the arbitrary-$\lambda$ Hamiltonian (\ref{Ham_l_k}). It would be interesting to explore this in future work. 

In this work we have focused on Krylov complexity for the infinite temperature TFD state. An important future direction is to explore other states, in DSSYK as well as in the triple-scaling limit. A class of such states are the thermal states.

The calculations in this work provide an explicit matching between K-complexity and bulk length in an instance of lower-dimensional holography. It is legitimate to wonder how generic this is, and what lessons may be extracted for higher dimensions. We advocate for an optimistic point of view and claim that this low-dimensional model may be reductionist for some aspects but not for the ones of interest: we investigated the relation between K-complexity and bulk length, using a model of holography where the gravity theory possesses a phase space consisting exclusively of the two-sided length variable (and its conjugate momentum); in this sense, the model is the simplest it can be while still preserving the ingredients relevant for the question at hand. Furthermore, some higher-dimensional scenarios involving near-extremal black holes may be reduced to a two-dimensional gravity problem when analysed in the near-horizon regime (see e.g. \cite{Sarosi:2017ykf} and references therein), which would make direct contact with JT gravity and therefore with our results. It would be interesting to work this out in detail. Another avenue to explore K-complexity in higher dimensional scenarios may consist of studying black hole microstates in microcanonical windows at fixed energy, whose Hilbert space dimension will be finite if the theory is defined on a compact manifold; the finiteness of such a Hilbert subspace should allow us to probe non-perturbative effects such as the saturation of complexity \cite{Kar:2021nbm}.

\section*{Acknowledgements}
It is a pleasure to thank Vijay Balasubramanian, José Barb\'on, Micha Berkooz, Dami\'an Galante, Vladimir Navrolansky and Edward Witten for insightful discussions.
This work has been supported in part by the Fonds National Suisse de la Recherche Scientifique (Schweizerischer Nationalfonds zur F\"orderung der wissenschaftlichen Forschung) through Project Grant 200020\_182513, and the NCCR 51NF40-141869, The Mathematics of Physics (SwissMAP). ER would like to thank the special fund for high energy physics of the PBC for partial support of this work. RS would like to thank Vladimir Narovlansky for a discussion about chord diagrams some years ago. We would like to thank Jos\'e Barb\'on for some insightful correspondence on the continuum limit of Krylov space.
\appendix 

\section{Coupling variance choices} \label{App:coupling_variance}

For reference, this appendix discusses the relation between the various choices of SYK coupling variance in the literature. Different choices are suitable depending on the specific limit in which the system is studied, as they will produce bounded results for observables such as expectation values of operators and correlation functions.

In \cite{Maldacena:2016hyu}, the SYK model is studied analytically at large $N$, first at fixed $p$ and later on at also large $p$. The Hamiltonian in that article is:

\begin{equation}
    \label{SYKHam_Malda_fixed_q}
    H^{(1)} = i^{p/2}\sum_{1\leq i_1<...<i_p\leq N} J^{(1)}_{i_1...i_p} \chi_{i_1}...\chi_{i_p},
\end{equation}
where the Majoranas are normalized such that they square to $\frac{1}{2}\mathbb{1}$, that is:
\begin{equation}
    \label{Majoranas_anticom_equals_1}
    \left\{\chi_i,\chi_j\right\}=\delta_{ij}.
\end{equation}
For the sake of studying the model in the large-$N$ limit at fixed $p$, the coupling variance used in \cite{Maldacena:2016hyu} is
\begin{equation}
    \centering
    \label{Variance_Malda_large_N_fixed_q}
    \left\langle \left( J_{i_1...i_p}^{(1)} \right)^2 \right\rangle = \frac{(p-1)! J^2}{N^{p-1}}.
\end{equation}

When \cite{Maldacena:2016hyu} turns to studying SYK at large $N$ and large $p$, the Hamiltonian keeps the same structure,
\begin{equation}
    \label{SYKHam_Malda_large_q}
    H^{(2)} = i^{p/2}\sum_{1\leq i_1<...<i_p\leq N} J^{(2)}_{i_1...i_p} \chi_{i_1}...\chi_{i_p},
\end{equation}
but the coupling variance is chosen to have an extra $p$-dependent scaling to get a more uniform limit:
\begin{equation}
    \label{Variance_Malda_large_N_large_q}
    \left\langle \left( J_{i_1...i_p}^{(2)} \right)^2 \right\rangle = \frac{2^{p-1}(p-1)!}{p N^{p-1}}J^2.
\end{equation}
The origin of the $2^{p-1}$ factor is related to the normalization of the Majoranas (\ref{Majoranas_anticom_equals_1}): Since they square to $\frac{1}{2}\mathbb{1}$, they will give an additional suppression of $2^{-p}$ every time a Wick contraction pairs two Hamiltonians inside of a trace, which needs to be compensated by the coupling variance in order to obtain finite results in the large $q$ limit. More explicitly, redefining the Majoranas as
\begin{equation}
    \label{Majoranas_redef}
    \chi_i = \frac{1}{\sqrt{2}} \psi_i,
\end{equation}
we obtain operators $\psi_i$ that are normalized in a way such that they square to one:
\begin{equation}
    \label{Majoranas_anticom_equals_2}
    \left\{ \psi_i,\psi_j \right\}=2\delta_{ij}.
\end{equation}
In terms of these operators, the Hamiltonian (\ref{SYKHam_Malda_large_q}) reads:
\begin{equation}
    \label{SYKHam_Malda_large_q_normalized_Majoranas}
    H^{(2)} = i^{p/2}\sum_{1\leq i_1<...<i_p\leq N} J^{(2)}_{i_1...i_p} 2^{-\frac{p}{2}}\psi_{i_1}...\psi_{i_p}\equiv i^{p/2}\sum_{1\leq i_1<...<i_p\leq N} \widetilde{J^{(2)}}_{i_1...i_p} \psi_{i_1}...\psi_{i_p},
\end{equation}
so that the effective coupling strength variance gets modified to
\begin{equation}
    \label{Variance_Malda_large_q_effective}
    \left\langle \left( \widetilde{J^{(2)}}_{i_1...i_p} \right)^2 \right\rangle=\left\langle \left( J^{(2)}_{i_1...i_p} \right)^2 \right\rangle 2^{-p} = \frac{(p-1)!}{2pN^{p-1}}J^2.
\end{equation}

When turning to double-scaled SYK, extensively studied in \cite{Berkooz:2018jqr}, it is more enlightening to work with normalized Majoranas $\psi_i$ satisfying (\ref{Majoranas_anticom_equals_2}) since, as mentioned in section \ref{Subsection_Background_DSSYK}, the monomials $\psi_I$, where $I$ denotes a collective index $i_1...i_p$, square to a phase, and therefore the combinatorial factor coming from the sum in the Hamiltonian can be directly compensated by the normalization of the coupling strength in order to yield bounded moments of the Hamiltonian. The Hamiltonian in \cite{Berkooz:2018jqr} is:
\begin{equation}
    \label{SYKHam_DSSYK_berkooz}
    H^{(3)}=i^{p/2}\sum_{1\leq i_1<...<i_p\leq N}J^{(3)}_{i_1...i_p}\psi_{i_1}...\psi_{i_p},
\end{equation}
and the coupling strength variance is given by
\begin{equation}
    \label{Variance_berkooz}
    \left\langle \left( J^{(3)}_{i_1...i_p} \right)^2 \right\rangle = {\binom{N}{p}}^{-1}J^2.
\end{equation}
As an example, this normalization ensures that $\left\langle \text{Tr}\left[ \left(H^{(3)}\right)^2 \right] \right\rangle = J^2$, even at finite and independent $N$ and $p$.

In the current article, we are interested in double-scaled SYK but we don't use the coupling strength variance (\ref{Variance_berkooz}). Instead, we follow \cite{Lin:2022rbf,goel2023semiclassical} and use a variance which, when taken in the double-scaling limit, has an asymptotic behavior that makes contact with (\ref{Variance_Malda_large_q_effective}) in a way about which we shall be precise below. Our Hamiltonian is:
\begin{equation}
    \label{SYKHam_Ours}
    H = i^{p/2}\sum_{1\leq i_1 <...<i_p\leq N} J_{i_1...i_p}\psi_{i_1}...\psi_{i_p}.
\end{equation}
We study this system in the double-scaling limit, taking $N$ and $p$ to infinity keeping the ratio
\begin{equation}
    \label{DS_thooft_coupling}
    \lambda=\frac{2p^2}{N}
\end{equation}
fixed. Given $\lambda$, the coupling strength variance is chosen to be:
\begin{equation}
    \label{Variance_ours}
    \left\langle J_{i_1...i_p}^2 \right\rangle=\frac{1}{\lambda}{\binom{N}{p}}^{-1}\,J^2.
\end{equation}
To compare with (\ref{Variance_Malda_large_q_effective}), we can perform an asymptotic analysis already within the double-scaling limit. That is, we take (\ref{Variance_ours}) and implicitly assume everywhere that $N\equiv N(p) = \frac{2p^2}{\lambda}$. Then, for large $p$ we find the asymptotic behavior:
\begin{equation}
    \centering
    \label{Variance_ours_asymp}
    \left\langle J_{i_1...i_p}^2 \right\rangle=\frac{1}{\lambda}{\binom{N(p)}{p}}^{-1}\,J^2\sim \frac{e^{\lambda/4}}{2p}\frac{(p-1)!}{{N(p)}^{p-1}}J^2 = \frac{(p-1)!}{2p {N(p)}^{p-1}} J^2 + \mathit{O}(\lambda).
\end{equation}
Where $f(p)\sim g(p)\leftrightarrow \lim_{p\to+\infty} \frac{f(p)}{g(p)}=1$.
This means that, in the double-scaling  limit, the variance (\ref{Variance_ours}) is equivalent to the choice (\ref{Variance_Malda_large_q_effective}) upon the additional assumption of small $\lambda$, which is precisely the regime on which papers like \cite{Lin:2022rbf,goel2023semiclassical} focus, since it zooms in near the ground state of SYK and yields the Liouville Hamiltonian which makes contact with Schwarzian dynamics and hence with an AdS bulk dual.

In view of this, the small-$\lambda$ regime of DSSYK can be seen as a controlled way to approach large-$N$ and large-$p$ SYK with the hierarchy $1\ll p \ll N$, as also discussed in \cite{mukhametzhanov2023large}.

\section{A note on the triple-scaling limit}\label{Appx:triple_scaling}

A priori, there is a whole family of triple scaling limits that one can take, which applied to the Hamiltonian (\ref{Ham_l_k}) yield Liouville-like Hamiltonians with a different relative weight between the kinetic term and the potential term. The family of limits is parametrized by some $a\in\mathbb{R}$ as follows:
\begin{equation}
    \centering
    \label{triple-scaling-family}
    \lambda\to 0,\qquad l\to\infty,\qquad(a\lambda)^{-2}e^{-\frac{l}{l_f}}=e^{-\frac{\tilde{l}}{l_f}}\,\text{fixed}.
\end{equation}
Applied to (\ref{Ham_l_k}), this limit yields:
\begin{equation}
    \centering
    \label{Ham_triple_scaled_family}
    \tilde{T} - E_0 = 2\lambda J\left( \frac{l_f^2 k^2}{2} + \frac{a^2}{2}e^{-\frac{\tilde{l}}{l_f}} \right)\;+\mathit{O}\left(\lambda^2\right).
\end{equation}
However, all the Hamiltonians in the family (\ref{Ham_triple_scaled_family}) are equivalent in the sense that they only differ by a finite translation of the length variable. Their density of states is therefore the same and the eigenfunctions of one Hamiltonian in the family can be obtained from those of any other by just a shift. In our work, we choose $a=2$ (instead of the $a=1$ convention in \cite{Lin:2022rbf}) because in this case the resulting triple-scaled Hamiltonian makes contact with the form of the Liouville Hamiltonian for JT gravity written explicitly in \cite{Harlow:2018tqv}, restated in \eqref{JT_Liouville_Ham}. Nevertheless, we stress that, since all the Hamiltonians in the family (\ref{Ham_triple_scaled_family}) are equivalent in the sense that we have just explained, changing $a$ in the definition of the triple-scaling limit (\ref{triple-scaling-family}) will not modify the gravity theory fundamentally. This ambiguity in the definition of $\tilde{l}$ may be seen as related to the choice of regularization scheme for the boundary divergence from the bulk perspective.

\section{Eigenvalues and eigenvectors of the effective Hamiltonian}\label{Appx:EigSysDSSYK}
The discussion in this appendix closely follows \cite{Berkooz:2018jqr, Berkooz:2018qkz} and is presented for the sake of completeness.

The symmetric version of $T$ given in (\ref{T_sym}) satisfies, in the chord basis,
\begin{equation} \label{T_chordB}
    T\,|l\rangle = \frac{J}{\sqrt{\lambda (1-q)}} \left( \sqrt{1-q^{l+1}}\,|l+1\rangle + \sqrt{1-q^{l}} \, |l-1\rangle \right) ~.
\end{equation}
To find the eigenvalues of $T$ we write down the eigensystem equation for the components of the eigenvectors of $T$, $\psi_l(E)=\langle l|E\rangle$:
\begin{equation} \label{q-eigsys1}
    E \,\psi_l(E) = \frac{J}{\sqrt{\lambda (1-q)}} \left( \sqrt{1-q^{l+1}}\,\psi_{l+1}(E) + \sqrt{1-q^{l}} \, \psi_{l-1}(E) \right)
\end{equation}
 We note that $E$ is required to be independent of the position label $l$ as (\ref{q-eigsys1}) is an eigenvalue problem; therefore, assuming there are no bound states, we can take the large $l$ limit of the above relation where $T$ becomes a tridiagonal Toeplitz\footnote{A Toeplitz matrix is a matrix with equal entries along the diagonals. For the tridiagonal case its eigenvalues are well known, see e.g. \url{https://de.wikipedia.org/wiki/Tridiagonal-Toeplitz-Matrix}.} matrix with eigensystem equation
\begin{equation}
     E \,\psi_l(E) = \frac{J}{\sqrt{\lambda (1-q)}} \left( \psi_{l+1}(E) + \psi_{l-1}(E) \right)~.
\end{equation}
The eigenvalues of such an $L$-dimensional  tridiagonal Toeplitz matrix are given by:
\begin{equation}
    E_k = \frac{2J}{ \sqrt{\lambda(1-q)}}\cos \left(\frac{k\pi}{L+1} \right), \quad k=1,\dots, L ~.
\end{equation}
When $L$ is large the argument of the cosine, $\theta =\frac{k\pi}{L+1}$, becomes a continuous variable between $0$ and $\pi$ and the spectrum has values between $E_{min}=-\frac{2J}{\lambda \sqrt{1-q}}$ and $E_{max}=\frac{2J}{\lambda \sqrt{1-q}}$. The eigenvalues are then a function of the continuous variable $\theta$\footnote{When $l$ becomes continuous, $\theta$ can be interpreted as momentum.}:
\begin{equation} \label{eigvals1}
    E(\theta) = \frac{2J \, \cos \theta}{ \sqrt{\lambda(1-q)}}, \quad 0\leq \theta \leq \pi ~.
\end{equation}
Note that as $L\to \infty$, the variable $\theta$ becomes continuous in a uniform way and thus summing over eigenvalues $\sum_{k=1}^{L}$ becomes $\int_0^\pi d\theta$ with uniform density $\rho(\theta)=1$. This fact will be important for later results. 

To find the eigenvectors, we go back to (\ref{q-eigsys1}) using the result (\ref{eigvals1}) and defining for convenience $\mu=\cos \theta$ (we will use $v_l$ instead of $\psi_l$, reserving the latter to the final normalized result):
\begin{equation} \label{eigsys1}
    2\mu\,v_l(\mu) = \sqrt{1-q^{l+1}}\,v_{l+1}(\mu)+ \sqrt{1-q^l}\, v_{l-1}(\mu) ~.
\end{equation}
This equation can be identified once $v_l(\mu)$ is redefined as follows
\begin{equation}
    v_l(\mu)=\frac{\tilde{v}_l(\mu)}{\sqrt{(q;q)_l}}
\end{equation}
where $(a;q)_n \equiv \prod_{k=0}^{n-1}(1-a q^k)$ is the $q$-Pochhammer symbol\footnote{\url{https://en.wikipedia.org/wiki/Q-Pochhammer\_symbol}.}. Plugging this redefinition into (\ref{eigsys1}) and multiplying both sides of the equation by $\sqrt{(q;q)_l}$ gives:
\begin{equation}
    2\mu \tilde{v}_l(\mu) = (1-q^l)\tilde{v}_{l-1}(\mu)+\tilde{v}_{l+1}(\mu)
\end{equation}
which now can be identified as the recurrence relation for the \textit{q-Hermite polynomials} (see appendix \ref{Appx:qHermite}), hence the solution to (\ref{eigsys1}) is
\begin{equation}
    v_l(\mu)=\frac{H_l(\mu | q)}{\sqrt{(q;q)_l}}, \quad \mu=\cos(\theta)~.
\end{equation}
Using the identity (\ref{qH_orthogonality}) and the fact that $0\leq \theta\leq \pi$, it can be shown that the norm squared of these eigenvectors is given by
\begin{equation}
     \sum_{l=0}^\infty |v_l(\mu)|^2 = \sum_{l=0}^\infty \frac{H_l(\mu | q)H_l(\mu|q)}{(q;q)^l}=\frac{2\pi }{|(e^{2i\theta};q)_\infty|^2(q;q)_\infty}
\end{equation}
and the normalized eigenvector is given by
\begin{equation}
    \psi_l(\mu) = \sqrt{(q;q)_\infty} |(e^{2i\theta};q)_\infty| \frac{H_l(\mu|q)}{\sqrt{2\pi (q;q)_l}}, \quad \mu=\cos \theta~.
\end{equation}
We note that 
\begin{equation} 
    \psi_0(\mu)=  \sqrt{(q;q)_\infty} |(e^{2i\theta};q)_\infty| \frac{H_0(\mu|q)}{\sqrt{2\pi (q;q)_0}} =  \sqrt{\frac{(q;q)_\infty}{2\pi}} |(e^{2i\theta};q)_\infty|
\end{equation}
where we used $H_0(x|q)=1$ and $(q;q)_0=1$.  We thus can write
\begin{equation} 
\label{Teigenvectors1}
    \psi_l(\mu)= \psi_0(\mu) \frac{H_l(\mu|q)}{\sqrt{ (q;q)_l}} , \quad \mu=\cos \theta.
\end{equation}
The orthogonality relations for these eigenvectors are shown in appendix \ref{App:Orthogonality}.

\subsection{q-Hermite Polynomials} \label{Appx:qHermite}
The q-Hermite polynomial is defined as\footnote{\url{https://en.wikipedia.org/wiki/Continuous\_q-Hermite\_polynomials}.} 
\begin{equation}
    H_n(\cos \theta|q)= e^{in\theta} {}_2 \phi_0\Big[ \begin{matrix}
        q^{-n}, 0 \\ - 
    \end{matrix};\; q,q^ne^{-2i\theta} \Big]
\end{equation}
where ${}_2 \phi_0$ is a q-hypergeometric function, defined by\footnote{\url{https://en.wikipedia.org/wiki/Basic\_hypergeometric\_series}.}
\begin{equation}
    {}_j \phi_k\Big[ \begin{matrix}
        a_1, a_2, \dots, a_j \\ b_1, b_2, \dots, b_k
    \end{matrix};\; q,z \Big] = \sum_{n=0}^\infty \frac{(a_1;q)_n\, (a_2;q)_n\, \dots (a_j;q)_n}{(b_1;q)_n\, (b_2;q)_n\, \dots (b_k;q)_n} \Big[ (-1)^n q^{n(n-1)/2}\Big]^{1+k-j} \frac{z^n}{(q;q)_n} ~.
\end{equation}
Using this definition we have
\begin{equation}
    H_n(\cos \theta|q)=  \sum_{k=0}^\infty \frac{(q^{-n};q)_k}{(q;q)_k} (-1)^k q^{nk-\frac{k(k-1)}{2}} e^{i(n-2k)\theta}
\end{equation}
Now note that, $(q^{-n};q)_k=(1-q^{-n})(1-q^{-n+1})(1-q^{-n+2})\dots (1-q^{-n+k-1}) $ so for $k=n+1$ we have $(q^{-n};q)_{n+1}=(1-q^{-n})(1-q^{-n+1})(1-q^{-n+2})\dots (1-1) =0$. We thus find that for $(q^{-n};q)_{k>n+2}=0$ and the sum becomes truncated at $k=n$:
\begin{equation}
    H_n(\cos \theta|q)=  \sum_{k=0}^n \frac{(q^{-n};q)_k}{(q;q)_k} (-1)^k q^{nk-\frac{k(k-1)}{2}} e^{i(n-2k)\theta}~.
\end{equation}
From this expression we can get another useful expression for $H_n(\cos \theta|q)$, as follows:
\begin{align*}
    (q^{-n};q)_k &= (1-q^{-n})(1-q^{-n+1})(1-q^{-n+2})\dots (1-q^{-n+k-1}) \\
    &= (-1)^k (q^{-n}-1)(q^{-n+1}-1)(q^{-n+2}-1)\dots (q^{-n+k-1}-1) \\
    &= (-1)^k\, q^{-n}(1-q^n)\,q^{-n+1}(1-q^{n+1}))\dots q^{-n+k-1}(1-q^{n-k+1}) \\
    &= (-1)^k\, q^{-nk+\frac{k(k-1)}{2}} \frac{(q;q)_n}{(q;q)_{n-k}}
\end{align*}
and thus, another expression for the q-Hermite polynomial is
\begin{equation} \label{qHermite_exp2}
    H_n(\cos \theta|q)=  \sum_{k=0}^n \frac{(q;q)_n}{(q;q)_k\,(q;q)_{n-k}} e^{i(n-2k)\theta}~.
\end{equation}
\subsection{Orthogonality relations}
A useful identity for q-Hermite polynomials with $x=\cos \theta, y=\cos \phi$ (see for example \cite{qHermite_formula}) is:
\begin{equation}
    \sum_{l=0}^\infty H_l(x|q)H_l(y|q) \frac{t^l}{(q;q)_l} = \frac{(t^2;q)_\infty}{(te^{i(\theta+\phi)};q)_\infty (te^{i(\theta-\phi)};q)_\infty(te^{-i(\theta-\phi)};q)_\infty (te^{-i(\theta+\phi)};q)_\infty},
\end{equation}
which for $t=1$ and $x=y$ is exactly the needed identity.
Noting that \\
$(te^{i(\theta+\phi)};q)_\infty (te^{-i(\theta+\phi)};q)_\infty = \prod_{k=0}^\infty (1-te^{i(\theta+\phi)}q^k)(1-te^{-i(\theta+\phi)}q^k)=|(te^{i(\theta+\phi)};q)_\infty|^2$ and similarly for $e^{i(\theta-\phi)}$, we have:
\begin{equation*}
    \frac{(t^2;q)_\infty}{|(te^{i(\theta+\phi)};q)_\infty|^2 |(te^{i(\theta-\phi)};q)_\infty|^2 } = \frac{(1-t^2)(qt^2;q)_\infty}{|1-te^{i(\theta+\phi)}|^2|1-te^{i(\theta-\phi)}|^2|(qte^{i(\theta+\phi)};q)_\infty|^2 |(qte^{i(\theta-\phi)};q)_\infty|^2},
\end{equation*}
where it is made clear that for $t=1$ the expression vanishes for all $\theta, \phi$ except for $\theta=\pm \phi$. Taking  $t=1-\epsilon$, in the limit $\epsilon \to 0$ $\theta \to \phi$ we have
\begin{align*}
    |1-te^{i(\theta-\phi)}|^2 &= 1-t[e^{i(\theta-\phi)}+ e^{-i(\theta-\phi)}] +t^2 \\
    &= 1-t[1+i(\theta-\phi)-\frac{(\theta-\phi)^2}{2}+ 1-i(\theta-\phi)-\frac{(\theta-\phi)^2}{2}+\dots] +t^2\\
    &= 1-(1-\epsilon)[2-(\theta-\phi)^2+O((\theta-\phi)^4)] +(1-\epsilon)^2 \\
    &= \epsilon^2 + (\theta-\phi)^2 + O(\epsilon (\theta-\phi)^2)+\dots
\end{align*}
and thus
\begin{align*}
    &\frac{(1-t^2)(qt^2;q)_\infty}{|1-te^{i(\theta+\phi)}|^2|1-te^{i(\theta-\phi)}|^2|(qte^{i(\theta+\phi)};q)_\infty|^2 |(qte^{i(\theta-\phi)};q)_\infty|^2} \\
    &\to \frac{2\epsilon}{\epsilon^2 + (\theta-\phi)^2} \frac{(q;q)_\infty}{|(e^{2i\theta};q)_\infty|^2|(q;q)_\infty|^2}~.
\end{align*}
In the limit $\epsilon \to 0$, the first term is recognized as a delta function $\delta(t)=\lim_{\epsilon\to 0} \frac{1}{\pi} \frac{\epsilon}{\epsilon^2+t^2}$ and taking into account also the possibility $\theta \to -\phi$ it is found that
\begin{equation} \label{qH_orthogonality}
    \sum_{l=0}^\infty \frac{H_l(x|q)H_l(y|q)}{(q;q)_l} = \frac{2\pi [\delta(\theta-\phi)+\delta(\theta+\phi)]}{|(e^{2i\theta};q)_\infty|^2(q;q)_\infty}~.
\end{equation}

\subsection{Orthogonality of the eigenvectors of DSSYK} \label{App:Orthogonality} 
In this appendix we show that the eigenvectors (\ref{Teigenvectors1}) satisfy orthogonality relations in energy and in chord number.  Firstly we check orthogonality in energy:
\begin{equation}
    \sum_{l=0}^\infty \psi_l(\cos \theta) \psi_l(\cos \phi) = \frac{(q;q)_\infty}{2\pi} |(e^{2i\theta};q)_\infty|^2 \sum_{l=0}^\infty   \frac{H_l(\cos \theta|q)H_l(\cos \phi|q)}{ (q;q)_l} = \delta(\theta-\phi)~,
\end{equation}
where we used the result (\ref{qH_orthogonality}) together with the knowledge that both $\theta$ and $\phi$ get values between $0$ and $\pi$. 
Secondly, we check orthogonality in chord number:
\begin{equation}
    \int_0^\pi d\theta \, \psi_n(\cos \theta) \psi_m(\cos \theta) = \frac{(q;q)_\infty}{\sqrt{ (q;q)_n (q;q)_m}} \int_0^\pi  \frac{d\theta}{2\pi} |(e^{2i\theta};q)_\infty|^2 H_n(\cos \theta|q)H_m(\cos \theta|q)~.
\end{equation}
To compute this we need the following identity:
\begin{equation} \label{exp_qH_identity}
    \int_0^\pi  \frac{d\theta}{2\pi} |(e^{2i\theta};q)_\infty|^2 H_n(\cos \theta|q)H_m(\cos \theta|q) = \frac{(q;q)_n}{(q;q)_\infty}\delta_{nm}~,
\end{equation}
which gives immediately
\begin{equation}
    \int_0^\pi d\theta \, \psi_n(\cos \theta) \psi_m(\cos \theta) = \delta_{nm} ~.
\end{equation}

\section{Continuum approximation of Krylov space}\label{Appx:Cont_approx}

This section provides a derivation of the continuum approximation of Krylov space developed in \cite{Barbon:2019wsy} in order to understand better its applicability and its relation to ballistic propagation.

In the case in which there are no diagonal Lanczos coefficients, $a_n=0$, it is useful to redefine the wavefunctions as
\begin{equation}
    \centering
    \label{Phi_redefinition_i_n}
    \phi_n(t)=i^n\varphi_n(t),
\end{equation}
so that $\varphi_n(t)\in\mathbb{R}$, satisfying the recurrence relation:
\begin{equation}
    \centering
    \label{recurrence_varphi_n}
    \dot{\varphi}_n(t) = b_n\varphi_{n-1}(t)-b_{n+1}\varphi_{n+1}(t),
\end{equation}
for all $n=0,...,K-1$ and with initial condition $\varphi_n(0)=\delta_{n,0}$. Note that, for the sake of rigor, we start by considering a finite Krylov space of dimension $K$; we shall eventually consider the limit of large $K$. For consistency, $b_0\equiv 0$ and $\varphi_{-1}(t)\equiv 0$.

We will start by performing some exact manipulations that amount to rewriting (\ref{recurrence_varphi_n}) in a more suggestive form. Since Krylov space is discrete, position is labelled by $n=0,...,K-1$ and the lattice spacing is constant and equal to $1$. Thus, strictly speaking, it does not go to zero in any sense. In order to make a continuum limit possible, we use a change of variables $\xi = \frac{n}{K-1}\equiv \varepsilon n$, where we identify the small parameter $\varepsilon=\frac{1}{K-1}$. As a result, position in Krylov space is denoted by a rescaled variable $\xi$ which takes discrete values contained in the interval $[0,1]$:
\begin{equation}
    \centering
    \label{xi_domain}
    \xi \in \left\{ \frac{n}{K-1},\; n = 0,1,...,K-1 \right\} = \left\{ \varepsilon n,\;n=0,1,...,\frac{1}{\varepsilon} \right\}.
\end{equation}
The spacing between the allowed values of $\xi$ is $\varepsilon=\frac{1}{K-1}$, and we note that in the limit $K\to+\infty$ they accumulate to $\xi\in [0,1]$, hence yielding a continuum limit. But let's still keep $K$ fixed.

An important technicality has to be considered before passing to the continuum limit. We need to assume that the Lanczos coefficients $b_n$ are given by the evaluation at discrete values of the domain of a certain smooth function that we loosely denote $b(n)$. This has to be assumed, since analytic continuation cannot be invoked because for it to apply one needs to have a function defined over at least an interval, not a discrete domain. Our Lanczos coefficients for DSSYK (\ref{Lanczos}) certainly satisfy this assumption, but there are examples in the literature of Lanczos sequences that do not fulfill it: the works \cite{Dymarsky:2021bjq,avdoshkin2022krylov,Camargo:2022rnt} have studied systems where the Lanczos sequence shows staggering, the even and odd coefficients following different profiles. For such systems, the Lanczos sequence does not have a well-defined continuum limit.

Upon the previous variable rescaling, we have:
\begin{equation}
    \centering
    \label{b_n_xi}
    b_n = b(n) = b\big( (K-1)\xi \big) =: w(\xi),
\end{equation}
where we have introduced the function $w$, defined over the rescaled compact domain (\ref{xi_domain}). 

Having assumed that the Lanczos coefficients $b_n$ are given by evaluating a smooth function $b(x)$ at discrete values $x=n$, it is now justified to make the same assumption for $\varphi_n(t) = \varphi(t,n)$. Then again, upon the variable rescaling:
\begin{equation}
    \centering
    \label{varphi_n_xi}
    \varphi_n(t) = \varphi(t,n) = \varphi\big(t,(K-1)\xi\big) =: f(t,\xi),
\end{equation}
where the domain for the second variable of $f$ is (\ref{xi_domain}). We can now rewrite (\ref{recurrence_varphi_n}) as:
\begin{equation}
    \centering
    \label{recurrence_f_xi}
    \partial_t f(t,\xi) = w(\xi) f(t,\xi - \varepsilon) -w(\xi + \varepsilon) f(t,\xi + \varepsilon).
\end{equation}
We emphasize that (\ref{recurrence_f_xi}) is an exact rewriting of the recurrence (\ref{recurrence_varphi_n}), with $\varepsilon=\frac{1}{K-1}$. A solution $f(t,\xi)$ of (\ref{recurrence_f_xi}) gives a solution for (\ref{recurrence_varphi_n}) by evaluating $\varphi_n(t)=f(t,\frac{n}{K-1})$.

Expanding (\ref{recurrence_f_xi}), which we can do since we have assumed that $w(\xi)$ is smooth and we have argued that $f(t,\xi)$ can also be taken to be so, we arrive at
\begin{equation}
    \centering
    \label{Chiral_wave_eqn_approx}
    \partial_t f(t,\xi) + v(\xi) \partial_\xi f(t,\xi) + \frac{1}{2}v^\prime (\xi) f(t,\xi) = \mathit{O}(\varepsilon^2),
\end{equation}
where we have defined $v(\xi):=2\varepsilon w(\xi)$. The leading term of (\ref{Chiral_wave_eqn_approx}) is a chiral wave equation with velocity field $v(\xi)$ and mass term $\frac{1}{2}v^\prime (\xi) f(t,\xi)$. This equation can be solved using a change of variables $\xi \mapsto \chi$ such that $\chi=0$ when $\xi = 0$ and $\frac{d\xi}{v(\xi)}=d\chi$. Since $v(\xi)$ is of order $\varepsilon$, this change of variables unwraps the compact domain of $\xi$ and has again a non-compact domain, but the discussion about the compact domain was technically necessary in order to explain in what sense there is a continuum limit, and why $\varepsilon=\frac{1}{K-1}$ controls the deviation from it. Introducing additionally the function 
\begin{equation}
    \centering
    \label{g_chi_def}
    g(t,\chi):= \sqrt{v(\xi(\chi))} f(t,\xi(\chi)),
\end{equation}
equation (\ref{recurrence_f_xi}) simplifies further into:
\begin{equation}
    \centering
    \label{wave_eqn_g_chi}
    \left(\partial_t + \partial_\chi \right)g(t,\chi) = \mathit{O}(\varepsilon^{5/2}),
\end{equation}
where we remind that $\varepsilon=\frac{1}{K-1}$. In the limit $K\to\infty$ equation (\ref{wave_eqn_g_chi}) becomes exactly a chiral wave equation with unit velocity for the wave function $g$ with coordinates $t,\,\chi$. However, this does not imply that the continuum approximation of Krylov space becomes exact and indistinguishable from the discrete problem in the large-$K$ limit: The intermediate steps of rescaling $n$ to a compact domain $\xi = \varepsilon n$ fail strictly speaking when $\varepsilon=0$, and the continuum approximation need not apply in general. It may be applicable for propagating wave packets whose typical length scale is larger than the lattice spacing, but in our cases of interest the initial condition for the discrete wave packet is $\varphi_n(t=0)=\delta_{n0}$, which therefore probes the discreteness of the Krylov chain at early times.

Interestingly, studying (\ref{wave_eqn_g_chi}) we note that this continuum approximation behaves as a classical limit, because it results in ballistic propagation, as we shall explain. Given an initial condition $g(0,\chi)=g_0(\chi)$, which in our case will be proportional to a delta function centered at $\chi=0$, the generic solution of (\ref{wave_eqn_g_chi}) simply propagates this packet to the right, without spreading it:
\begin{equation}
    \centering
    \label{g_solution}
    g(t,\chi)=g_0(\chi - t).
\end{equation}
Since the packet $g(t,\chi)$ is a propagating delta function, K-complexity coincides with the position of its peak, which propagates at velocity $1$ in $\chi$-space. Undoing the change of variables:
\begin{equation}
    \centering
    \label{packet_peak_chi_xi}
    t = \int_0^{\chi(t)}d\chi = \int_{0}^{\xi(t)}\frac{d\xi}{v(\xi)},
\end{equation}
and using $v(\xi)=2\varepsilon w(\xi)= 2 \varepsilon b(\xi / \varepsilon)$, together with $\xi = \varepsilon n$, we reach the equation\footnote{We thank Jos\'e Barb\'on for pointing out this expression to us.}: 
\begin{equation}
    \centering
    \label{n_peak_cont_estimate}
    t = \int_0^{n(t)}\frac{dn}{2b(n)},
\end{equation}
which can be used as an estimate for the position of the peak $n(t)$ in $n$-space, in an approximation in which $n$ is directly promoted to be a continuous variable.

We remark that the continuum approximation gives ballistic propagation as an output, rather than it being an additional, independent assumption. In this sense, this approximation can be thought of as a classical approximation, because the propagation of the wave packet has turned into the problem of the propagation of a localized particle with well-defined position that travels driven by a velocity field $2 b(n)$ in $n$-space.

\section{Wavefunctions and Krylov complexity for specific values of q} \label{App:Wavefunctions}
In this appendix we evaluate the exact wavefunctions, $\phi_n(t)$, for $q=0$ and for $q=1$. For $q=0$ we can use the result (\ref{phin_closedform}) by plugging $q=0$ directly into the equation.  For $q=1$ we  start from the eigensystem equation with Lanczos coefficients $b_n \sim \sqrt{n}$ and find the eigenvectors in order to compute $\phi_n(t)$.  The analysis of $q\to 1$ is done in the main text.
\subsection{Exact wavefunctions and K-complexity for q=0} \label{App:WFq0}

For $q\to 0$ the Lanczos coefficients are given by $b_n\sim\frac{J}{\sqrt{\lambda}}\equiv b$ ($\to 0$) for all $n=0,1,2,\dots$. When the tridiagonal matrix with these Lanczos coefficients above and below the diagonal is a Liouvillian, the result for $\phi_n(t)$ was found in \cite{Barbon:2019wsy} to be
\begin{equation} \label{phin_q0}
    \phi_n(t) = i^n \varphi_n(t)= i^n\frac{(n+1)}{bt} J_{n+1}(2bt) ~,
\end{equation}
whereas in our case (dynamics in the Schrödinger picture) we need to replace $t\to-t$.
We can arrive at this result from (\ref{phin_closedform}) by setting $q=0$:
\begin{align}
    (0;0)_\infty &= \prod_{k=0}^\infty (1-0\cdot 0^k)=1 = (0;0)_n \\
    (e^{2i\theta};0)_\infty &= 1-e^{2i\theta} \Rightarrow |(e^{2i\theta};0)_\infty|^2 = 4 \sin^2\theta \\
    H_n(\cos\theta|0) &= \sum_{k=0}^n \frac{(0;0)_n}{(0;0)_k(0;0)_{n-k}} e^{i(n-2k)\theta} = \sum_{k=0}^n e^{i(n-2k)\theta} = \frac{\sin[(1+n)\theta]}{\sin \theta}~,
\end{align}
where we used (\ref{qHermite_exp2}) to evaluate $H_n(\cos \theta|0)$. Plugging all this into (\ref{phin_closedform}) we have
\begin{align}
    \phi_n(t) &= \frac{2}{\pi} \int_0^\pi d\theta e^{-2ibt \cos \theta} \sin \theta \sin [(1+n)\theta] \nonumber\\
    &= \frac{1}{\pi} \int_0^\pi d\theta\, e^{-2ibt \cos \theta} \cos(n\,\theta) - \frac{1}{\pi} \int_0^\pi d\theta\, e^{-2ibt \cos \theta} \cos[(n+2)\,\theta] \nonumber\\
    &= i^n J_n(-2bt) + i^{n+2} J_{n+2}(-2bt) = i^n\Big[ J_n(-2bt)+J_{n+2}(-2bt)\Big] \nonumber\\
    &= i^n \frac{(n+1)}{(-bt)}J_{n+1}(-2bt) ~,\label{Bessel_func}
\end{align}
where in the second line we used the trigonometric identity $\sin \alpha \sin \beta = \frac{1}{2} [\cos(\alpha-\beta)-\cos(\alpha+\beta)]$, in the third line we used one of the integral representations of the Bessel function of the first kind\footnote{\url{https://dlmf.nist.gov/10.9}.} $J_n(z) = \frac{{i^{-n}}}{\pi} \int_0^\pi d\theta \, e^{i z \cos \theta} \cos(n\theta)$, and in the fourth line we used the Bessel function identity\footnote{\url{https://functions.wolfram.com/Bessel-TypeFunctions/BesselJ/17/01/01/}.} $J_n(z)+J_{n+2}(z)=\frac{2(n+1)}{z}J_{n+1}(z)$.  We managed to recover the result (\ref{phin_q0}) with $t\to -t$ as required.

With the wavefunction given by (\ref{Bessel_func}) the result for Krylov complexity can be evaluated exactly as follows:
\begin{align}
    C_K(t) &= \sum_{n=0}^\infty n |\phi_n(t)|^2 = \frac{1}{(bt)^2} \sum_{n=0}^\infty n (n+1)^2 J^2_{n+1}(-2bt)=\frac{1}{(bt)^2} \sum_{n=1}^\infty (n-1)n^2J_n^2(2bt) \nonumber\\
    &=\frac{16}{3}(bt)^2[J_0^2(2bt)+J_1^2(2bt)]+J_0^2(2bt)+\frac{1}{3}J_1^2(2bt)-\frac{8}{3}btJ_0(2bt)J_1(2bt)-1.
\end{align}
Using the asymptotic forms of $J_0$ and $J_1$,
\begin{eqnarray}  
J_0(x)&\approx& \sqrt{\frac{2}{\pi x}}\cos(x-1/4\pi)~,\\
J_1(x)&\approx& \sqrt{\frac{2}{\pi x}}\sin(x-1/4\pi)~,
\end{eqnarray}
gives
\begin{eqnarray}\label{KC_linear}
    C_K(t) &\approx& \frac{16}{3\pi}bt+\frac{4}{3\pi}\cos(4bt)-1+O(1/t)~,
\end{eqnarray}
which is linear in $t$ for large $t$.

\subsection{Exact wavefunctions for q=1}

For $q\to 1$ the Lanczos coefficients are given by $b_n\sim J \sqrt{\frac{n}{\lambda}}$, and the eigensystem equation is $T|v(E)\rangle=E |v(E)\rangle$. Re-defining the energy variable as $E=\frac{J}{\sqrt{\lambda}}\varepsilon$, the equation can be written component-wise as
\begin{equation}
    \varepsilon\,v_l(\varepsilon) = \sqrt{l} \, v_{l-1}(\varepsilon)+\sqrt{l+1} \, v_{l+1}(\varepsilon)~.
\end{equation}
In a similar manner to the general $q$ case, we redefine the eigenvector components as
\begin{equation}
v_l(\varepsilon)=\frac{\tilde{v}_l(\varepsilon)}{\sqrt{l!}}~,
\end{equation}
which leads to the equation
\begin{equation}
    \varepsilon\, \tilde{v}_l(\varepsilon) = l\, \tilde{v}_{l-1}(\varepsilon) + \tilde{v}_{l+1}(\varepsilon) ~.
\end{equation}
This equation is solved by the (regular) Hermite polynomials given by $H_{e_n}(x)=2^{-n/2} H_n(x/\sqrt{2})$. We thus have 
\begin{equation}
    v_l(\varepsilon)= \frac{H_{e_l}(\varepsilon)}{\sqrt{l!}}= \frac{1}{2^{l/2}\sqrt{l!}}H_l\left(\frac{\varepsilon}{\sqrt{2}}\right) ~.
\end{equation}
As in the q-Hermite case, we will use Hermite identities to normalize the eigenvectors.
\begin{equation}
   \sum_{l=0}^\infty v_l(\varepsilon_1) v_l(\varepsilon_2) = \sum_{l=0}^\infty \frac{H_l(\varepsilon_1/\sqrt{2})H_l(\varepsilon_2/\sqrt{2})}{2^l \, l!} = \sqrt{2\pi} e^{\frac{\varepsilon_1^2}{2}}\delta(\varepsilon_1-\varepsilon_2),
\end{equation}
where we used the identity\footnote{\url{https://functions.wolfram.com/Polynomials/HermiteH/23/02/0011/}.}: $\sum_{n=0}^\infty \frac{H_n(x)H_n(y)}{2^n \, n!}=\sqrt{\pi} e^{\frac{1}{2}(x^2+y^2)}\delta(x-y)$ with $x=y=\varepsilon/\sqrt{2}$ (using $\delta((\varepsilon_1-\varepsilon_2)/\sqrt{2})=\sqrt{2}\delta(\varepsilon_1-\varepsilon_2)$). The normalized eigenvectors are then given by
\begin{equation}\label{evecs_q_1}
    \psi_l(\varepsilon)= \frac{1}{(2\pi)^{1/4}2^{l/2}\sqrt{l!}}\, e^{-\varepsilon^2/4}\, H_l\left(\frac{\varepsilon}{\sqrt{2}}\right)~.
\end{equation}
This result can be used directly to find the result (\ref{Heisenberg_Weyl_Wave_Fn}) for the wavefunction in the case $q=1$:

\begin{align}
     \phi_l(t) &= \int_{-\infty}^\infty d\varepsilon\, e^{-i\frac{J}{\sqrt{\lambda}}t\varepsilon} \psi_l(\varepsilon)\psi_0(\varepsilon) = \frac{1}{\sqrt{2\pi}\, 2^{l/2} \sqrt{l!}}  \int_{-\infty}^\infty d\varepsilon\, e^{-i\frac{J}{\sqrt{\lambda}}t\varepsilon} \, e^{-\frac{\varepsilon^2}{2}} \, H_l\left(\frac{\varepsilon}{\sqrt{2}}\right) \\
     &= e^{-\frac{(\gamma t)^2}{2}} \frac{(-i\,\gamma t)^l}{\sqrt{l!}},
\end{align}
where $\gamma = \frac{J}{\sqrt{\lambda}}$. Here, the identity\footnote{\url{https://functions.wolfram.com/Polynomials/HermiteH/21/02/01/0002/}.} $\int_{-\infty}^\infty dx \, e^{-(z-x)^2}H_n(x)=2^n \sqrt{\pi} z^n$, was used. We have also assumed that the density of states is flat in $\varepsilon$; this has not been proved here, but it can be checked a posteriori by noting that the orthonormal eigenvectors (\ref{evecs_q_1}) indeed satisfy the completeness relation with respect to the flat measure $d\varepsilon$.

\bibliography{references}

\end{document}